\RequirePackage{fix-cm}
\documentclass[twocolumn,epjc3]{svjour3}  
\smartqed  
\pdfoutput=1 

\usepackage[T1]{fontenc} 
\usepackage{color}   
\usepackage{graphicx,float}
\usepackage{dcolumn}
\usepackage{bm}
\usepackage{slashed}
\usepackage{float}
\usepackage{amsmath}
\usepackage{amssymb}

\journalname{{}}
\begin{document}

\def\bb    #1{\hbox{\boldmath${#1}$}}
 \def\oo    #1{{#1}_0 \!\!\!\!\!{}^{{}^{\circ}}~}  
 \def\op    #1{{#1}_0 \!\!\!\!\!{}^{{}^{{}^{\circ}}}~}
\def\sD{D \!\!\!\!/}
\def\sd{\partial \!\!\!\!/}
\def\qcd{{{}^{\rm QCD}}}   
\def\qed{{{}^{\rm QED}}}   
\def\2d{{{}_{\rm 2D}}}         
\def\4d{{{}_{\rm 4D}}}         

\def\qcdu{{{}_{ \rm QCD}}}   
\def\qedu{{{}_{\rm QED}}}   
\def\qcdd{{{}^{ \rm QCD}}}   
\def\qedd{{{}^{\rm QED}}}   
\def\sg#1{ {\rm \,sign}(#1)\, } 

\title{\boldmath   Dynamics of quarks and gauge fields in the lowest-energy
  states in QED and QCD interactions} 



\titlerunning{quarks in QED and QCD interactions}        

\author{Cheuk-Yin Wong$^{{\rm 1,a,b}}$,  Andrew V. Koshelkin$^{\rm 2}$}

\institute{Physics
  Division, Oak Ridge National Laboratory, Oak
  Ridge, Tennessee 37831, USA \\
$^{\rm 2}$National Research Nuclear University MEPhI, 
 Moscow 115409, Russia}

\thankstext{t1}{Cheuk-Yin\,Wong's research  has been supported  in part by
  UT-Battelle, LLC, under contract DE-AC05-00OR22725 with the US
  Department of Energy (DOE). The US government retains and the
  publisher, by accepting the article for publication, acknowledges
  that the US government retains a nonexclusive, paid-up, irrevocable,
  worldwide license to publish or reproduce the published form of this
  manuscript, or allow others to do so, for US government
  purposes. DOE will provide public access to these results of
  federally sponsored research in accordance with the DOE Public
  Access Plan (http://energy.gov/downloads/doe-public-access-plan),
  Oak Ridge, Tennessee 37831, USA}
\thankstext{e1}{e-mail:
  wongc@ornl.gov (corresponding author)}


\date{}

\maketitle

\begin{abstract}
{ We examine the dynamics of quarks and gauge fields in the lowest
  energy states in the QED and
  QCD interactions  by combining Schwinger's longitudinal
  confinement in (1+1)D with Polyakov's transverse confinement in
  (2+1)D in a ``stretch (2+1)D'' flux tube model in (3+1)D.  For such QED and QED systems in the flux tube configuration with cylindrical symmetry, we separate out
  the transverse and longitudinal degrees of freedom,
approximate the
  non-Abelian QCD in the quasi-Abelian approximation, and  solve the
  derived equations to study the collective excitations.  We find
  stable collective QED and QCD excitations showing up as confined QED
  and QCD mesons, in support of previous theoretical 
  studies and recent observations of the  anomalous hypothetical X17 and E38 particles.
 Future theoretical lattice gauge calculations of QED in (3+1)D with the  inclusion of the Schwinger 
 longitudinal confinement mechanism
and  experimental confirmation of  the  hypothetical  X17 and E38 particles 
will shed definitive light
on quark confinement in the QED interaction  in (3+1)D.
 }
\end{abstract}



\maketitle \flushbottom

\setcounter{footnote}{0}

\section{Introduction}

Recent observations of possible neutral bosons with masses in the
region of many tens of MeV have generated a great deal of 
interests.  Specifically, (i) the observation of the
anomalous soft photons with $p_T \lesssim 60$ MeV/c  in high-energy hadron-hadron collisions
\cite{Chl84,Bot91,Ban93,Bel97,Bel02pi,Bel02,Per09} and high-energy
$e^+e^-$ collisions \cite{Per09,DEL06,DEL08,DEL10}, (ii) the
observation of the hypothetical  X17 particle at  $\sim$17 MeV in the $e^+e^-$ decay of low-energy proton-fusion 
of light $\alpha^n$ nuclei
\cite{Kra16,Kra19,Fir20,Kra21,Sas22,Kra22,x17Kra,Kra23,Tra23}, and (iii) the observation of resonance-like 
structures of 
diphoton invariant masses  at $\sim$17 and  $\sim$38 MeV \cite{Abr12,Abr19,Abr23} in $pA$ and $dA$ collisions at $p_{\rm lab}$ of a few GeV/c per nucleon, point to the possible 
existence  of neutral particles with masses  of many tens of
MeV.   As they lie 
 outside the known
families of the Standard Model, 
they may be  anomalous particles with unknown physical origins.
If these anomalous  particles  are further confirmed by future  experimental investigations, 
 they may have important implications on the physics
beyond the Standard Model and on Cosmology. 

The observations of  the anomalous soft photons
refer to the occurrence  of excess $e^+e^-$ yields associated with hadron production
in high-energy collisions. 
The yield of    $e^+e^-$ pairs with $p_T \lesssim 60$ MeV/c, in excess  of Standard Model QED bremsstrahlung predictions by a factor of about 4,
 was first observed in 1984 by the WA27 Collaboration at
CERN, using the BEBC bubble chamber in $K^+ p$ collisions at $p_{\rm
  lab}(K^+)$=70 GeV/c \cite{Chl84}.  
  Since then, it has been confirmed by many exclusive measurements in high-energy $K^+$$p$
\cite{Bot91}, $\pi^+ p$ \cite{Bot91}, $\pi^- p$
\cite{Ban93,Bel97,Bel02pi}, $pp$ collisions \cite{Bel02}, and
$e^+$$e^-$ collisions \cite{Per09,DEL06,DEL08,DEL10}.
The series of  DELPHI measurements in $e^+$$e^-$ annihilations at the $Z^0$ resonance energy  showed quantitatively 
  that whenever hadrons are produced, anomalous
soft photons in the form of excess $e^+e^-$ pairs with $p_T \lesssim 60$ MeV/c  are proportionally produced,  and   when
hadrons are not produced, these anomalous soft photons are also not
produced \cite{DEL08,Per09,DEL10}.  
The transverse momenta of these
excess $e^+$$e^-$ pairs lie in the range of a few MeV/c to many tens
of MeV/c, which indicate that significant components of  anomalous soft photons  are likely   massive neutral bosons with rest masses in the region of  many tens
of MeV.  They have been interpreted as arising from a cold quark gluon
plasma \cite{Van89,Lic94,Kok07}, pion condensate \cite{Bar89}, pion
reflection \cite{Shu89}, 
corrections to bremsstrahlung \cite{Bal90}, color flux tube particle production \cite{Czy94}, 
stochastic QCD vacuum \cite{Nac94,Leb22},
ADS/CFT supersymmetric Yang-Mills theory \cite{Hat10}, Unruh radiation
\cite{Dar91}, closed quark-antiquark loop \cite{Sim08}, QED-confined
$q\bar q$ composite particles 
\cite{Won10,Won11,Won14,Won20,Won22,Won22a,Won22b,Won22c,Kos22}, and
induced currents in the Dirac sea \cite{Kha14}.

  In the  QED-confined
$q\bar q$ composite particle
description of the anomalous soft photons 
in  \cite{Won10,Won11,Won14,Won20,Won22,Won22a,Won22b,Won22c,Kos22},
 it was proposed that  
a quark and an antiquark can form massive QED-confined neutral bosons
 through their mutual QED interactions by   the Schwinger confinement mechanism in (1+1)D QED \cite{Sch62,Sch63}.  These 
QED-confined
$q\bar q$ composite bosons  can be  called   ``QED mesons''
in analogy with the QCD-confined  QCD mesons.   The QED mesons may be simultaneous produced in conjunction with  QCD mesons, and 
the $e^+e^-$ decays of the produced QED  mesons 
may be the source of the observed anomalous soft photons associated with  QCD meson production.    The masses of  isoscalar and isovector 
QED mesons were  predicted  to be 12.8 MeV  and 38.6 MeV
respectively in the limit of massless quarks 
  (Table I of \cite{Won10}).  
 
In another series of experiments in search of possible candidate
particles for the axion \cite{Don78}, de Boer, Krasnarhokay, and
collaborators studied the $e^+ e^-$ spectrum in low-energy  proton-fusion production  of   light
$\alpha^n$ nuclei with  $n$=1, 2 and 3.  They reported the observation of a neutral
particle at 12 $\pm$ 2.5 MeV in the decay of the excited 1$^+$
isovector 17.64  $^8$Be  state \cite{Vit08}.  Subsequently,
Krasznarhokay and collaborators at ATOMKI continued the search with an
improved $e^+e^-$ spectrometer.  They reported the observation of a
hypothetical neutral ``X17'' boson with a mass of
16.70$\pm$0.35(stat)$\pm$0.5(syst) MeV    (i) in the decay of  the   18.15 MeV
$I(J^\pi)$=$0(1^+) $ excited   $^8$Be state to the $^8$Be ground
state \cite{Kra16},    (ii)  in the decay of the  18.05 MeV $I(J^\pi)$=$0(0^-)$ excited
$^4$He state to the  $^4$He  ground state \cite{Kra19,Kra21},  (iii) in  the decay of the 
off-resonance excited $^8$Be states  
to the $^8$Be ground state  \cite{Sas22}, and  (iv) in the decay of the
17.23 MeV $I(J^\pi)$=$1(1^-)$ excited  $^{12}$C state to the $^{12}$C ground state
\cite{Kra22}. 
In a recent preliminary measurement with a new $e^+$-$e^-$ spectrometer to study the decay of the excited $^8$Be 18.15 MeV  state 
in the proton-fusion of $^7$Li,  
 the HUS Collaboration reported the observation of 
a significant anomaly ($>4\sigma$) indicating  the decay of a hypothetical boson with a mass of 16.7$\pm 0.4$ MeV into $e^+e^-$, in agreement with the earlier ATOMKI results published in 2016 for the hypothetical X17 particle \cite{Tra23}.
 An additional support for the hypothetical  X17 particle comes from
the $p_T$ spectrum of the anomalous soft
photons in $pp$ collisions at
$p_{\rm lab}(p)$= 450 GeV/c \cite{Bel02},  in the
thermal model of the transverse momentum distribution \cite{Won20,Dyo21,Dyo21a}.
There is a resonance-like structure in the mass
region of 10-20 MeV, with a peak to valley ratio of about 3, in the
diphoton invariant mass spectrum obtained by the COMPASS Collaboration
in the $p p \to p \pi^-\pi^+ (\gamma \gamma) p$ reaction at $p_{\rm
  lab}(p)$=190 GeV/c \cite{Ber11,Ber14,Sch11,Sch12,Ber12}.  The 
 resonance-like structure suggests it useful to re-examine whether it may be a
signal for the hypothetical X17 particle.  Other observations of resonance structures in 
$e^+$$e^-$  invariant masses between 3 to 20 MeV have been
reported in the collisions of high-energy nuclei with emulsion
detectors \cite{El88,El96,Jai07,deB11}.  

The ATOMKI observation of such a hypothetical  X17 particle with a mass of about 17
MeV has generated a great deal of interest \cite{Won10,Won11,Won14,Won20,Won22,Won22a,Won22b,Won22c,Kos22,X1722,Zha17,Alv18,Fen16,Bat15,For17,Bor19,Bor19a,Cha22,Ros17,Kub22,Ell16,Ban18,Pad18,Viv22,Viv22a}.  Although the 
mass of the hypothetical X17 particle  was close to the  
isoscalar QED meson  predicted earlier in \cite{Won10}, the  hypothetical X17 boson led to many speculations inside and outside of the
Standard Model, as discussed in the Proceedings of the Workshop on
``Shedding lights on the X17'' \cite{X1722}.  
Among the proposed models
were 
the
QED  meson
\cite{Won10,Won11,Won14,Won20,Won22,Won22a,Won22b,Won22c,Kos22},
the axion \cite{Alv18}, the fifth force of Nature \cite{Fen16}, 
a dark photon \cite{Bat15}, new
physics particles \cite{For17}, 
the Framed Standard Model \cite{Bor19,Bor19a,Cha22}, 
Higgs doublet \cite{Ros17},
a 12-quark-state \cite{Kub22},
 and a
light pseudoscalar \cite{Ell16},
  Because the hypothetical X17 may be a new
fundamental particle beyond the Standard Model with usual properties,
the confirmation of the  hypothetical X17 particle is being actively pursued by many
laboratories \cite{X1722}, including ATOMKI \cite{x17Kra,Kra23}, Dubna
\cite{x17Abr}, HUS \cite{Tra23}, STAR \cite{x17STAR}, MEGII \cite{x17MEG}, TU Prague
\cite{x17Prague,Luz23}, NTOF \cite{x17NTOF}, NA64 \cite{x17NA64}, INFN-Rome
\cite{x17INFNRome}, NA48 \cite{x17NA48}, Mu3e \cite{x17Mu3e},
MAGIX/DarkMESA \cite{x17MAGIX}, JLAB PAC50 \cite{x17JLAB,x17JLAB1},
PADME \cite{x17PADME,Rag23}, DarkLight \cite{ARIEL,Tre22,Viv22V}, LUXE
\cite{Hua22}, FASER \cite{Fen23}, and ANU/UM \cite{Kib23}.

In  another set of independent experiments,\break Abraamyan and
collaborators at Dubna have been using the two-photon decay of a
neutral boson to study the resonance structure of the lightest hadrons
near their energy thresholds.  Upon the suggestion of van Beveran and
Rupp \cite{Bev11,Bev12,Bev12a,Bev20}, the Dubna Collaboration undertook a search
for a hypothetical E38 particle using the two-photon decay channel.  The search
was carried out in $d$(2.0 GeV/n)+C, $d$(3.0 GeV/n)+Cu and $p$(4.6
GeV)+C reactions with internal targets at the JINR Nuclotron.  They
observed that the invariant masses of the two-photon distribution in
these reactions exhibit a resonance structure at around 38 MeV
\cite{Abr12,Abr19,Abr23}.  
In a recent analysis in the diphoton spectrum extended down to the lower invariant mass region, 
 the Dubna Collaboration reported the observation of resonance-like structures
both  at $\sim$17 and  $\sim$ 38 MeV in the same experimental set-up,  in support of earlier ATOMKI observation of the hypothetical X17 particle and the 
earlier Dubna observation of the 
hypothetical E38 particle  \cite{Abr23}.
A  supporting indirect signal for the hypothetical E38
particle may come from the $p_T$ spectrum of anomalous soft photons in
$e^+e^-$ annihilation at $Z^0$ resonance energy of $\sqrt{s}=91.18 $
GeV, which is consistent with the production of a neutral boson with a
mass of about 38 MeV in the thermal model of the transverse momentum
distribution \cite{Won20,Dyo21,Dyo21a}.  Another possible signal may be the prominent 
resonance structure at 38 MeV in the diphoton invariant mass spectrum
in PbPb collisions at $\sqrt{s_{_{\rm NN}}}=2.76$ TeV (Fig.\ 5.6 in Ref. \cite{Sno12}) obtained by the CMS
Collaboration \cite{CMS13}.   
 It will be of great interest to examine whether 
the peak structure at 38 MeV may be a signal for the E38 particle,
as the corrections from the conversion of the  two $\pi^0$-decay photons
 do not appear to exhibit a  peak structure in the  invariant mass spectrum
 at the mass region of 38 MeV (as indicated in Fig.\ 6.6 of 
\cite{Sno12}).
Another possible E38 signal may be
the resonance structure of the diphoton invariant mass at 38 MeV in the $pp \to
pp \pi^+ \pi^-(\gamma \gamma) $ reaction at $p_{\rm lab}(p)$= 190
GeV/c \cite{Ber11,Ber14,Sch11,Sch12} and the $\pi^- p \to \pi^- p_{\rm
  slow} \pi^+ \pi^-(\gamma \gamma) $ reaction at $p_{\rm lab}(\pi^-)$=
190 GeV/c, obtained by the COMPASS Collaboration, as pointed out by
\cite{Bev11,Bev12,Bev12a,Bev20}.  
Theoretically, the masses of the hypothetical X17 and E38 particles are
close to the mass of the QED-confined $q\bar q$ isoscalar and isovector composite
particle predicted earlier in Table I of \cite{Won10} and discussed in
\cite{Won11,Won14,Won20,Won22,Won22a,Won22b,Won22c,Kos22}.  The hypothetical E38
particle was considered in the theory of SO(4,2) conformal symmetric
model with anti-De-Sitter background geometry
\cite{Bev11,Bev12,Bev12a,Bev20}.

While many different theoretical interpretations have been presented
for the anomalies, the only theoretical interpretation that
may link these three anomalies together in a single framework is the
open string $q\bar q$ meson model presented in
\cite{Won10,Won11,Won14,Won20,Won22,Won22a,Won22b,Won22c,Kos22}.  Such
a meson model for the lowest-energy QCD and QED $q\bar q$ systems is
based on the Schwinger confinement mechanism \cite{Sch62,Sch63} in
which a quark and an antiquark are confined and bound by QCD and QED
interactions as $q\bar q$ open strings in (1+1)D.  
QCD and QED mesons in (1+1)D may represent physical QCD and QED mesons in (3+1)D 
when the radius of the flux tube is properly taken into account \cite{Won10,Won20}. 
Such a 
phenomenological open string meson model was found to describe  adequately
$\pi^0, \eta$, and $\eta'$ in the QCD sector.  By extrapolating into
the $q\bar q$ QED sector in which a quark and an antiquark interact
with the QED interaction alone, an open string isoscalar
$I(J^\pi)$=$0(0^-)$ QED meson state was found to be located at about
17 MeV and an isovector $(I(J^\pi)$=$1(0^-), I_3$=0) QED meson state
at about 38 MeV \cite{Won20}.  The predicted masses of the isoscalar and isovector
QED mesons are close to the masses of the anomalous soft photons and
the hypothetical X17 and E38 particles observed experimentally, making
them good candidates to describe these particles.

Although the open-string model of QCD and QED $q\bar q$ mesons appears
to be phenomenologically successful in describing the spectrum of the
lowest-energy neutral QCD and QED mesons \cite{Won20}, the theoretical foundations
for the QCD string on the one hand and the QED string  on the other hand  rest on different grounds with 
different strengths of theoretical
supports.  The QCD open-string description in (3+1)D has a long history since
the early development of the Veneziano amplitude \cite{Ven68}, Nambu-Goto string model
\cite{Nam70,Nam74,Got71}, the non-Abelian gauge theory
\cite{Gro73,Pol73},  the lattice gauge theory \cite{Wil74}, and the compactification of QCD from (3+1)D to (1+1)D in the large $N_c$ limit \cite{tho74}.    It
stands on firm theoretical grounds and needs not be belabored again.
However, the QED string in (3+1)D receives supports from the Schwinger
confinement mechanism in (1+1)D QED  \cite{Sch62,Sch63} and its phenomenological success to describe  the observed
spectrum of the anomalous particles \cite{Won10,Won20}.  A strong theoretical
foundation for the QED $q\bar q$ string in (3+1)D is, however, sorely lacking.  
On the contrary   from the viewpoint of lattice gauge calculations, the proposed confinement of   $q\bar q$ in QED in (3+1)D 
appears to
contradicts the well-known result that   quarks  interacting in QED
belong to the weak-coupling deconfinement regime
 in compact QED in (3+1)D in lattice gauge
calculations
\cite{Wil74,Cas74,Kog75,Man75,Pol77,Pol81,Pol87,Ban77,Gli77,Pes78,Dre79,Gut80,Kon98,Arn03,Lov21,Mag20}.
According to lattice gauge calculations,  a quark and an
antiquark are deconfined in (3+1)D  if they interact in the QED interaction alone.

In the presence of the above contradicting lattice gauge predictions on quark deconfinement  in QED in (3+1)D on the one hand, 
and the 
phenomenological  description of  
the observed anomalous particles 
as QED-confined $q\bar q$ states in (3+1)D  \cite{Won10,Won20}
on the other hand,
we wish to inquire first of all whether a quark and an antiquark could ever be produced and to interact
in QED alone to allow possible production of  QED-confined $q\bar q$ states.
As it turns out that a quark and an antiquark pair could in principle be produced and could interact
in  QED alone, we 
question next whether the lattice
gauge results of a quark and an antiquark deconfined in compact QED in
(3+1)D be consistent with the experimental absence of isolated
fractional charges.  We would like to consider further whether there may
be important elements missing in the present-day lattice gauge
calculations for which the conclusion of quark deconfinement in QED in
(3+1)D has been reached.  We would like to explore  subsequently a
``stretch (2+1)D'' model of the creation and the interaction of a quark and antiquark in
(3+1)D to study whether the inclusion of the missing element of the
Schwinger confinement mechanism may lead to $q\bar q$ confinement in QED
in (3+1)D.  The present  study with the constructed stretch (2+1)D model may not solve the problem completely, but it may 
bring us one step closer to answer, at least partially, the central
question whether a quark and an antiquark are confined in QED in
(3+1)D.

Although our search is focused predominantly on the theoretical
foundation of quark confinement in the QED interaction in (3+1)D, such
a study will benefit from the analogous study in the QCD interaction
and vice versa.  We shall therefore include the QCD interaction also
in our investigation when it is appropriate to do so.
For brevity of phraseology, we shall use the shorter  phrase 
``in QCD''
for 
``in the QCD interaction'' and ``in QED'' for  ``in the QED interaction'', respectively.

Recently, there has been renew interest in generalizing the Schwinger
and the Sommerfeld models in (1+1)D to multiflavor massive fermions
and vector fields \cite{Geo19,Geo19a,Geo20,Geo22,Geo22a}.  The Schwinger model with two flavors
and small and opposite fermion masses in (1+1)D has been found to be
a non-trivial example of unparticle physics that may provide an
interesting (1+1)D laboratory to study the physics of the interacting
unparticle stuff \cite{Geo22}.  An important part of the unparticle
physics is its connection to the Standard Model whose particles reside
in (3+1)D.  The connection between the (1+1)D dynamics and the (3+1)D
dynamics examined here may help facilitate a similar connection of
Georgi's generalized Schwinger toy model of unparticle physics in
(1+1)D and the (3+1)D in the Standard Model.

The detail mechanism how the dynamics of a quark and an antiquark in
the (3+1)D may be connected to the dynamics in (1+1)D may also be of
interest when we compare and contrast similar connections in quantum
vortex strings where the spectrum in (3+1)D resembles the spectrum in
(1+1)D \cite{Dor99,Han04,Ton08}.  The connection may also be related
to the study of the effective string theory with long strings in
(3+1)D where the quark confinement is assumed from the outset
\cite{Lus81,Pol91,Pol92,Aha09,Aha11,Aha13,Dub12}.  If the quark
confinement occurs not only in the QCD interaction but also in the QED
interaction in (3+1)D as indicated by the experimental observation of
the hypothetical X17 and the E38 particles, the effective string theory may be
further extended to include effective strings from the QED
interaction.

As discussed in detail in \cite{Won22}, quark confinement is a rather
peculiar property.  The peculiarity of the quark confinement property
requires that  in order for a quark and an antiquark to be observable
and subject to examination for low energy $q\bar q$ states, the quark and the antiquark must be in one
of the confined and bound eigenstates at the eigenenergies of the $q\bar
q$ system.  At all other energies, the system of a quark and an
antiquark does not exists, neither as a quark-antiquark bound state
nor a continuum state of an isolated quark and antiquark.  We are
therefore required to limit our attention only to those $q \bar q$
systems that are already known to be confined and bound or could potentially be
 confined and bound at appropriate eigenenergies.  
Such a logical
circularity arises because of the peculiar property of quark
confinement.  
As bound states are involved,
we shall consider the QCD and QED interactions between the quark and
the antiquark to be implicitly non-perturbative in nature.

A brief discussion of the present work was presented at the International Conference on High Energy Physics, Bologna, Italy, July 2022 \cite{Kos22}.  
The  contents of the  present manuscript are organized  as follows.  In Section 2, 
we discuss the circumstances in which a quark and an antiquark
may be produced and may interact nonperturbatively in QED alone.  It is therefore
reasonable to examine the question of quark confinement in QED 
 from the viewpoint of lattice
gauge calculations.  
Past lattice gauge calculations reveal  that quarks in QED in  (3+1)D are not
confined.   The deconfinement of quarks in QED appears to 
 contradict the experimental absence of fractional
charges.  
 It is therefore suggested that the Schwinger longitudinal confinement
mechanism  may need to be included in future lattice gauge calculations for light quarks in QED in (3+1)D.
In order to
investigate the effect of the Schwinger longitudinal confinement for light quarks in QED in (3+1)D,
we study the Schwinger longitudinal confinement for QED in (1+1)D
  in conjunction with  Polyakov's transverse QED
confinement in (2+1)D.
We therefore propose in Section 3 a ``stretch (2+1)D'' flux tube model in (3+1)D, for the
production and the interaction  of a  light quark  $q\bar q$ pair in compact QED.
For our investigation, we introduce the dynamical
variables and the action integral in (3+1)D  in a flux tube configuration in Section 4 and treat the
QCD and QED gauge interactions in a broken U(3) framework.  We apply
Polyakov's result of transverse confinement in QED in (2+1)D
\cite{Pol77,Pol87} and study in Section 5 the quark field part of the
action integral.  We separate the action integral into the transverse
(2+1)D$_{\{x^1,x^2,x^0\}}$ Lagrangian and the longitudinal
(1+1)D$_{\{x^3,x^0\}}$ Lagrangian, for the system with cylindrical
symmetry in the stretch (2+1)D flux tube model.  We follow methods similar to those used
by Wang, Pavel, Brink, Wong and others
\cite{Wan88,Sai90,Sch90,Pav91,Won95} to separate the
longitudinal and transverse degrees of freedom of the Dirac equation,
with details given in Appendix A.  The relations between various
quantities in 2D and 4D are given in Appendix B.  In Sections 6, we
study the gauge field part and the total sum of the action integral.  In
Section 7, we write down the action integral in the idealized
(1+1)D$_{\{x^0,x^3\}}$ space-time which can be the starting point for
the description of the longitudinal dynamics in the flux tube
environment.  In Section 8, we examine the proposed ``stretch (2+1)D''
flux tube model for the production of a $q\bar q$ pair and study the
Landau level dynamics.  The solutions are then used to calculate the
quark currents to generate the self-consistent gauge fields.  We use the
Maxwell equation and obtain the masses of bound mesons as a function
of the coupling constant.  In Section 9, we compare our results of
many QED and QCD excitations in quark-QCD-QED systems with low-energy
experimental data.  In Section 10, we present our 
summary,  discussions, and  conclusions.

\section{Questions on quark confinement in  lattice gauge
calculations  in compact QED in (3+1)D}

It is well-known that a $q\bar q$ pair is confined in a QCD meson in
(3+1)D, and we do not need to re-examine the question of quark
confinement in QCD in (3+1)D again.  We can focus our attention on the
question of quark confinement in the QED interaction in (3+1)D.  However,
the suggestion in 
\cite{Won10,Won11,Won14,Won20,Won22,Won22a,Won22b,Won22c,Kos22}
 that a quark and an antiquark interacting in QED may be
confined as a QED meson in (3+1)D is
subject to serious questions.  Is it ever possible for a quark and an antiquark be produced and interact in QED alone?  Our common perception is
that a quark and an antiquark interact simultaneously in QCD and QED,
with the QCD interaction as the dominant interaction and the QED
interaction as a perturbation.  The interaction of a quark and an
antiquark in QED alone, without the QCD interaction, may appear
impossible at first sight.  However, as  discussed in detail 
in Section 2 of  Ref.\ \cite{Won22c} and explained briefly here,
there are physical circumstances in which a
quark and an antiquark pair can be produced  and can  interact in QED alone.

Quarks and antiquarks carry color and electric charges and  they interact in the QCD interaction and the QED interaction with the exchange of a gluon or a photon.
The QCD interaction and the QED interaction are independent interactions.
There is no theorem nor basic
physical principle that forbids a quark and an antiquark to interact non-perturbatively
in QED alone.  What is not forbidden is allowed, in accordance with
Gell-Mann's Totalitarian Principle \cite{Gel56}.  So, it is allowed that a quark and an antiquark can interact in QED alone. 
Experimentally,  there are circumstances in which a
quark and an antiquark pair can be produced  with a center-of-mass energy $\sqrt{s}$ below the pion mass gap, $m_q + m_{\bar q} <
\sqrt{s} < m_\pi$, 
where the sum of the rest masses of the light quark and light
antiquark is of order a few MeV and $m_\pi\sim 135$ MeV \cite{PDG19}.  In this range of energy below the pion mass gap, the quark and the antiquark cannot interact in the QCD interaction to form a QCD-bound state, nor can they interact in the QCD interaction to form an isolated quark and an isolated antiquark  continuum state because quarks are confined in QCD.   We are left with the only possibility of the quark and the antiquark interacting in the QED interaction alone 
in this energy range.  The quark and antiquark, if 
 produced in this energy range below the pion mass gap in
hadron-hadron, $AA$, $e^+e^-$, and $e^-A$ collisions,  will probe the property of the quark and the antiquark interacting in the QED interaction alone.
 A $q\bar q$
pair will be produced at the eigenenergy of a QED-confined $q\bar q$ eigenstate as a QED meson, if there exists such an
eigenstate  in this energy range.

In the production of a quark and an antiquark by  a process such as 
$e^+$+$e^-$$\xrightarrow{\gamma^*{\rm or}\,\gamma^*\gamma^*} $ $q$+$\bar q$
with $m_q + m_{\bar q}<  \sqrt{s_{e^+e^-}}< m_\pi$, the reaction process will proceed only if 
 the produced $q$ and $\bar q$ will  interact dynamically between themselves and with the vacuum 
 to generate a confined and bound $q\bar q$ system with a mass  matching  the eigenenergy of a  QED meson eigenstate, $\sqrt{s_{e^+e^-}}$=$m_{_{\rm QED}\,{\rm meson}}$,  if such a confined and bound state exists.  At all other  energies 
for which $\sqrt{s_{e^+e^-}}$$\ne$ $m_{_{\rm QED}\,{\rm meson}}$,
the production  process cannot occur because 
the produced quark and antiquark cannot form a confined and bound state, nor can they  exist as a pair of isolated  quark and antiquark propagating to asymptotically infinite separations.  The reaction cross section for the process  $e^+ $+$ e^-$$\xrightarrow{\gamma^*{\rm or}\,\gamma^*\gamma^*}$\,$q+\bar q$ contains the density of state 
delta-function factor,   $\delta ( \sqrt{s_{e^+e^-}}-\! m_{_{\rm QED}\,{\rm meson}})$, which limits the process to occur only for 
$\sqrt{s_{e^+e^-}} = m_{_{\rm QED}\,{\rm meson}}$.
In our formulation, we envisage that at the QED meson eigenenergy,  the produced valence quark and valence antiquark  
interact between themselves, and they also interact with  the vacuum, consisting of quarks under the Dirac sea.   The interaction between the quark and the antiquark can be well represented by the linear interaction of the QED gauge field and such an interaction gives the contribution of the first term of the mass formula in Eq. (115).   For the interaction of the quarks with the vacuum,  we envisage that prior to the production of the valence $q$ and $\bar q$ pair, the vacuum consist of quarks under the negative energy Dirac sea possessing chiral symmetry and the quarks possess their bare masses.   The presence of a valence  quark and an valence antiquark with their non-perturbative QED interaction turning on will lead to 
spontaneous chiral symmetry breaking and 
a QED quark condensate with non-vanishing value of $\langle \bar \psi  \psi \rangle_{_{\rm QED}}$.
As a consequence, there is dynamical modification of the quark-antiquark mass arising the change of the vacuum.   Their contribution to the $q\bar q$ mass  square as proportional to $(m_q+ m_{\bar q})\langle  \bar \psi \psi \rangle_{_{\rm QED}}$ is represented by the second term in Eq. (115).   
Together  with both the first and the second contributions, they  lead to the masses of confined and bound QED meson $q\bar q$ states
 as those listed in Table 1 in Section 9 below.   

Having settled on the question of the possibility of quarks interacting in QED alone, 
we proceed to examine  the question of
 quark confinement in QED in (3+1)D in lattice gauge calculations.
It has been known for a long time since the advent of Wilson's lattice
gauge theory that a fermion and an antifermion in (3+1)D in compact
Abelian U(1) QED interaction (which corresponds to a quark and an
antiquark with a single color and a single flavor in (3+1)D in compact
QED interaction) have a strong coupling confinement phase and a weak
coupling deconfined phase \cite{Wil74}.  The same conclusion was
reached subsequently by Kogut, Susskind, Mandelstam, Polyakov, Banks,
Jaffe, Drell, Peskin, Guth, Kondo and many others
\cite{Wil74,Cas74,Kog75,Man75,Pol77,Pol81,Pol87,Ban77,Gli77,Pes78,Dre79,Gut80,Kon98,Arn03,Lov21,Mag20}.
Lattice gauge calculations in compact QED in (3+1)D show that the
critical coupling constant $\beta_c $, below which opposite static
charges are not confined in (3+1)D, has been determined to be\break  $\beta_c$=$ 1/e_c^2$=1.0111331(21), or $e_c^2 $=0.988989481
\cite{Arn03,Lov21}.  As $e_\qed^2=1/137$ and $ e_\qed^2 \ll e_c^2 $,
the QED interaction between a quark and an antiquark belongs to the
weak-coupling deconfined regime.  This means that according to lattice
gauge calculations, a quark and an antiquark are deconfined in compact
QED in (3+1)D.

The deconfined solution for a quark and antiquark in lattice gauge
calculations in (3+1)D poses a serious contradiction to experimental
observations.  If a pair of  quark and  antiquark are deconfined when they interact in the QED interaction alone in
the physical world of (3+1)D, then they will appear as isolated quark
and antiquark  with fractional charges at energies below the pion mass gap.
However, in contradiction to such a prediction of deconfined quark and
antiquark, no such fractional charges have ever been observed.

It is important to point out however that the deconfined solution for
a static quark and a static antiquark pair in compact QED in (3+1)D in
lattice gauge calculations comes mainly from calculations in which the
quark and the antiquark are considered only as static external quark
probes represented by a Wilson loop, given in terms of the product of
link variables at fixed static spatial locations along time-like world
lines
\cite{Wil74,Cas74,Kog75,Man75,Pol77,Pol81,Pol87,Ban77,Gli77,Pes78,Dre79,Gut80,Kon98,Arn03,Lov21,Mag20}.
In such a formulation of quark confinement in terms of the Wilson
loop, the quarks are in essence static and infinitely heavy objects in
fixed spatial locations.  They have not been treated as dynamical
quark fields.  These calculations have not included the Schwinger
confinement mechanism when light quarks are treated as quanta in a
dynamical quark field.

On the other hand, as applied to light quarks approximated as
massless, the Schwinger confinement mechanism shows that a light quark
and a light antiquark are confined in QED in (1+1)D as an open string
\cite{Sch62,Sch63}.  It has been further shown by Coleman, Jackiw, and
Susskind that the Schwinger confinement mechanism persists even for
massive quarks in (1+1)D \cite{Col75,Col76}.  Furthermore, from the viewpoint of
phenomenology, a stable QED-confined $q\bar q$ system in (3+1)D
would be a likely occurrence because quarks cannot be isolated and
they reside predominantly in (1+1)D as no fractional charged particles
have ever been observed.  The phenomenological open-string QCD and QED
meson model with the hypothesis of a confined $q\bar q$ pair in
(3+1)D leads to QCD and QED meson spectra  in agreement with
experimental data \cite{Won10,Won20}.
 
In view of the above, the present-day static lattice gauge
calculations for compact QED in (3+1)D may not be as complete and
definitive as it may appear to be because the important Schwinger
confinement mechanism has not been included.  Future lattice gauge
calculations with the inclusion of the Schwinger confinement mechanism
in compact QED interactions in (3+1)D will be of great value in
clarifying the question of quark confinement in QED.  In this regard,
there have been many recent advances in efficient methods of lattice
gauge calculations in compact QED with dynamical fermions (which can
be taken to be quarks) in (3+1)D using the tensor network
\cite{Mag20}, dual presentation \cite{Ben20}, magnetic-field
digitization \cite{Bau21}, regulating magnetic fluctuations
\cite{Kap20}, or other efficient methods.  There are also other recent
advances in the studies of $q\bar q$ flux tubes in lattice gauge
theories in compact U(1) QED and SU(3) QCD
\cite{Bal05,Ama13,Car13,Cos17,Bic17,Bic18}.  They may be utilized to study
the question of confinement of dynamical quarks in compact U(1) QED in
a flux tube.

\section{ The stretch (2+1)D model for $q\bar q$ production
  and confinement in compact QED in (3+1)D }

Whatever the theoretical predictions on quark confinement in
QED in (3+1)D may be, in the final analysis, the question whether a
$q\bar q$ pair is confined in QED in (3+1)D can only be settled by
experiments.
 
In the meantime, there is the perplexing agreement of the experimental
anomalous particle spectrum with the theoretical predictions of the
phenomenological open string QED meson model in
\cite{Won10,Won11,Won14,Won20,Won22,Won22a,Won22b,Won22c,Kos22}.  In
the presence of the two opposing conclusions on quark confinement in
QED in (3+1)D, it is possible that both conclusions can still be
consistent with each other, if the confinement conclusion arises from
the inclusion of the Schwinger confinement mechanism
\cite{Sch62,Sch63} in the QED meson model in
\cite{Won10,Won11,Won14,Won20,Won22,Won22a,Won22b,Won22c,Kos22}, while
the deconfinement conclusion arises from the absence of the Schwinger
confinement mechanism in static lattice gauge calculations of
\cite{Wil74,Cas74,Kog75,Man75,Pol77,Pol81,Pol87,Ban77,Gli77,Pes78,Dre79,Gut80,Kon98,Arn03,Lov21,Mag20}.
Therefore, we wish to construct a plausible flux-tube model to
demonstrate such a possibility by showing that a $q\bar q$ system in a
``stretch (2+1)D'' configuration can attain confinement in QED in
(3+1)D at a confined $q\bar q$ eigenstate.

We would like to introduce the essential concepts of the ``stretch
(2+1)'' flux tube model.  We note first of all that there are two
different types of QED U(1) gauge interactions possessing different
confinement properties \cite{Pol77,Pol87,Dre79}.  There is the compact
QED U(1) gauge theory in which the gauge fields $A^\mu$ are angular
variables with a periodic gauge field action to allow transverse
photons to self-interact among themselves.  The gauge field action in
the compact QED U(1) gauge theory, in the lattice gauge units and
notations of Ref.\ \cite{Pol77,Pol87,Dre79}, is
\begin{eqnarray}
S=\frac{1}{2g^2}\sum_{x,\alpha \beta} (1-\cos F_{x,\alpha \beta}),
\label{1}
\end{eqnarray}
where $g$ is the coupling constant and the gauge fields $F_{x, \alpha \beta}$ are 
\begin{eqnarray}
F_{x, \alpha \beta}\!=\!A_{x,\alpha}\!\!+\!\! A_{x+\alpha,\beta}\!\! - \!\! A_{x+\beta,\alpha} \!\! - \!\! A_{x,\beta},{\rm with} -\!\!  \pi\!\!  \le A_{x,\alpha} \!\! \le\!\!  \pi.~
\label{eq2}
\end{eqnarray}
There is also the non-compact  QED U(1) gauge theory with the gauge
field action \cite{Pol77,Pol87,Dre79}
\vspace*{-0.1cm}
\begin{eqnarray}
S=\frac{1}{4g^2}\sum_{x,\alpha \beta} F_{x,\alpha \beta}^2,  ~~~{\rm with}~~~~- \infty  \le A_{x,\alpha} \le  +\infty  .
\label{3}
\end{eqnarray}
In non-compact QED gauge theories, the transverse photons do not
interact with other transverse photons.  

Even though the compact and
the non-compact QED gauge theories in Eqs.\ (\ref{1}) and (\ref{3})
have the same continuum limit, they may have different confinement
properties.   The QED confinement property  depends on 
the number of spatial dimensions of the quark and gauge  fields.

In (1+1)D, Schwinger showed that massless dynamical fermions are
confined in QED for all strengths of the coupling constant
\cite{Sch62,Sch63}, for which there is no distinction between the
compact and the non-compact QED.  

In (2+1)D,
Polyakov showed that static opposite charges are confined in compact QED for all strengths
of the coupling constant, but are deconfined in non-compact QED
\cite{Pol77,Pol87}.

In (3+1)D, static opposite charges are confined in compact QED only
for strong coupling, but they are deconfined for weak coupling and
 in non-compact QED \cite{Pol77,Pol87,Dre79}.  

We need to ascertain the type of the QED U(1) gauge interaction
between a quark and an antiquark in a QED meson.  As pointed out by
Yang \cite{Yan70}, the quantization and the commensurate properties of
the interacting electric charges imply the compact property of the
underlying QED gauge theory.  Because (i) quark and antiquark electric
charges are quantized and commensurate, (ii) quarks and antiquarks
cannot be isolated, and (iii) there are pieces of experimental
evidence for possible occurrence of confined $q\bar q$ QED meson
states as we mentioned in the Introduction, it is therefore reasonable
to propose that quarks and antiquarks interact in the compact QED U(1)
interaction.

In compact QED in (2+1)D$_{\{x^1,x^2,x^0\}}$ in lattice gauge
theories, Polyakov \cite{Pol77,Pol87} showed previously that a pair of
opposite electric charges and their gauge fields are confined and that
the confinement persists for all non-vanishing coupling constants, no
matter how weak.  As explained in detail by Drell and collaborators
\cite{Dre79}, such a confinement in (2+1)D${}_{\{x^1,x^2,x^0\}}$
arises from the angular-variable property of $A_\phi$ and the
periodicity of the gauge field action as indicated in Eqs.\ (1) and
(2).  The gauge action periodicity in the neighborhood of the produced
opposite electric charges leads to self-interacting transverse gauge
fields.  The transverse gauge fields interact among themselves,
they do not radiate away, and they join the two opposite electric
charges and their associated gauge fields by a confining interaction.
The $q\bar q$ confinement in (2+1)D can be described as the quark and
the antiquark being effectively magnetic monopole and magnetic
anti-monopole \cite{Pol74,tho74,Man75}, linked together by a magnetic flux
tube.  In the presence of quantum fluctuations, the magnetic monopole
confinement forces tend to counterbalance the disruptive quantum fluctuation
forces.

In lattice gauge calculations in  compact QED in (2+1)D,  the magnetic monopole confinement forces overcomes
the quantum fluctuation forces, and the quark confinement persists for
the QED interaction of all strengths in (2+1)D \cite{Pol77,Pol87}.

In lattice gauge calculations in compact QED in (3+1)D,  however, a pair of opposite electric charges (such as a $q$
and an $\bar q$ pair) and their gauge fields have two different  phases,
depending on the strengths of the coupling.  With a strong coupling,
the magnetic monopole confining forces between the charged pair
overcome the quantum fluctuation forces to lead to the confining phase
in (3+1)D.  As the electric charged pair is confined already in (2+1)D
by way of Polyakov transverse confinement in (2+1)D, the strong
strength of the gauge field between the charged pairs in the case of strong coupling facilitates the
longitudinal confinement, leading to the complete transverse and
longitudinal confinements in (3+1)D.  However, for a charged pair with
a weak coupling (such as a quark and an antiquark interacting in QED)
and a random disposition of the intermediary gauge fields, the weak
magnetic monopole confining forces in (3+1)D cannot overcome the
quantum fluctuation forces for longitudinal confinement.  As a
consequence, the strength of the QED interaction between a quark and
an antiquark places them in the weak-coupling deconfinement regime in
(3+1)D, and the quark and the antiquark are not confined in QED in
(3+1)D.

As we discussed earlier, the lattice gauge result of deconfinement of
a quark and an antiquark in QED in (3+1)D contradicts the
observational absence of fractional charges and the successful phenomenological 
description of the pair as a QED meson \cite{Won10,Won20}.  In lattice gauge calculations
up to now, the Schwinger confinement mechanism is however missing.  It is
possible that when the Schwinger confinement mechanism is
properly taken into account, there can be the longitudinal confinement  between the quark and the antiquark in the stretch (2+1)D configuration in QED.  Together with the additional transverse confinement
made possible by the Polyakov's transverse confinement in QED in
(2+1)D$_{\{x^1,x^2,x^0\}}$, there can be a complete confinement of the
quark and the antiquark in QED in (3+1)D$_{\{x^1,x^2,x^3,x^0\}}$. 

 To
include the Schwinger confinement mechanism into the dynamics in
(3+1)D, we consider light quarks which can be approximated as
massless, in contradistinction to quarks on time-like world lines as static and
infinitely-heavy external probes in conventional  lattice gauge calculations.   We also judiciously position and
prepare the quark fields and gauge fields such that the Schwinger
confinement mechanism can be operational.  The Schwinger
confinement mechanism will be operational if the quark and the
antiquark are linked by a flux tube.   The flux tube can be prepared
and maintained by stretching the Polyakov's transverse confinement of
QED in (2+1)D along the longitudinal direction.  We therefore
construct a ``stretch (2+1)D" flux tube model \cite{Won22b} by
combining the Schwinger (longitudinal) confinement with
Polyakov transverse confinement to examine the creation and the
interaction of a quark $q$ and an antiquark $\bar q$ in a QED-confined
$q\bar q$ eigenstate.  We wish to study how the quark and the
antiquark can be produced and be confined at the $q\bar q$ eigenstate
in QED in (3+1)D.
\begin{figure} [h]
\centering
\vspace*{-0.0cm}\hspace*{-0.3cm}
\resizebox{0.53\textwidth}{!}{
\includegraphics{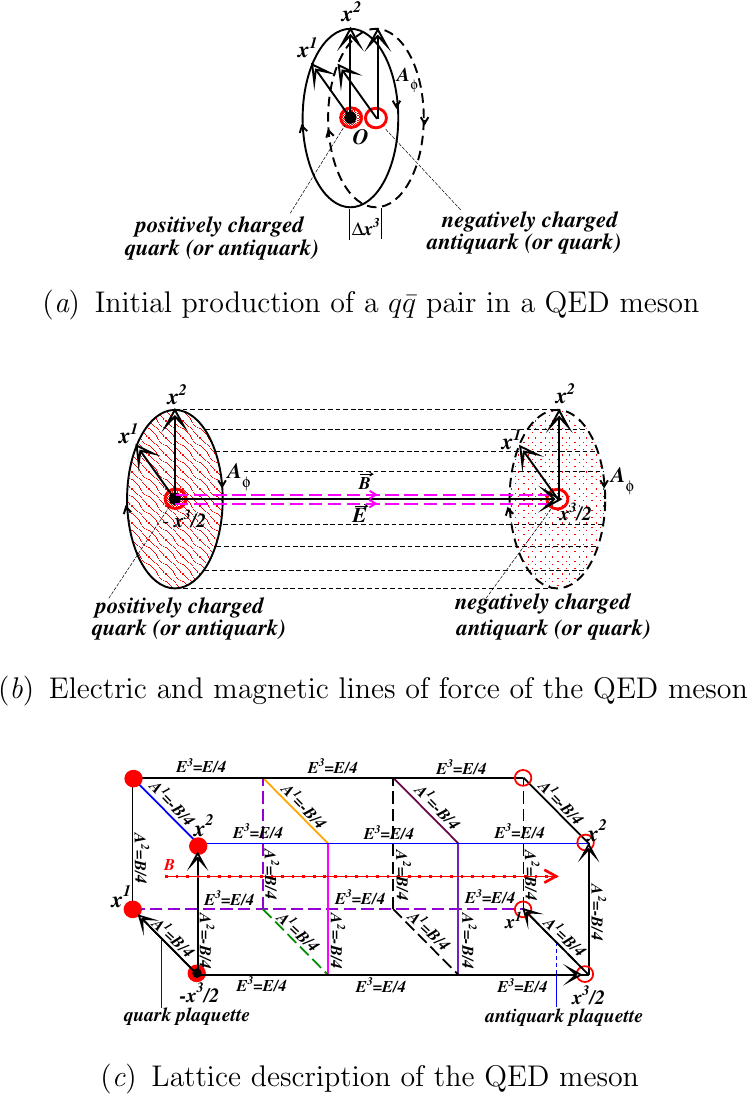}}
\caption{ Fig.\ ($a$) depicts the initial production of the $q\bar q$
  pair of a QED meson, with an infinitesimal longitudinal separation
  $\Delta x^3$.  Fig.\ ($b$) gives a snapshot of the configuration of
  the quark charge, the antiquark charge, gauge fields $\bb A, \bb E$,
  and $\bb B$=$\nabla $$\times$$ \bb A$ during the dynamical
  longitudinal yo-yo motion of a $q\bar q$ pair interacting in compact
  QED in a QED meson, staring from a (2+1)D transversely confined
  $q\bar q$ system.  Fig.\ ($b$) is the ``stretch (2+1)D''
  configuration.  Fig.\ ($c$) is the corresponding lattice
  transcription of Fig.\ ($b$), following the Hamiltonian formulation
  of Drell $et~al.$ \cite{Dre79}. }
\vspace*{-0.3cm}
\label{fig1}
\end{figure}
\vspace*{-0.1cm} 

In the stretch (2+1)D flux tube model, we envisage the production of
the nascent $q$$\bar q$ pair at the origin, $(x^1,x^2,x^3)=0$, in the
center-of-mass system in  Fig. \ref{fig1}($a$), at the eigenstate with
eigenenergy $\sqrt{s}(q\bar q)$ of a QED meson\footnote{Because quarks
and antiquarks cannot be isolated, the production of a $q\bar q$ pair can
occur only at certain CM eigenstate energies, $\sqrt{s }(q\bar q)$, of
confined and bound $q\bar q$ systems, such as in a QED meson or a QCD
meson.  At other energies different from confined and bound eigenstate
energies, the $q$ and $\bar q$ do not exist.  Hence, a quark and an
antiquark cannot be produced at energies that are not eigenenergies of
bound QCD or QED mesons.}.  We take for simplicity the quark
(electric) charge to be positive, which can be easily generalized to
other cases of negatively-charged quark and flavor mixing.  At birth,
the quark $q$ and the antiquark $\bar q$ possess equal and opposite
momenta in the center-of-mass system because the $q$ and the $\bar q$
reside in a bound eigenstate with non-zero kinetic energies.  The quark
momentum defines the longitudinal direction from which we can define
the transverse coordinates $(x^1,x^2)$ and the transverse planes at
$x^3$=(constant).

As depicted in Fig.\ \ref{fig1}($a$), the quark and the antiquark
produced at birth are separated by an infinitesimal longitudinal and
transverse separations, $(\Delta x^3, \Delta {\bb r}_\perp)$.  At the
birth of the $q$-$\bar q$ pair, there exists Polyakov's transverse
confinement of the quark, the antiquark and the gauge fields in
compact QED in (2+1)D on the transverse $(x^1,x^2)$-plane at $x^3\sim
0$.  The QED-confinement of the created $q$ and $\bar q$ pair in
(2+1)D necessitates the associated creation of their gauge fields $\bb
A$, $\bb E$, and $\bb B=\nabla\times {\bb A}$, which by causality can only
be in the neighborhood of the created charges initially, with the
created $\bb E$ and $\bb B$ fields lying along the longitudinal $x^3$
direction.  Because the created $q$ and $\bar q$ pair are constituents of a
confined eigenstate, the produced quark and the antiquark will execute
stretching and contracting ``yo-yo'' motion along the longitudinal
$x^3$ direction, appropriate for the QED meson bound state in
question.  The light masses also means that they moves with speed
close to the speed of light.  As the quark and antiquark stretch
outward in the longitudinal $x^3$ directions, we can construct a
longitudinal tube structure of gauge fields in the stretch (2+1)D
configuration by duplicating longitudinally the transversely-confined
gauge fields that exist on the transverse $(x^1,x^2)$-plane at $x^3
\sim 0$ initially at their birth, for the longitudinal region between
the stretching quark and antiquark.
By duplicating  the initial transversely-confined gauge fields along
the longitudinal direction, we obtain a longitudinal tube with a
cylindrical symmetry in (3+1)D.  The property of the transverse
confinement at one longitudinal coordinate at birth can thereby be
extended to the whole tube, and the longitudinal tube will likely be a
transversely-confining flux tube.

A snap shot of the stretch (2+1)D flux tube configurations at an early
moment in the longitudinal stretching motion is shown in
Fig. \ref{fig1}($b$).  We show the transcription of
Fig.\ \ref{fig1}($b$) in terms of the lattice link and plaquette
variables in Fig.\ \ref{fig1}($c$), by following the Hamiltonian
formulation and the notations of Drell $et~al.$ \cite{Dre79}.
Specifically, in the $A^0$=0 gauge we specify the canonical conjugate
gauge fields $\bb A$ and $\bb E$ at the links in Fig.\ 1($c$), where we
display only the $A^1,A^2$ and $E^3$ values of the conjugate gauge
fields.  The $\bb B$ and $\bb E$ fields are aligned along the
longitudinal $x^3$ axis and directed from the positively-charged
$q$  at $x^3/2$  to the negatively-charged  $\bar q$ at $-x^3/2$ (Fig.\ \ref{fig1}($c$)).  
These $\bb B$ and $\bb E$  fields are present to 
lead to Polyakov's transverse confinement of the quarks and the gauge
fields that reside on the transverse planes of $q$ at $-x^3/2$ and $\bar q$ at $x^3/2$.    Specifically, the magnetic field $\bb B$ sends the quark and antiquark
charges into the appropriate Landau level orbitals to execute confined
transverse harmonic oscillator zero-point motions on their respective
$\{x^1,x^2\}$ transverse planes.  At the transverse zero mode, 
the anticipated Schwinger longitudinal confining motion along with 
the electric field
$\bb E$ and the magnetic field $\bb B$ along the longitudinal $x^3$
direction send the quark and the antiquark in a 
longitudinal stretching and contracting ``yo-yo'' motion.  The
electric charge densities obey the Gauss law associated with the
divergence of the electric field $\bb E$.  Consequently, the positive
electric quark charge fractions (solid circles in Fig.\ \ref{fig1}($c$))
reside at the $-x^3/2$ plaquette vertices and the negative electric
antiquark charge fractions (open circles in Fig.\ \ref{fig1}($c$)) at
the antiquark plaquette vertices at their $+x^3/2$ planes, at the two
ends of the flux tube.  The transverse gauge fields $\bb A$ on the
transverse links are copies of those on the quark and the antiquark
plaquettes at $x^3(q)$=$-x^3/2$ and at $x^3(\bar q)=x^3/2$
respectively, and they are unchanged in $x^3$ along the stretching
motion, while the longitudinal links are all $E^3$=$|\bb E|/4$.

We shall show in Section 8.1 in the stretch (2+1)D model that the
longitudinal $\bb B$ field that is present initially to confine quarks
and antiquarks on the transverse plane at their birth continues to
confine quarks and antiquarks transversely in the stretched
configuration, because of the Landau level dynamics.  The cloud of
transverse gauge fields continue to interact with each other to
maintain the transverse confinement on their transverse planes.  As a
consequence, quarks, antiquarks, and gauge fields will continue to be
transversely confined, as if the quark and the antiquark are
effectively magnetic monopoles \cite{tho74,Pol74,Man75} linked together by a
flux tube in the transverse degree of freedom as shown in Fig.\ 1(b).

 With the attainment of transverse confinement and $\bb E$ and
$\bb B$ aligned along the longitudinal  direction in the flux tube,
 it remains necessary to examine the question of longitudinal confinement.  
If we can show that the system 
possesses both transverse and longitudinal
confinements as in an open-string, the quark and the antiquark will be
confined and bound in a QED meson in (3+1)D.

Although the above model has been constructed specifically with quarks
interacting in QED in our mind, Polyakov's transverse-confinement effect 
and Schwinger's  longitudinal confinement effect
apply also to quarks interacting in QCD  in the quasi-Abelian approximation of the QCD interaction.  For this reason, we shall
include both QED and QCD in the same footing in our subsequent
investigations.

\section{ The action integral in a  flux tube model 
in (3+1)D }

We proceed now to study in detail the dynamics of quarks, antiquarks,
and gauge fields in QCD and QED interactions in 
 the above cylindrically-symmetric stretch (2+1)D model of \cite{Won22}.
 Here, the dynamical field variables
are the quark fields $\Psi_f^i$, the QCD gauge field $A^\mu_\qcd$, and
the QED gauge field $A^\mu_\qed$.  The quark fields are $\Psi_f^i$
where $i$=1, 2, 3 are the color indices and $f$ is the flavor index.
The QCD gauge fields are $A^\mu_\qcd$=$\sum_{a=1}^8 A^\mu_a t^a $,
where $a$=1,...,8 are the SU(3) generator index,
$\{t^1,t^2,t^3,...,t^8\}$ are the generators of the SU(3) color-octet
subgroup, and $\mu=0,1,2,3$ are the indices of the space time
coordinates $x^\mu$ with the signature $ {g}^{\mu \mu}$=(1, -1, -1,
-1).  The QED gauge field is $A^\mu_\qed$=$ A^\mu_0 t^0 $ where $t^0$
is the generator of the color-singlet U(1) subgroup,
\begin{eqnarray}\label{q1}
t^0=\frac{1}{\sqrt{6}}
\left ( \begin{matrix}
                1 & 0 & 0 \cr
                0 & 1 & 0 \cr
                0 & 0 & 1
        \end{matrix}   \right ).
    \end{eqnarray}    
We shall  use the summation convention over repeated indices,
except when the summation symbols are needed to resolve  ambiguities.

In our problem of quarks interacting with the QCD and the QED
interactions in the stretch (2+1)D model, it is useful to treat the QED and QCD gauge interactions
in a single framework by considering an enlarged U$(3)$ group,
consisting of the union of the QCD SU(3) color-octet subgroup and the
QED U(1) color-singlet subgroup of generators \cite{Won10}.  This can
be achieved by adding $t^0$ onto the eight standard generators of the
SU$(3)$ subgroup to form the nine generators $\{ t^0,t^1,...t^8\}$ of
the U(3) group, satisfying $2 \,{\rm Tr}\{ t^a t^ b \} = \delta^{ab} $
with $a,b=0,1,..,8$ and $A^\mu$=$\sum_{a=0}^8 A^\mu_a t^a $. The SU(3)
and the U(1) subgroups of the U(3) group differ in their coupling
constants and communicative properties.  We consider quarks with
masses $m_f$ with the number of flavors $N_f^\lambda$ where
$f$=$u,d,s$=1,2,3 are the flavor labels, and the superscript $\lambda$
is the interaction label with $\lambda$=0 for the QED color-singlet
U(1) subgroup, and $\lambda$=1 for QCD color-octet SU(3) subgroup.
Because of the different mass scales associated with different gauge
fields, we can choose $N_f^{\qed}$\!=$N_f^0$=2 and
$N_f^{\qcd}$\!=$N_f^1$=3.  The coupling constants\footnote{We adopt
here the notations that unless stated otherwise, $g^\lambda$ is
actually $g_\2d^\lambda $, the coupling constant in 1+1 dimensions,
and $g_\4d^\lambda $ is the coupling constants in 3+1 dimensions.  The
superscript $\lambda$ specifies the interaction, with $\lambda=0$ for
QED and $\lambda=1$ for QCD.  The strong coupling constant is
$\alpha_\qcd$=$(g^\qcdu_\4d)^2/4\pi$ and the fine structure constant is
$\alpha_c$=$\alpha_\qed$=$(g_\4d^\qedu)^2/4\pi$=$(-e_\4d^\qedu)^2/(\hbar
c)$=1/137.  As pointed out in \cite{Won09,Won10,Won20,Kos12} and Eq.\ (\ref{76}),
$g_\2d^\lambda $ and $g_\4d^\lambda $ are approximately related by the
flux tube radius $R_T$ as $(g_\2d^\lambda)^2=(g_\4d^\lambda)^2/(\pi
R_T^2)$, when the confining flux tube  of radius $R_T$ is idealized as an open string
without a structure.  }  $g_{\4d f}^a$ are given explicitly by
\begin{subequations}
\label{eq2cc} 
\begin{eqnarray}
\label{qedcc} 
&&\hspace*{-0.0cm}g_{\4d u}^0\!=\!-Q_u^\qed\!\!g_{\4d}^{\qed}\!\!\!\!,\!
~~~~~~~g_{\4d d}^0\!=\!-Q_d^\qed\!\!g_{\4d}^{\qed} \nonumber\\
&&{\rm~~~~~~~~~~for~the ~QED~color}{\rm-singlet~subgroup},
\\ &&\hspace*{-0.0cm}g_{\4d {\{u,d,s\}}}^{\{1,..,8\}}=Q_{\!\{u,d,s\}}^{^{\qcd}}g_{\4d}^\qcd
\nonumber\\
&& {\rm~~~~~~~~~~for~the~QCD~color}{\rm-octet~subgroup},
\label{qcdcc}
\end{eqnarray}
\end{subequations}
where we have introduced the charge numbers $Q_u^{\qed} $\!\!\!=2/3,
$Q_d^{\qed}$\!\!\!=$-$1/3, $Q_u^{\qcd}$\!\!\!=$~Q_d^{\qcd}
$\!\!\!=$~Q_s^{\qcd} $\!\!\!=~1.  For brevity of notations, the color
and flavor indices $\{a$,$f\}$ and the interaction labels $\{\lambda$,
QCD, QED\} in various quantities are often implicitly understood
except when they are needed to resolve ambiguities.

We start with the 3+1 dimensional space-time $x^\mu$, with the
dynamical variables of the quark fields $\Psi_f^i$ and the U(3) gauge
fields $A_\mu^a $, the invariant action integral $\cal A$
\cite{Pes95}, as modified for compact gauge
interactions\footnote{There is a factor 2 on the left-hand side of the
trace relation, $2{\rm tr} \{t^a t^b\}=\delta^{ab}$, for the generator
$t^a$ and $t^b$.  For the convenience of notation, we shall define the
operation ``$Tr$ trace'' over the color space as $Tr_{\rm color} (t^a
t^b)$$ \equiv $2tr$ \{t^a t^b\}=\delta^{ab}$, when we calculate the
trace of the product of generators $t^a t^b$ in the color space.  }
\cite{Pol77,Pol87,Dre79}, is given by
\begin{eqnarray}\label{eq2}
\hspace*{-0.2cm} {\cal A}_\4d  \! = \!\!\int d^4 x ~Tr \Bigg\{
{\bar \Psi} (x) \gamma^\mu   \Pi_\mu   \Psi (x)
   - {\bar \Psi}   (x) m  \Psi  (x) 
   + {\cal L}_A \Bigg\},~
\end{eqnarray}
where 
\begin{subequations}
\begin{eqnarray}\label{eq3}
&&\gamma^\mu \Pi_\mu =i\sD =
\gamma^\mu iD_\mu 
\nonumber\\
&&~~~~~~~=
\gamma^\mu (i \partial_\mu +g_\4d A_\mu)
\!=\!
\gamma^\mu ( p_\mu + g_\4d A_\mu)
,
 \\
 &&{\cal L}_A = -\frac{1}{2\pi^2 R_T^4g_\4d^2}    [ 1 - \cos ( \pi R_T^2 g_\4d  F_{\mu \nu} (x))],
 \label{11b}
 \\
&&F_{\mu \nu}= \partial_\mu A_\nu - \partial_\nu A_\mu
-i g_\4d [A_\mu, A_\nu],
\label{4b}
\\
&&F_{\mu \nu}=F_{\mu \nu}^a t_a,
\\
&&A_{\mu}=A_{\mu }^a t_a.
\end{eqnarray}
\end{subequations}
Here the subscript label of `4D' in $g_\4d$ and ${\cal A}_\4d$ is to
indicate that $g_\4d$ is the coupling constant in 4D space-time,
${\cal A}_\4d$ is the action integral over the 4D space-time
coordinates of $x^0, x^1, x^2$, and $x^3$, and $m$ is the quark mass.
The quantity $\sqrt{\pi} R_T$ in Eq.\ (\ref{11b}) is a transverse
length scale which has been chosen to be the square root of the flux
tube area.

The degrees of freedom for a  quark system  with cylindrical symmetry 
 in (3+1)D  can be
 separated into the transverse and longitudinal degrees of
freedom.  The cylindrical symmetry allows the description of the (3+1)D
dynamics as the coupling of the (2+1)D$_{\{x^1,x^2,x^0\}}$ transverse
dynamics and the (1+1)D$_{\{x^3,x^0\}}$ longitudinal dynamics linked
together by the parametric time $x^0$ or other input parametric
quantities.  The simplification of the dynamics from the higher (3+1)D to the lower
dimensional (1+1)D space-time requires the appropriate transverse confinement and
the idealization of the cylindrical flux tube as a one-dimensional
string.  
Accordingly, we begin with Polyakov's result that electric charges of opposite
signs are confined in compact QED in (2+1)D \cite{Pol77,Pol87} and we
construct the stretch (2+1)D model of \cite{Won22} as described in Fig.\ \ref{fig1}
of the last section.  Our present task is to show that quarks and
antiquarks will be transversely and longitudinally confined in such a
model.  We consider the compact gauge field interaction
in which the length scales $\sqrt{\pi}R_T$ in the transverse direction
is small compared to those scales along the longitudinal $x^3$ and
temporal $x^0$ directions.  Because of the large length scales in the
longitudinal and temporal directions, it is reasonable to expand out
the cosine function in the continuum limit for all components except
the $F_{12}$ gauge field so that the compact gauge field Lagrangian in
Eq.\ (\ref{11b}) can be approximated as
\begin{eqnarray}
\hspace*{-0.2cm}{\cal L}_A =-\frac{ [1-\cos ( \pi R_T^2 g_\4d F_{12})] }{\pi^2  R_T^4 g_\4d^2}
- \frac{1}{4}\hspace*{-0.1cm}  \sum_{   \{\mu \nu\} \ne \{12,21\}}
\hspace*{-0.5cm}  F_{\mu \nu}F^{\mu \nu}.
\label{n12}
\end{eqnarray}

\section{Quark part of the action integral}

We first examine the quark part of the action integral, ${\cal A } _Q
$, in (\ref{eq2}) as given by
\begin{eqnarray}\label{eqq7}
 {\cal A }_Q   \!=\!    Tr \!\! \int\! d^4 x \Biggl\{ \!
{\bar \Psi}  (x)  \gamma^\mu i D_\mu   \Psi  (x)  \!-\! {\bar
\Psi} (x)  ~ m ~
 \Psi (x) \!\Biggr\},
\label{7}
\end{eqnarray}
where we choose to work with the 4D-Dirac matrices $ \gamma^\mu$ in
the Weyl representation \cite{Ber79}
 \begin{eqnarray}
&&\hspace*{0.7cm}  \gamma^0   = 
\begin{pmatrix}
0 & I \\
I  &  0  \\
\end{pmatrix} ,
 ~~~~ {\gamma }^j  = 
 \begin{pmatrix}
0  &- \sigma^j\\
\sigma^j  &  0  \\
\end{pmatrix}, ~~~j=1,2,3, 
\nonumber \\
&&\hspace*{-0.2cm}\text{and}
\hspace{0.3cm}\gamma^5   = 
 \begin{pmatrix}
-I & 0 \\
0 & I \\
\end{pmatrix} .
\label{n14}
\end{eqnarray}
In the dynamics of the quark-QCD-QED medium, the quark fields
$\Psi(x)$ depend on the gauge fields $A_\mu(x)$, which in turn depend
on the quark fields $\Psi(x)$, leading to a coupling problem of great
complexity.  Our problem is greatly simplified because we assume
cylindrical symmetry in the stretch (2+1)D model of \cite{Won22} in Figs.\ \ref{fig1}($a$)
and \ref{fig1}($b$), with
the alignment of the $\bb B$ and $\bb E$ fields along the longitudinal
direction.  The separation of the longitudinal and transverse degrees
of freedom in the Dirac equation can be carried out by following
methods similar to those used by Wang, Pavel, Brink, Wong, and many
others \cite{Wan88,Sai90,Sch90,Pav91,Won95}.   In the Weyl
representation \cite{Ber79}, we write the quark bispinor quark field
$\Psi(x)$ in terms of transverse functions $G_{1,2}({\bb r}_\perp)$
and longitudinal-temporal functions ${ f}_\pm(X)$ as
\begin{eqnarray}\label{eqq9}
\Psi_\4d  =\!\!\Psi ( x ) 
  =
\begin{pmatrix}
  G_1   ({\bb r}_\perp ) f_+ (X )  \\
  G_2  ({\bb r}_\perp)  f_- (X ) \\
 G_1  ({\bb r}_\perp)  f_- (X ) \\
-G_2  ({\bb r}_\perp)  f_+(X ) \\
\end{pmatrix}  ,
  \label{wf}
\end{eqnarray}
where $x$=$\{x^1,x^2,x^3,x^0\}$, $\bb r_\perp$=$\{x^1,x^2\}$, and
$X$=$\{x^3,x^0\}$.  The transverse functions $G_{\{1,2\}} ({\bb
  r}_\perp)$ are normalized as
\begin{eqnarray}\label{Norm-cond}
 \int \!\!d\bb r_\perp \{  (G_1^*(\bb r_\bot\!) G_1 (\bb r_\bot\!) \!+\!G_2^*(\bb r_\bot\!) G_2 (\bb r_\bot)\}  \!=\!1,
\end{eqnarray}
and the longitudinal functions $f_\pm (x^3,x_0)$ are normalized as
\begin{eqnarray}
&&\hspace*{-0.3cm}\int\!\! dx^3 dx^0 \{ f_+^*(x^3,x^0) f_+(x^{3},x^{0'})+f_-(x^3,x^{0'}) f_-^*(x^{3},x^0)\}
\nonumber\\
&&\hspace*{0.7cm}\times 
\delta (x^0-x^{0'})=1.
\label{eqfm}
\end{eqnarray} 
The quark Lagrangian density in the Dirac representation becomes
\begin{eqnarray}
{\cal L}_Q&&= \bar \Psi_\4d  \gamma^0 \Pi^0 \psi_\4d - \bar \Psi_\4d  \gamma^3 \Pi^3 \Psi_\4d 
\nonumber\\
&&\hspace*{0.1cm}
- \bar \Psi_\4d  \gamma^1   \Pi^1 \Psi_\4d  - \bar \Psi_\4d  \gamma^2 \Pi^2 \Psi_\4d  
- m  \bar \Psi_\4d  \Psi_\4d \nonumber\\
&&= ( G_1^* G_1 + G_2^* G_2 ) (  f_+^* \Pi^0f_+  + f_-^* \Pi^0f_-)
\nonumber\\
&&\hspace*{0.1cm}
 - ( G_1^* G_1 + G_2^* G_2 )(  f_+^*\Pi^3 f_+  -   f_-^*\Pi^3    f_-)
\nonumber\\
&&\hspace*{0.1cm}
 - ( G_1^* \Pi^1 G_2 + G_2^*\Pi^1  G_1 )(  f_+^* f_-  +   f_-^*    f_+)
\nonumber\\
&&\hspace*{0.1cm}
 -  ( -G_1^* ~i\Pi^2~G_2 + G_2^*~i\Pi^2~ G_1 )(  f_+^* f_-  +   f_-^*    f_+)
\nonumber\\
&&~~~ - m (G_1^*G_1 - G_2^*  G_2 )(  f_+^* f_-+   f_-^* f_+).
\label{10A}
\end{eqnarray}
As shown in Appendix A, the minimization of the action integral with
respect to $G_{\{1,2\}}^* f_\pm^*$ leads to the equations of motion
\begin{subequations}
  \begin{eqnarray}
&&\!\!\!\delta^2 {\cal L}/\delta f_+^* \delta G_1^*
\nonumber\\
&&\!\!=\!\!( \Pi^0\! -\!\Pi^3) G_1 f_+  \!-\! m G_1   f_- 
 \!-\!(\Pi^1\!-\! i\Pi^2) G_2  f_-\!\!=\!0,~~~~
\label{a}
\\
&&\!\!\!\delta^2 {\cal L}/\delta f_+^* \delta G_2^*
\nonumber\\
&&\!\!=\!\!( \Pi^0 \!-\!\Pi^3) G_2 f_+  \!+\! m G_2   f_- \!-\!  (\Pi^1\!+\!i\Pi^2) G_1  f_- \!=\!0,~~~~
\label{b}
\\
&&\!\!\!\delta^2 {\cal L}/\delta f_-^* \delta G_1^*
\nonumber\\
&&
\!\!=\!\!
  ( \Pi^0 \!+ \! \Pi^3)G_1    f_-  \!- \!m G_1  f_+  \!-\! (\Pi^1\!- \! i\Pi^2)G_2     f_+ \!=\!0 ,~~~~
  \label{c}
\\
&&\!\!\! \delta^2 {\cal L}/\delta f_-^* \delta G_2^*
\nonumber\\
&&
\!\!=\!\! (\Pi^0\!+\!\Pi^3) G_2    f_- \!+\! m   G_2  f_+ 
 \!-\!(\Pi^1  \!+\!  i\Pi^2) G_1      f_+ \!=\!0,~~~~~
 \label{d}
\end{eqnarray}
\end{subequations}
which agree with Eqs.\ (13a)-(13d) of \cite{Won09} using the Dirac
representation.  For gauge fields $A^1$ and $A^2$ (and consequently,
$\Pi^1$ and $\Pi^2$) independent of the longitudinal coordinates as we
assumed in the model of \cite{Won22} and Fig. 1, Appendix A shows that
the above set of four equations can be separated into a set of two
transverse equations for $G_{\{1,2\}}(\bb r_\perp) $ and another set
of two longitudinal-temporal equations for $f_\pm(X) $.  The set of
equations for the transverse degrees of freedom are
\begin{subequations}
 \begin{eqnarray}
&&\hspace*{-0.3cm} -  m_T G_1(\bb r_\perp) \!+\!  
  mG_1 (\bb r_\perp) + (\Pi^1- i\Pi^2) G_2 (\bb r_\perp) \!=\!0,
  \label{24a}
 \\
 &&\hspace*{-0.3cm}
   -  m_T G_2(\bb r_\perp)- m G_2 (\bb r_\perp) \!+\! (\Pi^1+i\Pi^2) G_1(\bb r_\perp) \!=\!0,
   \label{24b}
\end{eqnarray}
\label{n19}
\end{subequations}
\hspace*{-0.2cm}where 
we have introduced 
the transverse mass $m_T$ as the constant of the separation of
variables and the eigenvalue of the set of transverse equations.  They
lead to the transverse equations,
\begin{subequations}
\label{21}
\begin{eqnarray}
&& \left \{ \Pi_T^2 + i [\Pi^2,\Pi^1] + m^2 -  m_T^2\right \} G_2 (\bb r_\perp) =  0,
\label{21a} 
 \\
 &&
 \left \{ \Pi_T^2 + i [\Pi^1,\Pi^2]+ m^2 - m_T^2 \right \} G_1(\bb r_\perp)=  0,
 \label{21b}  
\end{eqnarray}
\end{subequations}
 where $\Pi_T^2=(\Pi^1)^2+(\Pi^2)^2$.  The longitudinal equations are
 \begin{subequations}
 \begin{eqnarray}
&&\int d\bb r_\perp ( |G_1(\bb r_\perp)|^2 + |G_2(\bb r_\perp)|^2 )( \Pi^0 -\Pi^3) f_+ (X)
\nonumber\\
&&\hspace*{3.5cm}
 - m_T f_-  (X) =0 , 
\label{13}
 \\
&&\int d\bb r_\perp ( |G_1(\bb r_\perp)|^2+ |G_2(\bb r_\perp)|^2) ( \Pi^0 +\Pi^3)f_-  (X) 
\nonumber\\
&&\hspace*{3.5cm}
- m_T f_+ (X)  =0 .
\label{14}
\end{eqnarray}
\end{subequations}
The operators $\Pi^\mu$ with $\mu=0,3$ in Eqs.\ (\ref{13}) and
(\ref{14}) are actually $\Pi_\4d^\mu $=$p^\mu + g_\4d A_\4d^\mu(\bb
r_\perp,X)$, where we have added the label 4D to indicate that the
gauge fields $A_\4d^\mu(\bb r_\perp,X)$ are functions of the (3+1)D
coordinates $(\bb r_\perp,X) $ and the constant $g_\4d$ is the
dimensionless coupling constant of the gauge fields in (3+1)D
space-time.  In contrast, in the idealized (1+1)D$_{\{x^0,x^3\}}$
space-time with a structureless string, the corresponding gauge fields
$A_\2d^\mu(X)$ are functions only of the coordinates $X$, and they
satisfy the (1+1)D Maxwell equation with a coupling constant $g_\2d$.

We would like to cast Eqs.\ (\ref{13}) and (\ref{14}) into the form of
a Dirac equation for the quark fields in the idealized
(1+1)D$_{\{x^0,x^3\}}$ space-time, by separating the transverse degrees
of freedom under a coupled treatment.
This can be achieved  by integrating over the
transverse coordinates $\bb r_\perp$ in Eqs.\ (\ref{13}) and
(\ref{14}) and identifying the term as as $A_\2d^\mu(X)$ in the (1+1)D
space-time with the the coupling constant $g_\2d$,
  \begin{eqnarray}\label{A4-2}
&&\!\!\!g_\4d \!\!\int \!\!d\bb r_\perp  \!(|G_1(\bb r_\perp\!)|^2 \!\!+\! |G_2(\bb r_\perp\!)|^2 )A_\4d^\mu(\bb r_\perp,X\!) 
\!\equiv \!g_\2d 
A_\2d^\mu\!\!(X),\nonumber\\
\label{AA}
\end{eqnarray}
for $\mu=0,3.$
As shown in Eq.\ (B10) in Appendix B, consistent gauge field solutions
of the Maxwell equations in (3+1)D and (1+1)D in the flux tube
environment are related by
\begin{eqnarray}\label{AA4-2}
 \!\!A_\4d^{\mu} \!(\bb r_\perp\!,\!X\!)\!=\!\frac{g_\4d}{g_\2d}\! \biggl [\!G_1^*\!(\bb r_\perp\!)G_1 \!(\bb r_\perp\!)\!+\! G_2^*\!(\bb r_\perp\!) G_2 \!(\bb r_\perp\!)\!\biggr ] \! A_\2d^\mu \!(\!X\!),~~
 \label{18A}
\end{eqnarray}
for $ \mu$=0,3.
As a consequence, the coupling constant $g_\2d$ and $g_\4d$ are
related by
\begin{eqnarray}\label{g4-2}
g_\2d^2\! =\!   g_\4d^2\! \int \! d \bb r_\perp \biggl  [G_1(\bb r_\perp)^*G_1(\bb r_\perp)\!  +\!  G_2(\bb r_\perp)^* G_2(\bb r_\perp) \biggr  ]^2\!\!\!,~~~~
\label{18}
\end{eqnarray}
which shows that the coupling constant $g_\2d$ now acquires the
dimension of a mass.  By such an introduction of $A_\2d^\mu(X)$ and
$g_\2d$, Eqs.\ (\ref{13}) - (\ref{14}) become
\begin{subequations}
\label{26a}
\begin{eqnarray}
&&( \Pi_\2d^0 -  \Pi_\2d ^3) f_+ (X) - m_T f_-(X)=0, 
\label{19a}
\\
&&( \Pi_\2d^0+  \Pi_\2d^3) f_- (X) - m_T f_+(X)=0,
\label{19b}
\end{eqnarray}
\end{subequations}
where  
\begin{eqnarray}
&&\Pi_\2d^\mu=p^\mu+g_\2d A_\2d^\mu(X),~~~~ \mu=0,3.
\end{eqnarray} 
These equations (\ref{19a}) and (\ref{19b}) can be cast further into
the following matrix equation
\begin{eqnarray}
&&\biggl \{ \begin{pmatrix} 0 & 1 \\ 1 & 0 \end{pmatrix} \Pi_\2d^0
- \begin{pmatrix} 0 & -1 \\ 1 & 0 \end{pmatrix} \Pi_\2d ^3 
\! - \!  m_T \biggr \}  \begin{pmatrix} f_+(X)\\ f_ - (X)\end{pmatrix}  \! =\!  0.~~
\label{16}
\end{eqnarray}
It is then useful to introduce 2D gamma matrices $\gamma_\2d^\nu$ as in
\cite{Wit84}
\begin{eqnarray}
&& \gamma_\2d^0= \begin{pmatrix} 0 & 1 \\ 1 & 0 \end{pmatrix},~~~
\gamma_\2d^3=\begin{pmatrix} 0 & -1 \\ 1 & 0 \end{pmatrix}, 
\nonumber\\
&&\gamma_\2d^5=\gamma_\2d^0\gamma_\2d^3=\begin{pmatrix}  1 &  0 \\ 0 & -1 \end{pmatrix}, 
\end{eqnarray}
and the longitudinal 2D wave function $\psi_\2d(X)$,
\begin{eqnarray}
\psi_\2d (X) = \begin{pmatrix} f_+(X)\\ f_ -(X) \end{pmatrix} ,
\label{22}
\end{eqnarray}
to rewrite Eq.\ (\ref{16}) as a Dirac equation for the longitudinal
(1+1)D motion, interacting with a gauge field with a coupling constant
$g_\2d$
\begin{eqnarray}
&&\{ \gamma_\2d^0 \Pi_\2d^0 
       - \gamma_\2d^3  \Pi_\2d ^3  - m_T \} \psi_\2d(X) =0 ,~~~ \mu=0,3,
       \label{24}
\end{eqnarray}
which can be re-written as   
\begin{eqnarray}
&&\hspace*{-0.2cm}\{ \gamma_\2d^\mu ( p_\mu + g_\2d A_\mu^\2d (X)) \!- \!m_T \}  \psi_\2d(X) \!=\!0 ,~~\mu=0,3.
\label{32}
\end{eqnarray}
It is important to note that in idealizing the dynamics from (3+1)D
to (1+1)D in the above manipulations, relevant information regarding
the transverse degrees of freedom is now subsumed and stored in the
quark transverse mass eigenvalue $m_T$ of the set of eigenvalue
equations, Eq.\ (\ref{21}), and $g_\2d$ which is related to $g_\4d$
and the transverse eigenfunctions $G_{\{1,2\}}(\bb r_\perp) $ by
Eq.\ (\ref{18}).  By such supplementary relations, the information on
the transverse degrees of freedom is packaged and passed on from the
(3+1)D to the idealized (1+1)D dynamics.  

The Dirac equation of motion in (1+1)D space-time corresponds to the
2D action integral for quarks in the form
\begin{eqnarray}\label{AF}
 &&\!\!{\cal A }_{\2d Q}  \!=\!    Tr \!\! \int\! \!d X \Biggl\{ \!
\bar \psi_{\2d}(X)  \gamma_{\2d}^\mu( p_\mu \! 
+ \!g_{\2d}  A_{\mu}  (X))\psi_{\2d}(X) 
\nonumber\\
&&~~~-
m_T {\bar
\psi}_{\2d}  (X) 
 \psi_{\2d}  (X) \Bigg\} ,
 \label{24A}
\end{eqnarray}
with the normalization condition (\ref{Norm-cond}) which can be
re-written in terms of $\bar \psi_\2d, \psi_\2d $ functions as
\begin{eqnarray}
\int\!\! dx^3 dx^0 \bar \psi_\2d (x^3,x^0) \gamma^0_\2d \psi_\2d ({x^{3}},{x^{0}}')
\delta (x^0 \!-\!{x^{0}}')\!=\!1.
\end{eqnarray}
For brevity of notation, the descriptors 2D and 4D are often omitted
unless otherwise needed to resolve ambiguities, as the 2D and 4D
nature of each function or gamma matrix can be inferred from the
arguments of the function or from the context.

\section{Gauge field part of the action integral}

The confinement properties of the quark-QCD-QED system consists of the
transverse confinement on the transverse plane and the longitudinal
confinement.  The confinement pertains not only to the quarks and
antiquarks, but also to the gauge fields which link the quarks and the
antiquarks.  That is, in compact QED and non-Abelian QCD, the gauge
fields also self-interact among themselves so that they do not readily
radiate away.  They will also retain the confinement properties when
the quark-QCD-QED system settles into one of the meson states.

At the moment of the creation of the quark and the antiquark pair for
a QCD or a QED meson state, the gauge fields are also created.  We can
apply Polyakov's result on the confinement in (2+1)D to infer the
presence of transverse confinement of the quarks, antiquarks, and the
gauge fields at the moment immediately after the creation of the
$q\bar q$ pair.

The occurrence of transverse confinement makes it convenient to
partition the gauge field actions into three pertinent parts,
\begin{eqnarray}
&&{\cal A}_A   =  {\cal A}_{A }^{\rm I}+   {\cal A}_A^{\rm II} +   {\cal A}_A^{\rm III},
\end{eqnarray}
where ${\cal A}_{A}^{\rm I}$ and ${\cal A}_{A}^{\rm II}$ pertains to
transverse gauge fields and transverse confinement, while ${\cal
  A}_{A}^{\rm III}$ is concerned mainly with longitudinal gauge field
dynamics and longitudinal confinement:
\begin{subequations}
\begin{eqnarray}
{\cal A}_A^{\rm I} &&\!=\!- \!\int \!d^4 x  \frac{1}{\pi^2R_T^4 g_\4d^2}~Tr [1-\cos ( \pi R_T^2 g_\4d F_{12})] ,
\\
{\cal A}_A^{\rm II}&&\!\!=\!\!-\! \!\int\! \!d^4 x  Tr \!\frac{1}{2}  \!\left [ \!F_{23}F^{23}\!\!+\!F_{13}F^{13}\!\!+\!F_{01}F^{01}\!\!+\!F_{02}F^{02}\!\! \right ]\!,\!~~~~~~
\label{36a}
\\
{\cal A}_A^{\rm III}&&=- \int d^4 x  Tr ~ \frac{1}{2}  F_{03}F^{03}.
\label{36cn}
\end{eqnarray}
\label{36}
\end{subequations}
We can alternatively express the gauge field tensors $F^{\mu \nu}$ in
terms of the gauge field vectors as $\bb E$=$\sum_{i=1}^3E^i \bb e^i$
and $\bb B$=$\sum_{i=1}^3B^i \bb e^i$.  The action partition ${\cal
  A}_A^{\rm I}$ contains the magnetic field $F^{12}$=$-B^3$, which
lies along the longitudinal direction perpendicular to the transverse
plane.  Such a magnetic field $\bb B$ confines the quarks and
antiquarks into their Landau level orbitals on the transverse plane,
as we shall demonstrate in Section 8.1.  The zero mode of the Landau
levels allows the quark and the antiquark to retain an effective mass
$m_T$ that is the same as the quark rest mass $m$.  Furthermore, As
discussed in detail in Drell $et~al.$ \cite{Dre79}, such a magnetic
field $\bb B$ also sets up the self-interaction between transverse
gauge $\bb A$ fields. The periodicity of the action cosine
function in ${\cal A}_A^{\rm I}$ allows the transverse photons to
interact among themselves to build up a confining interaction between
the quark, the antiquark, and the other transverse gauge fields
themselves in the transverse confinement in (2+1)D.

The second action partition ${\cal A}_A^{\rm II}$ pertains to
$F^{23}=-B^1$, $F^{31}$=$-B^2$, $F^{01}$=$-E^1$, $F^{02}$=$-E^2$ which
are the magnetic and electric field components lying on the transverse
$\{x^1,x^2\}$ plane, and they participate directly in the transverse
confinement of the quarks and the transverse gauge fields on the
transverse plane.

 In the proposed model in the stretch (2+1)D
configuration with the gauge fields that are copies of the gauge field
configuration of the initial quark and antiquark transverse
plaquettes, the transverse confinement of the gauge field will likely
be retained so that the system may be idealized as a one dimensional string for which 
 the Schwinger longitudinal confinement may be effective to  bring about 
 the longitudinal confinement of the quark and 
 the antiquark.
 
There is the third partition ${\cal A}_A^{\rm III}$ pertaining to the
action integral involving the gauge fields $F^{03}$.  Such a gauge
field participates in longitudinal dynamics with a linear interaction
between the quark and the antiquark in the form of a string.  It is
the essential ingredient in Schwinger's longitudinal confinement of
opposite charges in QED in (1+1)D.  As Schwinger's confinement of a
pair of opposite massless charges occur irrespective of the strength
of coupling constant, we expect that it will be the dominant driving
force for longitudinal confinement.  In the consideration of the
question of longitudinal confinement,  the ${\cal A}_A^{\rm I}$ and
${\cal A}_A^{\rm II}$ pertain to the quark-gluon part of the condensate.

As mentioned earlier in Section 3 and 
discussed in detail in Drell $et~al.$ \cite{Dre79}, the periodicity
of the angular function $\cos (\pi R_T^2 g_\4d F_{12})$ in the gauge
field action leads to the self-interaction of the transverse gauge
fields.  For strong coupling such as QCD, the strong self-interaction
among the transverse gauge fields leads to a linear confining
interaction joining the quark and the antiquark by a flux tube.  For
weak coupling such as the compact QED interaction, quantum
fluctuations occur which randomize the coherent self-interaction.  In
the case of compact QED in (3+1)D, such random quantum fluctuations disrupt the linear
confinement so greatly that without
dynamical-quark effects of the Schwinger longitudinal confinement, 
a static quark and a static antiquark with a weak coupling
would
not be longitudinally confined in compact QED 
\cite{Pol77,Pol87,Dre79}.  In contrast, in the case of compact QED in
(2+1)D,  however, 
such  quantum random quantum fluctuations  can never be sufficient to destroy
the linear QED gauge confinement effects in two-dimensional spatial space, no matter how
weak the compact gauge interaction coupling may be.  

For our problem with the
creation of a $q\bar q$ pair in compact QED in (3+1)D, we can take advantage of 
 such a  transverse confinement of
opposite charges in QED (2+1)D to prepare a  system  
transverse confinement after the moment of the creation of the quark and
the antiquark.  Thereafter, we stretch the incipient transversely
confined quark and antiquark longitudinally to execute the
longitudinal yo-yo motion appropriate for the $q\bar q$ at the QED
meson eigenenergy,  by
copying the configurations of the transverse links and the transverse
fields longitudinally from the incipient quark and antiquark position
$x^3(q),x^3(\bar q)\sim 0$ position to the region between 
the quark at   $x^3( q)=-x^3/2$ and the antiquark at $x^3(\bar q)=x^3/2$, in a
coherent manner as in Fig.\ 1(b).  Because the transverse links are
copies of those of the initial configuration, the transverse
confinement will be likely retained.  The remaining problem will be to
investigate whether there will be the possible longitudinal confinement in
compact QED in the stretch (2+1)D configuration as an idealized open string , as anticipated 
in the Schwinger confinement mechanism.

For the scratch (2+1)D single-plaquette configuration in Fig.\ 1(b), 
the integral in Eq.\ (\ref{36a}) will be
proportional to the longitudinal separation between the quark and the
antiquark, $|x^3(\bar q)-x^3(q)|$.  With $n$=$n_1n_2 n_3$ and $n_3
\sqrt{\pi} R_T = |x^3(\bar q) - x^3(q)|$, we have

\begin{eqnarray}
{\cal A}_A^{\rm I}
&& =\!\!- \!Tr\!\! \int \!dt\! \! \!\sum _{n_1 n_1 n_2} \!\!\!\frac{[1\!-\!\cos (\pi R_T^2 g_\4d \!F_{12}(n({\bb r}_\perp\!, X\!) ))]} { \pi  R_T^2 g_\4d^2} 
\nonumber\\
&&=\!\!- \!Tr \!\!\!\int \!dt  n_3\! \sqrt{\pi} R_T \!\!\!\sum _{n_1 n_2} \!\!\frac{
\![1\!-\!\cos (\pi R_T^2 g_\4d F_{12}(n({\bb r}_\perp, \!X\!) \!)\!)\!] } { \pi  R_T^2 g_\4d^2\!}
\nonumber\\
&&=  -  \int dt ~~\kappa_1  |x^3(\bar q) - x^3(q)|,
\nonumber\label{eq37}
\end{eqnarray}
where $\kappa_1$ is a function of $F_{12}$, 
\begin{eqnarray}
\kappa_1\! = \!\frac{n_3 (\sqrt{\pi} R_T) }{ \pi R_T^2 g_\4d^2} \!\!\!\sum _{n_1 n_2}\!\!\!Tr[1\!-\!\cos (\pi R_T^2 g_\4d F_{12}(n({\bb r}_\perp, X) ))],
\nonumber
\end{eqnarray}
which is evaluated at $X$=$\{x^0x^3\}$=$0$.
The above action gives a longitudinal linear interaction between the
quark and the antiquark arising from the gauge field $F^{12}$=$-B^3$
that is favorable for the longitudinal confinement of the quark and
the antiquark.

With the stretch (2+1)D configuration in our model, the transverse
gauge fields between the quark plaquette and the antiquark plaquette
are copies of those on the quark plaquette as in Fig. 1(b), the action
can be written as
\begin{eqnarray}
\!\!\!{\cal A}_A^{\rm II}&&\!\!\!=\! \!\!\int \!\!\!dt dx^3dx^1\! dx^2   \!Tr \frac{1}{2} \! \!\left [ F_{23}F^{23}\!\!+\!F_{13}F^{13}\!\!+\!F_{01}F^{01}\!\!\!\!+\!F_{02}F^{02} \!\right ]\!,
\nonumber\\
&&\hspace*{-0.3cm}\!\!\!\!=\!\! \int\! dt |x^3(\bar q)\! - \!x^3(q)| 
\nonumber\\
&& \!\!\!\!\!\!\times \! \! \!
\int \!\!dx^1 dx^2 Tr\! \frac{1}{2}\!
 \left [ \!(B^1)^2 \!\!+\! (B^2)^2 \!\!+\! (E^1)^2\! \!+ \!(E^2)^2  \!\right ]\! \!\biggr |_{X=\{x^0,x^3\}=0} 
\nonumber\\
&&\hspace*{-0.3cm}\!\!\!\!=- \int dt ~ \kappa_2 |x^3(\bar q) - x^3(q)|, 
\nonumber \label{eq39}
\end{eqnarray}
where $\kappa_2$ is 
\begin{eqnarray}
\kappa_2\!\! =\! \!\!\int \! \!\! dx^1 \!dx^2   \!Tr\! \frac{1}{2}\!\! \left [ (B^1\!)^2 \!\!+\! (B^2\!)^2 \!\!+ \!(E^1\!)^2 \!\!+ \!(E^2\!)^2  \right ]\! \biggr |_{X=\{x^0,x^3\}=0}\!\!\!.
\nonumber
\end{eqnarray}
The action ${\cal A}_A^{\rm II}$ also contribute a linear interaction that will facilitate longitudinal confinement.  

%

Having simplified the gauge field actions ${\cal A}_A^{\rm I}$ and
${\cal A}_A^{\rm II}$, we come to examine ${\cal A}_A^{\rm III}$ of
Eq.\ (\ref{36}) pertaining to $F^{03}$ of the action integral.  We
would like to find out what will be the form of the gauge action
${\cal A}_A^{\rm III}$ when the gauge fields $A_\mu(\bb r_\perp,X)$ in
(3+1)D and the gauge fields $A_\mu(X)$ in (1+1)D, with $\mu=0,3$, are
related to each other by Eq.\ (\ref{18A}).  Equations (\ref{4b}) and
(\ref{18A}) give $F_{03}({\bb r}_\perp,X)$ in four-dimensional
space-time as
\begin{eqnarray}
&&F_{03}({\bb r}_\perp,X) 
\nonumber\\
&& \hspace*{0.2cm}= \frac{g_{4D}}{g_{2D}}[{ |G_1({\bb
r}_\perp)|^2+|G_2({\bb
    r}_\perp)|^2}]^{} 
[\partial_0 A_3( X)-\partial_3 A_0( X)]
\nonumber\\
&& \hspace*{0.3cm}
 - i  \frac{g^2_{4D}}{g_{2D}}  [{ |G_1({\bb r}_\perp)|^2+|G_2({\bb
    r}_\perp)|^{}}] [A_0( X),A_3(X)]. \label{eq21}
\end{eqnarray}
On the other hand, the gauge field $F_{03}$ in (1+1)D$_{\{x^3,x^0\}}$
space-time is given, by definition, as
\begin{eqnarray}\label{eq22}
F_{03}(X)&=&\partial_0 A_3(X)-\partial_3 A_0(X)
-i{ g}_{2D} [A_0(X),A_3(X)].
\nonumber
 \end{eqnarray}
As a consequence, $F_{03}(X)$ in two-dimensional space-time and
$F_{03}({\bb r}_\perp, X)$ in four-dimensional space-time are related
by
\begin{eqnarray}\label{eq23}
&&F_{03}({\bb r}_\perp, X)
\nonumber\\
&&\hspace*{0.2cm}= \frac{g_{4D}}{g_{2D}} [{ |G_1({\bb
r}_\perp)|^2+|G_2({\bb
    r}_\perp)|^2}]^{}
    \nonumber\\
    &&\hspace{1.5cm} \times \{ F_{03}( X)+i{ g}_{2D} [A_0(X),A_3(X)]\}
\nonumber\\
&&\hspace*{0.3cm} - i  \frac{g^2_{4D}}{g_{2D}}   [{ |G_1({\bb
r}_\perp)|^2+|G_2({\bb
    r}_\perp)|^2}]^{} [A_0(X),A_3(X)].
\end{eqnarray}
The above equation can be re-written as
\begin{eqnarray}\label{eq24}
&&F_{03}({\bb r}_\perp, X) = \frac{g_{4D}}{g_{2D}} [{ |G_1
({\bb r}_\perp)|^2
+|G_2({\bb   r}_\perp)|^2}]
\nonumber\\
&&\hspace*{0.2cm}
\times \biggl \{ F_{03}( X)+\biggl [  i { g}_{2D} - i  \frac{g^2_{4D}}{g_{2D}}  [{ |G_1({\bb r}_\perp)|^2
+|G_2({\bb  r}_\perp)|^2}]^{}\biggr ] 
\nonumber\\
&& \hspace*{2.4cm}\times [A_0(X),A_3(X)] \biggr \}.
\end{eqnarray}
The term involving the product $F_{03} (x)F^{03}(x)$ in Eq.\ (\ref{36cn}) becomes
\begin{eqnarray}
&&\int d^4x F_{03}({\bb r}_\perp, X) F^{03}(x^0,x^3,{\bb
r}_\perp) 
\nonumber\\
&&=\int d^4x  [{ |G_1({\bb r}_\perp)|^2+|G_2({\bb
    r}_\perp)|^2}]\biggl \{ F_{03}( X)F^{03}( X)
\nonumber\\
&& \hspace*{1.2cm}+\biggl [ i { g}_{2D} - i  \frac{g^2_{4D}}{g_{2D}}  [{( |G_1({\bb
r}_\perp)|^2+|G_2({\bb
    r}_\perp)|^2})]^{}\biggr ]^2
    \nonumber\\
   && \hspace*{1.6cm} \times \biggl (  [A_0(X),A_3(X)]
 [A^0(X),A^3(X)] \biggr )
\biggr \}.
\end{eqnarray}
The action integral ${\cal A}_A^{\rm III}$ in Eq.\ (\ref{36cn})
involves the integration of the above quantity over $x_1$ and
$x_2$. Upon integration over $x^1$ and $x^2$, the second term inside
the curly bracket of the above equation, is zero,
\begin{eqnarray}
\label{second} &&\hspace*{-0.2cm}\int dx^1 dx^2 [{( |G_1({\bb
r}_\perp)|^2+|G_2({\bb
    r}_\perp)|^2})]
    \nonumber\\
    &&\hspace*{0.1cm}\times \biggl [i{ g}_{2D} \!-\! i \frac{g^2_{4D}}{g_{2D}} [{( |G_1({\bb r}_\perp)|^2+|G_2({\bb
    r}_\perp)|^2})]^{1/2}\biggr ] \!= \!0 ,~~~
\label{25}
\end{eqnarray}
where we have used the relation between ${ g}_{\2d}$ and ${ g}_{\4d}$
as given by Eq.(\ref{18}) and the normalization condition of
(\ref{Norm-cond}). As a consequence, the integral of $F_{03}
(x)F^{03}(x)$ in Eq.\ (\ref{36cn}) becomes
\begin{eqnarray}
&&\int d^4x F_{03}({\bb r}_\perp, X) F^{03}(x^0,x^3,{\bb
r}_\perp) 
\nonumber\\
&&=\int d^4x  [{ |G_1({\bb r}_\perp)|^2+|G_2({\bb
    r}_\perp)|^2}]\biggl \{ F_{03}( X)F^{03}( X)
\nonumber\\
&& \hspace*{1.2cm}+\biggl [ i { g}_{2D} - i  \frac{g^2_{4D}}{g_{2D}}  [{( |G_1({\bb
r}_\perp)|^2+|G_2({\bb
    r}_\perp)|^2})]^{}\biggr ]^2
    \nonumber\\
   && \hspace*{1.6cm} \times \biggl (  [A_0(X),A_3(X)]
 [A^0(X),A^3(X)] \biggr )
\biggr \}.
\end{eqnarray}
For the second term in the curly bracket, the integral over $dx^1$ and
$dx^2$ is
\begin{eqnarray}
&&\hspace*{-0.2cm}\int dx^1 dx^2  [{ |G_1({\bb r}_\perp)|^2+|G_2({\bb
    r}_\perp)|^2}]\nonumber\\
    &&\hspace*{0.2cm}\times \biggl [ i { g}_{2D} - i \frac{g^2_{4D}}{g_{2D}}  [{( |G_1({\bb
r}_\perp)|^2+|G_2({\bb
    r}_\perp)|^2})]^{1/2}\biggr ]^2,
\end{eqnarray}
which can be considered as an integral over ${ g}_{2D}$ in the form
\begin{eqnarray}
&&\frac{2}{g_{2D}} \int d(g^2_{2D})\int dx^1 dx^2  [{ |G_1({\bb
r}_\perp)|^2+|G_2({\bb
    r}_\perp)|^2}] 
    \nonumber\\
    &&\hspace*{0.6cm}\times [ i  g^2_{2D} - i g^2_{4D}  [{( |G_1({\bb
r}_\perp)|^2+|G_2({\bb
    r}_\perp)|^2})]^{} ]= 0.
\end{eqnarray}

Because of Eq.\ (\ref{25}), the above integral gives an irrelevant
constant which we can set to zero.  After these manipulations, we
obtain
\begin{eqnarray}\label{eq29}
&&\int d^4x F_{03}({\bb r}_\perp, X) F^{03}({\bb
r}_\perp,X) 
\nonumber\\
&&\hspace*{0.3cm}= \int d x^1 d x^2  dx^3 dx^0
 [{( |G_1({\bb r}_\perp)|^2+|G_2({\bb
    r}_\perp)|^2})] 
 \nonumber\\
&&  \hspace*{1.9cm}\times
    F_{03}(X)F^{03}(X)
\nonumber\\
&&\hspace*{0.3cm}= \int   dx^3 dx^0F_{03}(X)F^{03}(X).
\end{eqnarray}

We therefore obtain the longitudinal action  ${\cal A}_A^{\rm III}$ as 
\begin{eqnarray}
{\cal A}_A^{\rm III}
= Tr \int dX  \left \{ - {1\over 2 } 
F_{03}(X)F^{03 }(X)\right \}.
\label{51}
\end{eqnarray}

\section{Total action integral }

Summing up all terms above in Eqs.\ (\ref{eq37}), (\ref{eq39}), and
(\ref{51}), we have reduced the action integral ${\cal A}_\4d$ in (3+1)D,
Eq.\ (\ref{eq2}), 
 to the action integral ${\cal
  A}_\2d$ in (1+1)D given by
\begin{eqnarray}\label{2daction}
{\cal A}_{\2d}
 &&  \!=\!    Tr \!\! \int\! \!d X \Biggl\{ \!
\bar \psi_{}(X)  \gamma_{2D}^\mu( p_\mu \! 
+ \!g_{2D}  A_{\mu}  (X))\psi_{}(X) 
\nonumber\\
&&
\hspace*{1.0cm}-m_T~
{\bar \psi}_{}  (X)  \psi_{}  (X),
 - {1\over 2 }  F_{03}(X)F^{03 }(X)  \Biggl\}\nonumber \\
&& 
+ A_A^I + A_A^{II},
\label{eq34}
\end{eqnarray}
where $A_A^I$+$A_A^{II}$ are given by Eqs.\ (\ref{eq37}) and (\ref{eq39}) explicitly as
\begin{eqnarray}
A_A^I + A_A^{II}=- \int dt ~(\kappa_1+\kappa_2)  |x^3(\bar q) - x^3(q)|,
\end{eqnarray}
with 
\begin{eqnarray}
\kappa_1&&\!\!=\! \frac{n_3 (\sqrt{\pi} R_T) }{ \pi R_T^2 g_\4d^2} \!\sum _{n_1 n_2}Tr[1\!-\!\cos (\pi R_T^2 g_\4d B^3 (n({\bb r}_\perp, X) ))]
\nonumber\\
\kappa_2 &&\!\!=\!\int dx^1 dx^2 ~  Tr ~ \frac{1}{2} \left [ (B^1)^2 + (B^2)^2 + (E^1)^2 + (E^2)^2  \right ] ,
\nonumber\!
\end{eqnarray}
where $X$ is evaluated at $X$=$\{x^0x^3\}$=0.
The subscript label of `2D' is to indicate that this is the action
integral involving the integration over the 2D space-time coordinates,
$X$=($x^3$,$x^0$).  This action integral can be the starting point for
the study of problems in the longitudinal (1+1)D$_{\{x^3,x^0\}}$
space-time in which the properties of the gauge-field coupling
constant $g_\2d$, and the transverse mass $m_T$ are input properties
of the open-string parameters fixed by the initial choice of the class
of transverse states in question, representing the ingredients that
depend on the transverse degrees of freedom.  The
longitudinal-temporal space-time dynamics can be treated in the
simpler (1+1)D$_{\{x^3,x^0\}}$ space-time, to facilitate the
examination of the complex problem of the quark-QCD-QED system.
 
Previously, the Schwinger model of QED in (1+1)D has been considered
as an unrealistic ``toy'' model incapable of quantitative connection
to the realistic fermion or quark systems in (3+1)D space-time.  A
major difficulty arises because the coupling constant $g_\2d$ in\break
Schwinger's (1+1)D space-time has the dimension of a mass.  In the
physical (3+1)D space-time, the coupling constant $g_\4d$ is
dimensionless, but the flux tube has a structure with a radius $R_T$.
In contrast in (1+1)D, the open-string does not have a structure, but
the coupling constant has the dimension of a mass.  It was recently
realized that in reducing from the dynamics in (3+1)D space to the
dynamics in (1+1)D, the information in the structure of the flux tube
is in fact stored in the coupling constant $g_\2d$ and the transverse
mass $m_T$ in (1+1)D \cite{Won10}.  The coupling constants in
different dimensional space-time systems are connected by
Eq.\ (\ref{18}) or by the radius of the flux tube as given by
Eq.\ (\ref{76}) below \cite{Won09,Won10,Kos12}, and the transverse
mass $m_T$ is obtained by solving the eigenvalue equation in the
transverse plane Eqs.\ (\ref{n19}).  The (1+1)D action integral in
Eq.\ (\ref{eq34}) shows that the Schwinger model in (1+1)D can be viewed as a
realistic idealization of the system in (3+1)D space-time, capable of
semi-quantitative confrontation with experiment \cite{Won20} and
consistent with the string picture of hadron physics as developed in
\cite{Ven68,Nam70,Nam74,Got71,tho74a,tho74b,Cas74,Kog75,Man75,Pol77,Pol81,Pol87,tho81,Art74,And83,Won94},
when the flux tube radius and the transverse mass are properly taken into account.

\section{Dynamics of  the quark fields  and the gauge fields}

\subsection {  Transverse confinement   of quarks and antiquarks in 
a longitudinal magnetic field $\bb B$ } We shall discuss the
transverse confinement of quarks in this subsection and the
longitudinal confinement of quarks and gauge fields in the next
subsection.

For quarks interacting with the non-Abelian QCD gauge field leading to
bound hadrons with energies above the pion mass, the transverse
confinement arises from the strong non-perturbative self-interaction
among the gluons linking color charges.  The Feynman diagram loops
involving self-interacting gluons give rise to the dominance of planar
diagrams over non-planar diagrams, leading to the string description
of the interaction between color charges with both transverse and
longitudinal confinement, as described in \cite{tho74}.  We do not
need to examine the question of the transverse confinement in QCD
again.
 
Focusing on the question of the transverse confinement of quarks and
antiquarks in QED, we mentioned earlier in the proposed ``stretch
(2+1)D'' model \cite{Won22} in Section 3 that by the application of
Polyakov QED-generated transverse confinement in (2+1)D, we invoke the
QED-generated transverse confinement to created the
QED-transversely-confined quark, antiquark, and the gauge fields at
the creation of the $q\bar q$ pair.   The creation of the transversely-confined
$q\bar q$ pair and the $A^\mu$ fields of the QED meson necessitates
the creation of the longitudinal magnetic field
$\bb B$(=$\nabla $$\times $${\bb A})$ perpendicular to the transverse
$\{x^1,x^2\}$ plane, along the longitudinal $x^3$ directions.  It acts
as if the quark and the antiquark behave as effectively as magnetic
monopoles bound by a magnetic flux $\bb B$ between the quark magnetic
monopole and the antiquark magnetic monopole \cite{Pol74,tho74,Man75}.

Immediately after their creation, the light quark and the light
antiquark of opposite charges emerge from the creation point and
stretch outward longitudinally to execute a longitudinal yo-yo motion,
in accordance with the equation of motion appropriate for such a
confined QED meson, with the transverse gauge fields to be
longitudinal copies of the transverse gauge field at $x^3=0$.  We
would like to show explicitly here that the longitudinal $\bb B$ field
that exists at the moment of creation to confine the quark and the
antiquark also will continue to confine the quarks and the antiquarks
on the transverse planes because of Landau level dynamics as described below.

We examine  the case with an assumed cylindrical
symmetry in which a momentary snapshot of the quark, the antiquark,
and the gauge fields at a time shortly after creation is shown in
Fig.\ \ref{fig1}($b$).  We ascribe the  (Cartesian) transverse gauges fields $(A^1, A^2)$
or  (cylindrical)    $(A^{r_\perp},A^\phi)$
as arising from the magnetic field $\bb B$=$B\bb e^3$ along the
longitudinal direction, confined within the flux tube of radius
$R_T$,
\begin{eqnarray}
&&\hspace*{-0.2cm}\bb B = \bb e^3   B_z(r_\perp) =\bb e^3 \frac{  B_0}
{1+e^{(r_\perp-R_T)/ a_B}},~~  B^1=B^2=0,\nonumber\\
\label{eq44A}
\end{eqnarray}
where $R_T$ is the radius of the flux tube and $a_B$ is the
diffuseness parameter.  We shall assume that $R_T >>
a_B$ so that the flux tube is well defined with a sharp surface.  
For the cylindrical coordinate system with unit vectors  
$\bb e_{r_\perp}$, $\bb e_\phi$ and  $\bb e_z $, we have 
\begin{eqnarray}
&& \bb A\!=\! \bb e^{r_\perp}  A^{r_\perp} \! +\!   \bb e^\phi A^\phi  \!+\! \bb e^z  A^z, ~~~ \text{and} ~~  A_{r_\perp}  = 0, ~~ A_z=0, 
\nonumber\\
&& \nabla \times \bb A\! =\! 
       \bb e^{r_\perp}\!\left ( \frac{1}{r_\perp}  \frac{ \partial A^z  }{\partial \phi } 
                           - \frac{      \partial A^\phi }{\partial z}     \right )     
\!+\!      \bb e^\phi \left (   \frac{ \partial A^{r_\perp} }{\partial z } 
                           - \frac{      \partial A^z }{\partial r_\perp}     \right )   
\nonumber\\
&&\hspace*{1.2cm}  
+     \bb e^z \frac{1}{r_\perp} \left (   \frac{ \partial }{\partial r_\perp } (r_\perp A^\phi) 
                          \! -\! \frac{      \partial A^r }{\partial \phi}     \right )    .            
\end{eqnarray}
From the magnetic field $\bb B$ of (\ref{eq44A}), we can solve for $A^\phi$ and we get 
\begin{eqnarray}
&&A^\phi(r_\perp)= B_0 \biggl \{  \frac{r_\perp}{2}+
a_B \sum_{n=1}^\infty    \frac{1} {n}(-1 )^n  e^{-n R_T/ a_B}      
\nonumber\\
&&\hspace*{2.8cm}
\times \biggl [  e^{ nr_\perp/a_B }     +\frac{1-e^{ nr_\perp/a_B } }{nr_\perp/a_B } \biggr ]    \biggr \}.
\end{eqnarray}
We can carry out  a Taylor expansion about $r_\perp=0$.
We expand the second term in powers of $r_\perp$ at $r_\perp\sim 0$ and 
get 
\begin{eqnarray}
&&\hspace{-0.3cm}A^\phi(r_\perp)\!= \!B_0 \biggl \{  \frac{r_\perp}{2}\!+\!\!
\sum_{i=1} \frac{i}{i+1}~\frac{1}{i ! }  
\nonumber\\
&&\hspace*{1.6cm} \times  \biggl [ \sum_{n=1}^\infty    \frac{1} {n}(-1 )^n  e^{-n R_T/ a_B}      
      (n r_\perp/a_B)^i  
  \biggr ]   \biggr \}  ,
\end{eqnarray}
which is approximately
\begin{eqnarray}
 A^\phi(r_\perp,\phi) =     \frac{B_0 r_\perp}{2}   \Theta (R_T - r_\perp),
\end{eqnarray}
where $\Theta(  R_T - r_\perp  )$ is the step function. 
Because Eq.\ (\ref{n19}) is in the Cartesian coordinate system, we  
represented $A^\phi$ in terms of the Cartesian components $A^1$ and $A^2$ within the flux
tube of radius $R_T$, 
\begin{eqnarray}
(A^1,A^2)=\left (-\frac {B_0 x^2}{2}, \frac{B_0 x^1}{2} \right )  \Theta (R_T - r_\perp) .
\label{60}
\end{eqnarray}
Such a gauge field are defined in (3+1)D and we shall assume
that they are approximately independent of the longitudinal
coordinate $x^3$.  Such a transverse gauge field $A^1$
and $A^2$ will lead to the confinement of the quarks and
the antiquarks on the transverse plane.
For the equations of transverse motion for quarks in
Eq.\ (\ref{n19}), the momentum operator $\Pi^{\{1,2\}}$ is
\begin{eqnarray}
\Pi^{\{1,2\}}=p^{\{1,2\}} - e Q  A^{\{1,2\}},
\end{eqnarray}
where $e=g^\qed$=$g_\4d^\qed$ is the QED coupling constant, and $Q$ is the quark
electric charge,
\begin{eqnarray}
Q= \sg Q |Q|.
\end{eqnarray}
The magnitude of the quark electric charge $|Q|$ and the sign of the
charge,  $\sg Q$, are flavor-dependent.  In the magnetic field of Eq.\ (\ref{60}), Eq.\ (\ref{n19})
becomes
\begin{subequations}
\label{52}
\begin{eqnarray}
&& \hspace*{-0.3cm}  [ (p^1\!\!-\!eQ A^1)\!- \!i(p^2\!\!-\!eQA^2)]G_2(\bb r_\perp\!)\! =\!
 [  m_T  \!- \! m]G_1(\bb r_\perp\!) \!,\hspace*{0.75cm}
\label{eq52a}\\
 && \hspace*{-0.3cm}
   [(p^1\!\!-\!eQA^1)\! +\!i(p^2\!\!-\!eQA^2) ] G_1(\bb r_\perp\!)\! =\!
   [ m_T\! + \!m]G_2(\bb r_\perp\!)\! .\hspace*{0.75cm}
   \label{eq52b}
\end{eqnarray}
\end{subequations}
Manipulating these two equations, we obtain 
\begin{subequations}
\label{62}
 \begin{eqnarray}
  &&\hspace*{-0.4cm} [(p^1\!\!-\!eQA^1\!) \!+\!i(p^2\!\!-\!eQA^2\!) ]   [ (p^1\! \!-\! eQ A^1\!)\!- \!i(p^2\!\!-\!eQA^2\!)]G_2(\bb r_\perp\!) 
\nonumber\\
&& \hspace*{1.0cm}
 \!=\! [  m_T ^2\! -\!  m^2 ]G_2(\bb r_\perp) ,~~~~~~~~~
 \\
&&\hspace*{-0.4cm}[ (p^1\!\! -\!eQ A^1\!)\!- \!i(p^2\!\!-\!eQ A^2\!)]  [(p^1\!\!-\!eQ A^1\!) \!+\!i(p^2\!\!-\!eQ A^2\!) ]G_1(\bb r_\perp\!) 
\nonumber\\
&& \hspace*{1.0cm}
\!=\! [  m_T^2 \! -\!  m^2 ] G_1(\bb r_\perp) ,~~~~~~~~~
\end{eqnarray}
\end{subequations}
which are separate equations for $G_2(\bb r_\perp)$ and $G_1(\bb
r_\perp)$.  These two equations lead to
 \begin{subequations}
 \begin{eqnarray}
  &&\biggl [  p_T^2+ [(\frac{eQ B_0}{2})^2 r_\perp^2
   +(-eQ)B_0 L^3
   + eQ B_0    \biggr  ] G_2 (\bb r_\perp)
\nonumber\\
&& \hspace*{2.5cm} =  [  m_T ^2 -  m^2 ]G_2(\bb r_\perp) , 
   \\
  &&\biggl [  p_T^2+(\frac{eQB_0}{2})^2 r_\perp^2 
          +(-eQ)B_0 L^3
     -   eQ B_0    \biggr  ] G_1(\bb r_\perp) 
\nonumber\\
&& \hspace*{2.5cm} 
=  [  m_T ^2 -  m^2 ]G_1(\bb r_\perp) ,
  \end{eqnarray}
  \end{subequations}
where $p_T^2$=$(p^1)^2+(p^2)^2$, $L^3$ is the angular momentum along
the longitudinal 3-axis.  The eigenvalues and eigenfunctions for these two equations of
$G_2$ and $G_2$ can be obtained separately.  For $G_2$, we have
\begin{eqnarray}
\hspace*{0.cm}m_T^2&&\!\!\!=\!m_{T2}^2 
\nonumber\\
&&\hspace*{-0.6cm}= \!m^2  \!\!+\!\!2(\hbar \omega\!)^2 (2n_2\!\!+\!\!|\Lambda_2|\!+\!1\!-\!\Lambda _2\sg Q \! \!+\!\! \sg Q \!)\!,
\label{eq55}
\end{eqnarray}
 For $G_1$, we have
\begin{eqnarray}
\hspace*{0.0cm}
m_T^2&&\!\!\!=\!m_{T1}^2 
\nonumber\\
&&\hspace*{-0.6cm}=  \!m^2 \! \!+\! \!2(\hbar \omega\!)^2 (2n_1\!  \!+ \!\!|\Lambda_1| \! + \!1- \!\Lambda _1\sg Q \!  \!-\!  \!\sg Q \!)\!.
\label{eq56}
\end{eqnarray}
The eigenfunctions are
\begin{subequations}
\label{67}
\begin{eqnarray}
&&\hspace*{-0.2cm}G_{1} (\bb r_\perp)  
\!=\! C_1 e^{ i \Lambda_1 \phi} e^{-\omega^2r_\perp^2 /2} (\omega r_\perp)^{|\Lambda_1|}L_n^{(| \Lambda_1| )}(\omega^2r_\perp^2) ,
\label{eq57a}\\
&&\hspace*{-0.2cm}G_2   (\bb r_\perp) 
\!=\! C_2 e^{ i \Lambda_2 \phi} e^{-\omega^2r_\perp^2/2} (\omega r_\perp)^{|\Lambda_2|}L_n^{(| \Lambda_2| )}(\omega^2 r_\perp^2) ,
\label{eq57b}
\end{eqnarray}
\end{subequations}
where  $(\hbar \omega)^2$ =$|eQ B_0|/2$.  

Among the solutions of the transverse equations of motion, we are interested in the 
 Landau level zero mode which represents the state with the  
lowest transverse energy  $m_T^2=m^2$ or  $|m_T|$= $m$ of the quark-antiquark system.
For the quark with $Q=1$ and  $\sg Q=1$, the transverse eigenvalue equation (\ref{eq56}) gives 
the zero mode  as the state with 
$n_1=0$, $\Lambda_1$=0, $m_T^2$=$m$, and $m_T=m$.   The transverse eigenfunction is given by Eq. (\ref{eq57a}) as
\begin{eqnarray}
G_1({\bb r}_\perp) = C e^{-\frac{\omega^2 r_\perp^2}{2}}.
\end{eqnarray}
The other component of the transverse wave function $G_2(\bb r_\perp)$ can be obtained from 
$G_1({\bb r}_\perp^2) $ and Eq.\ (\ref{eq52b}).  Equation (\ref{eq52b}) gives for the quark at the zero mode   
the transverse wave function component
\begin{eqnarray}
G_2(\bb r_\perp)=0.
\end{eqnarray}

For the antiquark with $Q=-1$ and  $\sg Q=-1$, the transverse eigenvalue equation (\ref{eq55}) gives 
the zero mode  as the state with 
$n_2=0$, $\Lambda_2$=0, $m_T^2$=$m^2$,  and $m_T$=$-m$ for a (valence)  antiquark as a hole in the Dirac sea.   The transverse eigenfunction for the antiquark as given by Eq. (\ref{eq57b}) is
\begin{eqnarray}
G_2({\bb r}_\perp) = C e^{-\frac{\omega^2 r_\perp^2}{2}}.
\end{eqnarray}
The other component of the transverse wave function $G_1(\bb r_\perp)$ can  be obtained from 
$G_2({\bb r}_\perp^2) $ and Eq.\ (\ref{eq52a}).  Equation (\ref{eq52a}) gives for  antiquark  at the  zero mode 
the transverse wave function component
\begin{eqnarray}
G_1(\bb r_\perp)=0.
\end{eqnarray}

We can represent the transverse wave function on the $\{x^1,x^2\}$ plane as a two-component spinor
\begin{eqnarray}
G(\bb r_\perp)=\begin{pmatrix}
   G_2  (\bb r_\perp)   \\
G_1  (\bb r_\perp) \\
\end{pmatrix},
\label{66}
\end{eqnarray}
and introduce the spinors with quantum numbers $\sigma^3=\pm
1$, chosen to be quantized along the $x^3$ direction,
\begin{eqnarray}
\chi_{1/2}=\begin{pmatrix}
   1   \\
 0    \\
\end{pmatrix},~~~
\chi_{-1/2}=\begin{pmatrix}
  0   \\
 1    \\
\end{pmatrix},
\end{eqnarray}
 then ,
\begin{eqnarray}
\sigma^3 \chi_{\pm 1/2} = \pm  \chi_{\pm 1/2}.
\end{eqnarray}
At the zero mode, the transverse wave function  is the Gaussian wave 
function  given by 
\begin{eqnarray}
&&\hspace*{-0.2cm}G(\bb r_\perp ) 
\!=\! C e^{-\frac{\omega^2 r_\perp^2}{2}}\! \begin{cases}  \!\chi_{1/2}  ~~\text{ ~for ~quark,} \sg Q\!\! =\!\! 1, \\
                                                       \! \chi_{-1/2}  \text{ for~antiquark,} \sg Q \!\!=\! -1.
\end{cases}
\label{70}
\end{eqnarray}
Therefore, a quark and an antiquark can populate their transverse
zero-mode states in a longitudinal magnetic field with their spins aligned in  opposite directions
as shown in Fig. 2.
Their
transverse mass $|m_T|$ for longitudinal motion are the same as their
rest mass $m$.  The external magnetic field $\bb B$ has no effect on
their effective mass $|m_T|$ for longitudinal motion at the zero mode.  
\begin{figure} [h]
\centering
\vspace*{0.2cm}\hspace*{-0.3cm}
\resizebox{0.40\textwidth}{!}{
\includegraphics{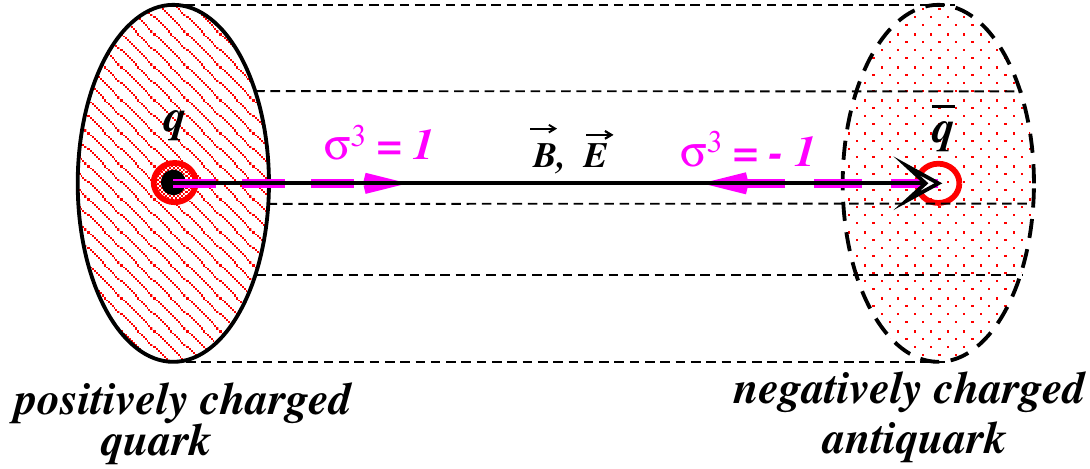}}
\caption{The lowest-energy Landau level zero-mode solution with
  $|m_T|$=$m$ for the quark and the antiquark with their spins and
  magnetic moments aligned with the magnetic field, with a total zero
  spin in a flux tube. }
\vspace*{-0.1cm}
\label{fig2}
\end{figure}

 The eigenvalue equations (\ref{eq55}) and  (\ref{eq56}) indicate that
the effective transverse mass $|m_T|$ for longitudinal motion increases with
increasing $n$ and $|\Lambda|$ quantum numbers.  For non-zero values
of $n$ and $|\Lambda|$ quantum numbers, the magnetic effect depends on
the quark rest mass $m$ and decreases as the quark mass increases. 
In all these solutions, the harmonic oscillator wave functions of
Eq.\ (\ref{67}) confine the quark and the antiquark on the transverse
plane.  The magnetic field $\bb B$ along the longitudinal direction
facilitates the transverse confinement of quarks and antiquarks on the
transverse plane.  As
the transverse quark current $j^1$ and $j^2$ are proportional to the
product $G_1^*G_2$ as well as $G_2^*G_1$ (see Eq.\ (\ref{b1})), the
transverse currents $j^1$ and $j^2$ are zero for quarks and antiquarks
in the zero mode, but are non-zero for the case when the products of
$G_1$ and $G_2$ are not zero.

Our central interest is on the lowest-energy states.  For this
purpose, it suffices to limit our attention to the transverse zero mode.  For such type of motion, the
transverse wave functions are Gaussian functions, with the quark and
the antiquark executing simple harmonic oscillator zero-point motion.
The quark the antiquark are transversely confined.  
We obtain the result that  the longitudinal $\bb B$ field
that exists at the moment of creation to confined the quark and the
antiquark  will continue to confine the quarks and the antiquarks
on the transverse planes because of Landau level dynamics.

In such a
zero mode, because of the alignment of the the magnetic field $\bb B$
and spins in the longitudinal direction, the quark transverse mass
$|m_T|$ is the same as the quark rest mass $m$.  Upon further
approximating light quarks to be massless, we have $|m_T|$=$m$=0 at the zero mode.  The
transversely confined massless quarks and antiquarks provide favorable
ingredients for the virtual pair production and vacuum polarization,
linking the valence quark, the valence antiquark, and gauge fields
with a confining interaction.

With transverse confinement well at hand and with $|m_T|$=$m$  approximated as massless, we can therefore idealizing
the flux tube in (3+1)D as a one-dimensional string and apply
Schwinger confinement mechanism  of massless fermions in QED in (1+1)D to our problem
of quarks.  Consequently, we can infer that the transversely confined
massless quarks are also longitudinally confined.  Possessing both
transverse and longitudinal confinement, such a quark-QED system will
be confined in (3+1)D space-time and may show up as a QED meson
\cite{Won10,Won20}.

From Eqs.\ (\ref{70}), the solution $G(\bb
r_\perp)$ of the transverse equations for the zero mode is in the form
of a Gaussian function which allows us to obtain the relation between
the 2D coupling constant $g_{2D}$ and the 4D coupling mass $g_\4d$,
according to Eq. (\ref{18}), we obtain
\begin{eqnarray}\label{2gmass}
&& g_{2D}^2 = g_\4d^2 \frac{\omega^2}{ 2{\pi}} = \frac{g_\4d^2}{ {\pi} R_T^2},
\label{76}
\end{eqnarray}
where we have introduced 
\begin{eqnarray}
R_T = \sqrt{2}/\omega = \sqrt{4 / (e|QB_0|)}.
\label{75}
\end{eqnarray}
The above relation (\ref{76}) coincides with previous results relating
$g_\2d$ and $g_\4d$ in \cite{Won09,Won10,Kos12,Won20}.  It is clear
from Eqs.\ (\ref{eq55}) and (\ref{eq56})  that in the zero mode, the effective quark mass
$|m_T|$ does not depend on the magnitude of the magnetic field, but the
effective flux tube radius $R_T$ is related to the magnitude of
magnetic field $B_0$ as given in Eq.\ (\ref{75}).  Upon assuming that
the radius of the flux tube is an intrinsic property of the quark, it
is reasonable to consider then that by association, the magnitude of
$B_0$ is also an intrinsic property of the quark and the antiquark.

Figure 2 indicates that along with the longitudinal magnetic field
$\bb B$, there exists a longitudinal electric field $\bb E$ between
the quark $q$ and the antiquark $\bar q$.  Thus, along the flux tube
because both the $\bb B$ field and the $\bb E$ fields are parallel to
each other, their divergence of the chiral current $j^{\mu 5}$ does
not vanish, as given by Eq.\ (19.45) of \cite{Pes95}
\begin{eqnarray}
\partial _\mu j^{\mu 5} = -  \frac{g^2}{16 \pi^2} \epsilon ^{\alpha \beta \mu \nu } F_{\alpha \beta} F_{\mu \nu }.
\end{eqnarray}
The presence of transverse confinement leads to the spontaneous breaking of chiral symmetry
and perhaps also the chiral condensate. 
It has been conjectured that chiral symmetry breaking is concomitant
with the presence of a chiral condensate and perhaps also of
confinement in QCD \cite{Cas80,Ban80,Oha21}.  

For our case with the
bonsonization of QED, there is a contribution of the quark mass term  
in Eq.\ (\ref{qed1111}) associated with non-zero quark rest masses and the chiral condensate. 
While the Schwinger  longitudinal confinement mechanism is associated with dynamical quarks, the longitudinal 
magnetic field $\bb B$ and longituidnal electric field $\bb E$ created
at the birth of the $q\bar q$ pair also facilitates longitudinal confinement, and  may be connected with the confinement effects associated  with the chiral condensate.

\subsection {Quark  current in the
  longitudinal (1+1)D space-time \label{sec4c}}

From the last subsection, we find that the quark-QED-QCD system under
consideration possesses transverse confinement in the form of a flux
tube which can be approximately idealized as a one-dimensional string
in (1+1)D space-time arena.  We wish to examine now the longitudinal
dynamics of such an idealized system.

For quarks in QCD, although QCD is a non-Abelian gauge theory  many features of the
 QCD mesons (such as quark confinement, meson states, and meson
 production) mimick those of the Schwinger model for the Abelian
 gauge theory in (1+1)D, as noted early on by Bjorken, Casher, Kogut,
 and Susskind \cite{Bjo73,Cas74}.  Such generic string feature in
 hadrons was first recognized even earlier by Nambu \cite{Nam70,Nam74}
 and Goto \cite{Got71} before the advent of the non-Abelian QCD gauge theory.  They indicate that in matters of confinement,
lowest energy  quark-antiquark bound states and hadron production, 
 an Abelian
 approximation of the non-Abelian QCD theory is a reasonable concept.
 Various nonlocal maximally Abelian projection methods to approximate
 the non-Abelian QCD by an approximate Abelian gauge theory have been
 suggested by t'Hooft \cite{tho81}, Belvedere $et~al.$ \cite{Bel79},
 Sekeido $et ~al.$ \cite{Sei07}, and Suzuki $et~ al.$ \cite{Suz08}.
 Suganuma and Ohata \cite{Sug21} investigate Abelian projected QCD in
 the maximally Abelian gauge, and find a strong correlation between
 the local chiral condensate and magnetic fields in both idealized
 Abelian gauge systems and Abelian projected QCD.

To search for the lowest energy confined $q\bar q$
states in QCD in (3+1)D with three colors, 
we shall first compactify 
QCD in (3+1)D to QCD in (1+1)D, and we then 
 adopt here a quasi-Abelian  approximation 
as carried out in \cite{Won10,Won20}.  
In QCD  in (3+1)D with an SU($N_{\rm color}$) gauge group approximated
as a U($N_{\rm color}$) in the large $N_{\rm color}$
limit,  t'Hooft showed that planar Feynman diagrams with quarks at the edges dominate, and
the QCD dynamics in (3+1)D can be well approximated as QCD in
(1+1)D \cite{tho74a,tho74b}.  Alternatively, we can justify  
the compactification of 
QCD dynamics from (3+1)D to (1+1)D 
for the
lowest energy $q\bar q$ states 
by 
the occurrence of flux tubes in QCD  lattice gauge calculations in (3+1)D, and  flux tubes can be idealized as one-dimensional strings.

For the dynamics of a $q\bar q$ pair in  (1+1)D QCD,
we need to
introduce current component   $j_a^\mu$ and gauge field components $A_a^\mu$ to describe  $j^\mu$ as $\sum_{a=1}^8  j_a^\mu t^a$ and  $A^\mu $ as $\sum_{a=1}^8  A_a^\mu t^a$ in SU(3).  
However, dynamical  variations   $\Delta j^\mu$
and $\Delta A^\mu$ in general do not commute respectively 
with  $j^\mu$ and
$A^\mu$ 
in color space, resulting in currents and gauge fields that are
complicated non-linear color admixtures.  In general,  it
is a difficult task  to look for the lowest stable boson states with these
currents and gauge fields in the 8 dimensional color generator $t^a$ space, as shown in \cite{Kut95,Arm98,Tri02,Kom20,Tri22}.

We can guide ourselves to  an easier task  of
finding the lowest-energy stable and confined quark-antiquark QCD boson states by noting that  the  color excitation  occurs at much higher energies
as revealed by the large estimated mass of the glueball.  A color excitation corresponds to a higher-order non-planar Feynman diagram in which a gluon comes from one part of the plane, propagates out of the plane, and attaches itself to another part of the plane, in the quark-antiquark space-time picture of t'Hooft \cite{tho74a,tho74b}.
We can therefore make the quasi-Abelian approximation of 
 freezing the color excitations in our search for the lowest-energy QCD states.
Consequently, as far as the lowest-energy confined   QCD $q\bar q$ states in  (1+1)D  are concerned,  it   suffices to 
consider the confinement dynamics in a single  randomly-oriented degree of freedom in the generator $t^a$ space, while we  freeze  out the other seven angular degrees of freedom.    Upon such a consideration of QCD dynamics, the color generator space comprises of only a single 
normalized unit generator 
${\tau}^1$=$\sum_{a=1}^8 n_a t^a$ with $n_a$=$2{\rm tr}(\tau^1 t^a)$
oriented in any direction of the eight-dimensional generator space,
\begin{eqnarray}\label{cosines}
 \tau^1=\sum_{i=1}^8 n_a t^a, ~~~\text{with}~~~\sqrt{n_1^1+n_2^2+...+n_8^2}=1,
\end{eqnarray}
where $n_a$ are the components of generator vectors $\tau^1$ in the
eight-dimensional $t^a$ space, with $n_a={\rm tr }\{\tau^1 t^a\}/2$
oriented randomly.  

For the quark-QCD system,
we can therefore represent the dynamics
of the QCD quark current $j^\mu$ and the QCD gauge fields $A^\mu$  
by restricting  our consideration
of the 8-dimensional $t^a$ generator space only in the 
subspace  of $\tau^1$
without varying the $\tau^1$ orientation. 
 Restricting the color dynamics variations to  the subspace of a single generator $\tau^1$  
 and fixing the orientation of the unit vector
$\tau^1$ 
is a quasi-Abelian approximation of the non-Abelian QCD dynamics
\cite{Won10,Won20}.
It is Abelian because the generator $\tau^1$ commutes with itself and also with $\tau^0$, and it is 
quasi-Abelian because it retains the 3$\times$3 matrix structure 
and it is  distinctly different from the Abelian QED interaction associated with the $t^0$ generator.
As Abelian gauge theory in (1+1)D is confining \cite{Sch62,Sch63}, 
 the dynamics of the quark-QCD systems in the 	quasi-Abelian approximation in (1+1)D will  also be confining.  It will lead 
to stable and confining collective excitations.

With the inclusion of the QED dynamics
in the quark-QED-QCD system, we then have 
\begin{eqnarray}
A^\mu (X) &=&  A^\mu_0 (X) \tau^0 + A^\mu _1 (X)\tau^1,
\nonumber\\
j^\mu (X) &=&  j^\mu_0 (X) \tau^0 + j^\mu _1 (X)\tau^1,
\label{2v}
\end{eqnarray}
where $\tau^0=t^0$ and the generators $\tau^0$ and $\tau^1$ satisfy
2{\rm Tr}$(\tau^\lambda \tau^{\lambda'}) =\delta^{\lambda {\lambda'}}
$, with $\lambda,{\lambda'}=0,1$.  Furthermore, because $\tau^0$ and $\tau^1$ commute, the
dynamics of the gauge fields in the $\tau^0$ and $\tau^1$ generator
subspace is Abelian.    In such a description in (1+1)D, the dynamics of QCD and QED
are analogous.  They are  represented by different generators, $\tau^0$ and $\tau^1$, each of   which has $C_F=C_V=1$.

The longitudinal dynamics obeys the equation of motion
Eq.\ (\ref{32}), which can also be obtained by minimizing the action
integral $A_{\2d}$ of Eq.\ (\ref{eq34}) with respect to the variation
of $\psi(X)$.  The solution of the 2D Dirac equation (\ref{32}) is in
the form
\begin{eqnarray}\label{dir-sol}
&&\hspace*{-0.2cm}  \psi (X) \!=\!\begin{pmatrix} f_+(X) \\ f_-(X) \end{pmatrix}
\nonumber\\
&&\hspace*{-0.2cm} = \{T_{l (M_0 ; M) }\exp\} \!\!\left\{  i g_{\2d}\!\!\!
\int\! \!dX^\mu\! A_\mu (X)\!\!\right\}
 \!\exp\!\left(  \! - i\omega x^0 \!\!+\! i p^3 x^3 \right)\! u ,
 \nonumber
\\
\label{eq77}
\end{eqnarray}
where  $u = \begin{pmatrix} u_+ \\ u_- \end{pmatrix}$,
and the symbol $\{T_{l (M_0 ; M) }\exp\}$ means that the integration
is to be carried out along the line on the light-cone from the point
$M_0$ to the point $M$ such that the factors in the exponent expansion
are chronologically ordered from $M_0$ to $M$.  Substituting
Eq.(\ref{eq77}) into the 2D Dirac equation (\ref{32}), we obtain
\begin{eqnarray}\label{111}
&& \hspace*{0.0cm}  \begin{pmatrix} 0 &i (\partial_0 - \partial_3 ) \\ i (\partial_0 + \partial_3 ) & 0\end{pmatrix}
\!\!\exp\left( - i\omega x^0 + i p^3 x^3 \right)\begin{pmatrix} u_+ \\ u_- \end
{pmatrix}\!
\nonumber\\
&&\hspace*{0.8cm} =\! m_T \exp\left( - i\omega x^0 + i p^3 x^3 \right)\!\! \begin{pmatrix} u_+ \\  u_- \end{pmatrix}\!,~~~~~
\end{eqnarray}
where $m_T=m$ is the quark transverse mass given by Eqs.\ (\ref{eq55}) and (\ref{eq56})
when the quark and the antiquark are in their zero-mode, and $u$ is
normalize as ${\bar u_{\pm p^3}} u_{\pm p^3}=\pm 2 m_T$=$\pm 2m$,
\begin{eqnarray}
&&   ( \omega  + p^3 ) u_- = m_T u_+, \nonumber \\
&& (\omega  - p^3 ) u_+ =  m_T u_-,
\end{eqnarray}
\begin{eqnarray}
&& m_T^2 - (\omega^2 - (p^3)^2) =0 ,
\\
&& \omega = \pm \varepsilon = \pm \sqrt {m_T^2 + (p^3)^2},
\end{eqnarray}
\begin{eqnarray}\label{UUU}
&& \omega =\varepsilon ; ~~ u_- = \frac{m_T}{ \varepsilon + p^3}u_+ ,  \\
&&\omega =- \varepsilon ; ~~ u_- = - \frac{m_T}{  \varepsilon - p^3 }u_+ .
\end{eqnarray}
Taking the normalization condition in the form 
\begin{eqnarray}
&&Tr \int d x_3 : (\psi^\dag_{2D} \psi_{2D}): = 1,
\end{eqnarray}
we find the general solution of the 2D Dirac equation 
\begin{eqnarray}\label{2dsol}
\psi_{} (X) &&= \sum\limits_{p^3} \{T_{l (M_0 ; M) }\exp\} \biggl \{ i
g_{2D} \int dX^\mu A_\mu (X)\biggr \}\nonumber
\\
&&\hspace*{0.1cm} \times\frac{1}{\sqrt{2\varepsilon }} \biggl [~ {\hat
  q}_{p^3} \begin{pmatrix} \sqrt{ \varepsilon +p^3}\\ \sqrt{
    \varepsilon -p^3}\end{pmatrix}\exp\biggl ( - iP X \biggr ) 
\nonumber\\
&&\hspace*{1.0cm} + ~ {\hat
  {\bar q} }^\dag_{p^3} \begin{pmatrix} \sqrt{ \varepsilon + p^3}\\-
  \sqrt{ \varepsilon - p^3}\end{pmatrix}\exp\biggl ( + iP X
\biggr )\biggr ],
\end{eqnarray}
where $P=(\varepsilon, p^3 )$, $X=(x^3, x^0 )$, and the hat symbols
${\hat q} _{p^3}$ and ${\hat q^\dag} _{p^3}$ denote the operators of
the creation and annihilation of quarks and antiquarks of longitudinal
momentum $p^3$, which satisfy the standard commutative relations
\cite{Pes95}.

\subsection{Self-consistent longitudinal quark currents and gauge fields }

The quark-QCD-QED vacuum is the lowest energy state of the
quark-QCD-QED system with quarks filling up the (hidden) Dirac sea and
interacting in QCD and QED interactions.  It is defined as the state
with no valence quark above the Dirac sea and no valence antiquark
below the Dirac sea.  A local disturbance in the form of the gauge
field $A^\mu (X)=\sum_{\lambda=0}^1 A^\mu_\lambda \tau^\lambda (X)$
will generate a collective excitation of the quark-QCD-QED medium.
Our task is to find out whether there can be stable and confined
collective QCD and QED excitations showing up as stable particles.

For such a purpose, we introduce the initial perturbing gauge field $A^\mu
(X)$=$\sum_{\lambda=0}^1 A^\mu_\lambda \tau^\lambda (X)$ that acts on
the quark fields. The subsequent motion of the quark fields will
generate a current which, through the Maxwell equation, will in turn
generate a gauge field.  The requirement of the self-consistency of
the initial and induced gauge fields and currents leads to the stable
collective excitations of the quark-QCD-QED system.  The collective
dynamics can sustain themselves indefinitely, if the decay channels
are assumed to be turned off for such an examination.  How do the
quark currents and gauge fields relate to each other when such
self-consistent collective excitations are attained?
 
The knowledge of the solution of the longitudinal quark field
$\psi(X)$ provides us the tool to obtain the quark current $
j_\lambda^\mu (X)$ as a function of the introduced gauge field
$A^\mu(X)$.  For simplicity, we shall consider first  the simple 
case with a single flavor $f$, whose label is sometimes implicitly
omitted for brevity of notation.  The case of many flavors and the
case with flavor mixture can be easily generalized and will be
considered in Section 9, as was carried out in \cite{Won10} and
\cite{Won20}.  The quark current is given by
\begin{eqnarray}
 && j_{\lambda f}^\mu (X) = Tr  \left\{ {\bar \psi_{f} ( X)} \gamma^\mu \tau^\lambda   \psi_{f}
 (X' )\right\}, 
\label{47a}
  \end{eqnarray}
where
$X^\prime \to X +0$,  $\lambda=0,1,$
and 
 the trace means summing with respect to colors (but not
flavors).  Substituting the $\psi$-function solution given by
Eq.(\ref{2dsol}) into the above formula, we obtain
\begin{eqnarray}
&& \hspace*{-0.3cm}j_{\lambda f}^\mu (X) \!\!\!= \!\! Tr   \!\!\!\int \!\! \frac{d p}{ (2\pi)  } \sum_{\sigma  }
\!\!\Bigg\{ \frac {P^\mu \tau^\lambda }{\varepsilon N_f} \!\! 
\biggl [n^{ \lambda}_q(\varepsilon, f\sigma)\exp\biggl ( \!\!\!- i P (X' \!- \!X ) \!\!\biggr ) 
\nonumber\\
&&\hspace*{1.8cm}+ n^{ \lambda}_{\bar q}(\varepsilon, f\sigma)\exp\biggl (  i P (X' - X ) \! \! \biggr )
\! \biggr ]
\nonumber \\
&&\hspace*{1.6cm}\times (T\exp)\biggl \{   i \sum_{\lambda'} ~g^{\lambda' }_{\2d f}     \int\limits_{X}^{X'}  A^{\lambda'}_\mu
\tau^{\lambda'}
dX^\mu \biggr \}\Bigg\}, 
\label{47aa} 
\end{eqnarray}
where $X^\prime \to X +0,~~\lambda,\lambda'=0,1,~\text{no sum in }\lambda$.
Here,  $n^{ \lambda}_q(\varepsilon, f\sigma) = \langle {\hat q}^\dag_{f
  \sigma, \lambda} ( p,\omega) {\hat q}_{f\sigma, \lambda} (p, \omega)
\rangle $ and $ n^{ \lambda}_{\bar q}(\varepsilon, f\sigma) =\langle
        {\hat{\bar q}}_{f\sigma, \lambda}^\dag(p, \omega ) {\hat {\bar
            q}}_{f \sigma, \lambda}(p, \omega) \rangle$ 
are the quark
        and antiquark occupancy numbers, respectively,  $\sigma$ is the spin along the longitudinal direction for massive quarks and is the helicity quantum number for massless quarks, and $N_f$ is the flavor
 number.  The above current (\ref{47a}) contains the Schwinger gauge-invariance factor  
\begin{eqnarray}
 &&
(T\exp) \left\{   i ~\sum_{\lambda'}     \int\limits_{X}^{X'}  g^{\lambda'}_{\2d f} A^{\lambda'}_\mu \tau^{\lambda'}
dX^\mu \right\},
\end{eqnarray}
where $g^0_{f}=g^0_{\2d\,f}=Q_f^\qedu g_{\2d f}^\qedu$ and
$g^1_f$=$g^1_{\2d\, f }$=$Q_f^\qcdu g_{\2d f}^\qcdu$ are the QED and
QCD coupling constants.  The quantities $Q_f^\qedu$ and $Q_f^\qcdu$
are the electric and color charge numbers as given by
Eq.\ (\ref{eq2cc}).  We expand the operator exponent in the last
equation as a series with respect to $( X' - X )\to$+0,
\begin{eqnarray}\label{eq57}
&&\hspace*{-0.2cm}(T\exp) \biggl \{  i g^{\lambda'}_{\2d f} \tau^{\lambda'}   \int\limits_{X}^{X'}  A^{\lambda'}_\mu dX^\mu  \biggr \}
~~~~~~~~\label{49a}\\
&&\hspace*{0.0cm}
=1 +  i g^{\lambda'}_{\2d f} \tau^{\lambda'}   (X' - X )^\mu A^{\lambda'}_\mu (\xi)
\nonumber\\
&&\hspace*{0.2cm}+ \frac{i}{2}g^{\lambda'}_{\2d f} \tau^{\lambda'}  (X' - X )^\mu (X' - X )^\nu
\partial_\nu
A^{ \lambda'}_\mu (\xi) \nonumber \\
&&\hspace*{0.2cm}- \sum_{\lambda''}  g^{\lambda'}_{\2d f} g^{\lambda'' }_{\2d f}(\tau^{\lambda'} \tau^{\lambda''} ) ( {\tilde X}' - {\tilde X} )^\mu ( X'
- X )^\nu
\nonumber \\
&&  \hspace*{3.8cm}\times A^{\lambda'}_\mu ({\tilde \xi } ) A^{\lambda''}_\nu (\xi) \Theta({\tilde
\xi } -  \xi  ), 
\nonumber
\end{eqnarray}
where ${\tilde \xi } \in [ {\tilde X} , {\tilde X}' ],$ and $\xi \in [
  X , X' ]$ and the step function $\Theta({\tilde \xi } - \xi )$ comes
from the $T$-exponent.  Taking the limits $( {\tilde X}' - {\tilde
  X})\to+ 0 $ and $( X' - X )\to +0 $, we have to take into account
the limit 
\begin{eqnarray}\label{eq58}
 && {( {\tilde X}' - {\tilde X} )\over  (  X' -
X )} \to+ 0,
\label{222}
\end{eqnarray}
due to the step function.  As a consequence, the last term in the
expansion in Eq.\ (\ref{49a}) is equal to zero.  

Upon placing the quark and the antiquark in a longitudinal magnetic
field along the transverse direction, we find earlier in the last
subsection that a quark and an antiquark in their lowest transverse
energy state lie at the zero-mode orbitals with a transverse mass
$|m_T|$=$m$.  

We consider the approximation to treat the quarks as
massless, with $m=0$. The idealized longitudinal dynamics corresponds
then to the dynamics of massless charges in QED as in Schwinger QED in
(1+1)D \cite{Sch62,Sch63}.  Taking into account that $P^\mu =
(\varepsilon = |p|, p)$, and substituting $(T\exp) \{ i \sum_{\lambda'
}g_{2D f}^{\lambda'} ~ \tau^{\lambda'} \int\limits_{X}^{X'} A_{\lambda
  ` f}^\mu dx_\mu \}$ given by Eq. (\ref{eq57}) into Eq.(\ref{47a}),
we obtain for $(X'-X)\to +0$
\begin{eqnarray}
j^\mu_{\lambda f} (X\!) &&\!\!= \!\!  \frac{2i g^\lambda_{\2d f}}{N_{f}} Tr \!\sum_{\sigma  }\!\!\!\int \limits_{-\infty}^{+\infty} \!\!\frac{P^\mu d p}{ (2\pi) \varepsilon  } 
\Biggl\{ \!
   \biggl ( \!- \!{\partial  \exp
\bigl ( \!- i P (X' \!- \!X ) \!\bigr ) \over \partial P^\nu }   \biggr ) 
\nonumber\\
&& \hspace*{-0.8cm}\times \biggl (n_q(\varepsilon, f\sigma) + n_{\bar q}(\varepsilon, f\sigma) \biggr )\nonumber \\
&& \hspace*{-0.8cm}\times \sum_{\lambda'} \biggl (\!\tau^\lambda  \tau^{\lambda'}  A_{\lambda' f}^\nu (\xi) \!+\!
 {1\over 2}\tau^\lambda  \tau^{\lambda'}
(X' - X )_\eta
\partial^\nu
A_{\lambda' f}^\eta (\xi)\! \biggr ) 
 \!\!\Biggr\} .
\nonumber
 \label{93}
  \end{eqnarray} 
To evaluate the term $(X' - X )_\eta \partial^{\mu}_{(X)}
A^\eta_{\lambda' f}(\xi )$, we note that when we apply the partial
derivative $\partial^2\equiv\partial_{(X)}^2$ onto $(X'-X)_\eta
\partial^{\mu}_{{(X)}} A_{\lambda' f}^\eta (\xi)$, we get
\begin{eqnarray}
&&\hspace*{-0.2cm}\partial_{(X)}^2 \lim \limits_{X' \to~
X} \biggl \{ (X' - X )_\eta \partial^{\mu}_{(X)} A_{\lambda' f}^\eta (\xi )
\biggr\} 
\nonumber\\
&&\hspace*{0.0cm}=\lim \limits_{X' \to~ X}
\partial_{(X)}^2 \biggl \{ (X' - X )_\eta
\partial^{\mu}_{(X)}  A_{\lambda' f}^\eta(X)
 \biggr \}
\nonumber\\
&&= \lim \limits_{X' \to~ X} \biggl \{ - 2
\partial_{\eta  (X)}
\partial^{\mu}_{(X)} A_{\lambda' f}^\eta  (X)
\nonumber\\
&&\hspace*{0.8cm} + (X' - X)_\eta
\partial_{\kappa(X)} ~ \partial_{(X)}^\kappa \{  \partial^{ \mu}_{ (X)} A_{\lambda' f}^\eta (X) \}\biggr \}. 
\label{94}
  \end{eqnarray}
Upon taking the limit $X' \to X+0$, the second term vanishes.
Therefore, we have in the limit of $X' \to X+0$,
\begin{eqnarray}\label{cc1}
\lim \limits_{X' \to~ X} \!\!\left( (X' - X )_\eta
\partial^\mu A_{\lambda' f}^\eta (\xi) \right) \!= \!- 2
\partial^\mu \frac{1}{\partial^2} \partial_\eta 
A_{\lambda' f}^\eta (X).  ~~~~
\end{eqnarray}
As a consequence, after integration by parts and calculating the trace
with respect to the color indices in Eq.\ (\ref{94}) we obtain 
\begin{eqnarray}\label{cc1}
j_{\lambda f}^\mu (X) =\left( A_{\lambda f} ^\nu (X) -  
\partial^\nu \frac{1}{\partial^2} \partial_\eta 
A_{\lambda f}^\eta (X)\right){\cal S}^\mu_\nu, 
\label{eq70}
\end{eqnarray}
\begin{eqnarray}
 {\cal S}^\mu_\nu \!=\! g^\mu_\nu   Q^\lambda_{\2d f} \!\! \sum_{\sigma} \!\!\! \int \limits_{-\infty}^{+\infty}
\!\!\frac{dp}{(2\pi)\varepsilon (p) N_{f}}  [ n_{ q}(\varepsilon,f \sigma)\!+ \! n_{\bar q}(\varepsilon, f\sigma)] , ~~~~
 \label{60a}
\end{eqnarray}
where $g^\mu_\nu$ is the space-time metric tensor and a bar over the
letters denotes antiparticles. To find the occupancy number we have
to specify additionally the state in which the averaging in the last
formula is carried out.

We consider first a single particle. Since it moves either along the
longitudinal axis or in the opposite direction, we have, taking into
account the normalization of the flavor states,
\begin{eqnarray}\label{60bb}
\hspace*{0.0cm}\frac{1}{N_{f}}  \sum_{\sigma} \int \limits_{-\infty}^{+\infty} \frac{d p}{\varepsilon(p)} ~ n_{ q}(\varepsilon, f\sigma)
\! &&=\! \frac{1}{N_{f}} \sum_{\sigma} \int \limits_{-\infty}^{+\infty} \frac{d p}{\varepsilon(p)} n_{\bar q}(\varepsilon, f\sigma)
\nonumber\\
&& = 1 . ~~~~
\end{eqnarray}
Substituting Eq.(\ref{60bb}) into Eq.(\ref{60a}) and in
Eq.(\ref{cc1}), we get
\begin{eqnarray}\label{cc11}
j_{\lambda f}^\mu (X) = \frac{g_{\2d f }^\lambda }{\pi}~\left( A_{\lambda f}^\mu (X) - 
\partial^\mu
\frac{1}{\partial^2} 
\partial_\nu
A_{\lambda f}^\nu (X)\right),
\label{77}
\end{eqnarray}
where $\lambda$=0 for QED, $\lambda$=1 for QCD.  We note specially
that in the above Eq.\ (\ref{77}), the induced current $j_{\lambda
  f}^\mu(X)$ of type $\lambda$ and flavor $f$ depends only on the
initial gauge field $A_{\lambda f }^\mu(X)$ of same type $\lambda$ and
flavor $f$, and it does not depend on the gauge field of the other
type, even though both types of gauge fields are present in the sum
over $\lambda'$ in the Schwinger gauge-invariant exponential factor.
Such a dependence (and independence) arises because in obtaining the
above result in Eq.\ (\ref{77}), we expand the Schwinger
gauge-invariant exponential factor in Eq.\ (\ref{49a}), and take the
trace over the color space involving the $\tau^{\lambda'}$ generator
from the Schwinger factor, and the $\tau^\lambda$ generator from
$T\langle \bar \psi \gamma^\mu \tau^\lambda \psi \rangle$.  Because of
the orthogonality of the color generators,
2tr$\{\tau^{\lambda'}\tau^{\lambda}\}$=$\delta^{\lambda' \lambda}$,
the contribution from the interaction of the other type is zero.

\subsection{Maxwell equation and the  longitudinal dynamics of 
  quark and gauge fields}

The equation of motion for the gauge fields $A^{\{0,3\}}_{\lambda f}(X)$
can be obtained by minimizing the action integral ${\cal A}_\2d$ in
Eq.\ (\ref{2daction}) with respect to the gauge fields $A_{\lambda
  f}^{\{0,3\}}(X)$ in question.  Such a minimization of the 2D action
integral ${\cal A_\2d}$ in (\ref{2daction}) gives the Maxwell equation for $\lambda =0,1$,
\begin{eqnarray}\label{Maxwell-1}
\partial_\nu \partial^\nu A_{\lambda f}^\mu (X)-   \partial_\nu \partial^\mu A_{\lambda f}^\nu  (X) =-
g^\lambda_{\2d f}  j_{\lambda f}^\mu (X) .
\label{100}
\end{eqnarray}
The coupled equations (\ref{77}) and (\ref{100}) provide us the tool
to determine the stable and sustainable collective current
oscillations of the quark-QCD-QED system.  To begin with, we consider
the introduction of a disturbance such as a density oscillation
$A_{\lambda f}^\mu$(initial) in space at an initial time $t_{\rm
  initial}$ as the input quantity on the right hand side of
Eq.\ (\ref{77}).  Such a disturbance will lead to an induced current
$j_{\lambda f}^\mu($induced), as given by the left hand side of
Eq.\ (\ref{77}).  After we obtain the induced current, $j_{\lambda
  f}^\mu($induced), we use it as the right hand side of
Eq.\ ({\ref{100}) as the current source to find out what the new gauge
  fields $\tilde A_{\lambda f}^\mu$(generated) that it generates.
  Stable and sustainable cycles of (current $\to$ gauge field $\to$
  current $\to$ gauge field) will be possible, if the generated gauge
  field $\tilde A_{\lambda f}^\mu$(generated) behaves and oscillates
  in the same way as the initial disturbance $ A_{\lambda
    f}^\mu$(initial).  That is, stable self-consistent collective
  dynamics of the quark field and the gauge fields are attained when
  the newly generated gauge fields $\tilde A^\mu_{\lambda
    f}$(generated) are the same as the initial applied gauge fields
  $A^\mu_{\lambda f}$(initial) to start with.  Such a self-consistency
  can be achieved by setting $\tilde A^\mu_{\lambda
    f}$(generated)=$A^\mu_{\lambda f}$(initial) and substituting
  Eq.\ (\ref{100}) into Eq.\ (\ref{77}).  We get the Klein-Gordon
  equation for the currents $j_{\lambda f} ^\mu$
\begin{eqnarray}
&&-\partial^\nu \partial_\nu j_{\lambda  f} ^\mu   = m_{\lambda f}^2 j_{\lambda  f} ^\mu ,
\label{92}
\end{eqnarray}
and similarly the Klein-Gordon equation for the gauge field  $A_{\lambda  f} ^\mu$
\begin{eqnarray}
&&-\partial^\nu \partial_\nu A_{\lambda  f} ^\mu   = m_{\lambda f}^2 A_{\lambda  f} ^\mu , 
\end{eqnarray}
which correspond to the occurrence of a boson of a stable and
independent collective excitation of the quark-QCD-QED medium, with a
mass $m_{\lambda f}$ given by
\begin{eqnarray}\label{mass_a}
&&m_{\lambda f}^2 = \frac{( g^{\lambda }_{\2d f})^2 }{\pi}
\begin{cases}
  \lambda=0 & \text{for QED}, \cr
  \lambda=1 &\text{for QCD}. \cr 
\end{cases}.
\label{b15}
\end{eqnarray}
From another perspective, the separation and the independence of the
color-singlet and the color-octet excitations are possible because the
gauge-invariant relations between the charge currents $j^\mu$ and the
gauge fields $A^\mu$ in (1+1)D in Eqs. (\ref{77}) and (\ref{100})) are each a linear
function of $j^\mu$ and $A^\mu$.  As a consequence, there is a
principle of the superposition of the currents and the gauge fields of
different color components in color-space.  Thus, the quark-QCD-QED
medium possesses stable and independent collective QCD and QED
excitations with different bound states masses $m_\lambda$, depending
on the coupling constants $g^\lambda _f$.  From Eq.\ (\ref{b15}), the
ratio of the QED meson mass to the QCD meson mass is of order
\begin{eqnarray}
\frac{m_{\qedu\,{\rm meson}}}{m_{\qcdu\,{\rm
    meson}} }
    \sim \frac{g^\qedu_f}{g^\qcdu_f} 
    \sim \!\sqrt{\frac{\alpha_{{}_\qedd}}{\alpha_{{}_\qcdd}}}
    \sim \!\sqrt{\frac{1/137 }{0.6 }}
     \sim \frac{1}{9}.
\label{b16}
\end{eqnarray}

\section{Low energy QED and QCD excitations in  quark-QCD-QED  systems}

Up to now, we have been dealing with a single flavor.
We proceed now to examine the case with two flavors.  How the flavor
degrees of freedom manifest themselves in QCD and QED mesons will
depend on their underlying symmetry of the system.  For the quark-QCD
system, the isospin symmetry is known to be a good symmetry.  We
shall first consider the system if there is isospin symmetry.

If isospin is a good symmetry, then the QCD and the QED meson states
$\Phi^\lambda_I$, with quantum numbers $I$=0 and 1, are given by
\begin{eqnarray}
\Phi^\lambda_I= \frac{1}{\sqrt{2}} \left ( |u \bar u\rangle^\lambda  + (-1)^I |d \bar d\rangle ^\lambda\right ) \equiv \sum_{f=u,d}  D^\lambda _{If}|f {\bar f}\rangle.
\label{106}
\end{eqnarray}
More explicitly, they are related by
\begin{eqnarray}
\begin{pmatrix}
 \Phi_0^\lambda\\ \Phi_1^\lambda
\end{pmatrix}
\!\!=\!\!\begin{pmatrix} \frac{1}{\sqrt{2} } &  \frac{1}{\sqrt{2} } \\
  \frac{1}{\sqrt{2} } &  -\frac{1}{\sqrt{2} } \\
\end{pmatrix}
\!\!
\begin{pmatrix}
 |u \bar u\rangle \\  |d \bar d\rangle
\end{pmatrix},
~~~
D^\lambda = \begin{pmatrix} \frac{1}{\sqrt{2} } &  \frac{1}{\sqrt{2} } \\
  \frac{1}{\sqrt{2} } &  -\frac{1}{\sqrt{2} } \\
\end{pmatrix}\!\!.~~~
\end{eqnarray}

As $|u \bar u\rangle$ arise from the quark currents given by\break
$j^\mu_{\lambda
  u}$=$\langle T(\psi_u \gamma^\mu \tau^\lambda \psi_u)\rangle)$ and
$|d \bar d\rangle$ by $j^\mu_{\lambda d}$=$\langle T(\psi_d
\gamma^\mu \tau^\lambda \psi_d)\rangle)$ of the up and down flavor
quarks,
the above relation  (\ref{106}) express also the relation
between the isospin
currents and the flavor currents,
\begin{eqnarray}
j^\mu_{\lambda I} =  \frac{1}{\sqrt{2}} \left (  j^\mu_{\lambda u}  + (-1)^I j^\mu_{\lambda d} \right ).
\end{eqnarray}
The coupling constant for the state with isospin $I$ is the weighted
sum of the coupling constants of the quark flavors given from the
above by
\begin{eqnarray}
g^\lambda_I = \frac{g^\lambda_u +(-1)^I g ^\lambda_d }{\sqrt{2}} .
\end{eqnarray}
Because $g^\lambda_f = g^\lambda Q^\lambda_f$ where $Q^\lambda_f$ is the charge number
of the quark with the flavor $f$, we have 
\begin{eqnarray}
g^{\lambda}_{ I} = \frac{ g^\lambda(Q^\lambda_u+ (-1)^I Q^\lambda_d )}{\sqrt{2}}.
\end{eqnarray}
The presence of the isospin symmetry makes it possible to project out
the currents and the gauge fields for  states with a  good isospin
symmetry.  Limiting ourselves to  neutral systems, we introduce generators
$\sigma^{\{0,1\}}$ operating in the flavor space
\begin{eqnarray}
\sigma^0=\frac{1}{\sqrt{2}}
\begin{pmatrix}  1  & 0 \cr
                                 0  &  1 \cr
                                  \end{pmatrix},  
  ~~~\text{and}~~~\sigma^1=\frac{1}{\sqrt{2}}
\begin{pmatrix}  1  & 0 \cr
                                 0  &  - 1\cr
                                  \end{pmatrix},                                                  
\end{eqnarray}                            
with ${\rm tr}(\sigma^I \sigma^{I'})=\delta^{I I'}$.  We can expand
the total currents and the gauge fields as a sum of the isospin
components as
 \begin{eqnarray}
&&j^\mu_\lambda(x) =\sum_{I=0}^1 j^\mu_{\lambda I}(x) \sigma^I ,
\\
&&A^\mu_\lambda(x) =\sum_{I=0}^1 A^\mu_{\lambda I}(x) \sigma^I .
\end{eqnarray}
We need to evaluate the isospin current $j_{\lambda I}^\mu (x)$ as a
function of the applied gauge fields $A^\mu(x)$ which has all the
components in $\lambda$ and $f$, $A^\mu(x)=\sum_{\lambda' f'}
A_{\lambda' I'}^\mu(x) \tau^{\lambda'} \sigma^{I'} $,
\begin{eqnarray}
&&j_{\lambda I}^\mu (x) 
\nonumber\\
&&
 \!\!= \!\!{1\over 2} \biggl \lbrace\!\! \lim_{{x^0 = {x^0}'} \atop
  {x^1={x^1}'-\epsilon}} \!\!\!+\!\!\! \lim_{{x^0 = {x^0}'} \atop
  {x^1={x^1}'+\epsilon}} \!\!\biggr \rbrace
Tr \biggl [e^{i\int_{x'}^x
    \sum_{\lambda' I' }^1 (-g^{\lambda' }_{I'})  A_{\lambda' I'}^\mu (\xi)\tau^{\lambda'}
   \!\! \sigma^{I'} d\xi_\mu} 
\nonumber\\
&&\hspace*{0.8cm}\times 
  \langle T({\bar \psi} (x')\gamma^\mu \tau^\lambda \sigma^I \psi (x) \rangle \biggr ].~~
\end{eqnarray}
To evaluate the above current, we carry out calculations similar to
the evaluation of $j^\mu_\lambda$ in Eq.\ (\ref{77}).  We expand out
the exponential Schwinger gauge-invariance factor.  We take the
generators $\tau^{\lambda '}$ and $\sigma^{I'}$ from the Schwinger
gauge-invariance factor, and $\tau^\lambda$ and $\sigma^I$ from
$T({\bar \psi} (x')\gamma^\mu \tau^\lambda \sigma^I \psi (x) \rangle$.
We then take the trace in color and flavor.  Because of the
orthogonality condition in taking the trace, we get $\delta^{\lambda'
  \lambda} \delta^{I' I}$ when we take the trace over the colors and
the flavors.  Consequently, upon evaluating the isospin-dependent
quark current $j^\mu_{\lambda I}$, we obtain for the color type
$\lambda$ and isospin $I$,
\begin{eqnarray}
j_{\lambda I}^\mu (x)=  \frac{g^{\lambda}_{ I}}{\pi}\left ( A_{\lambda I}^\mu(x)  
- \partial ^\mu \frac{1}{\partial^2}  \partial _\nu A^\nu_{\lambda I} (x) \right ).
\label{b18}
\end{eqnarray}
On the other hand, the Maxwell equation for the gauge field
$A_{\lambda I}^\mu$ for the quark current course $ j_{\lambda
  I}^\mu(x)$ is
\begin{eqnarray}\label{Maxwell-1}
\partial_\nu \partial^\nu A_{\lambda I }^\mu(x) -    \partial_\nu \partial^\mu A_{\lambda  I}^\nu (x) = - g^{\lambda}_{ I } j_{\lambda I}^\mu(x) .
\label{b20}
\end{eqnarray}
Stable self-consistent collective dynamics of the quark field and the
gauge fields of different isospin can be obtained by substituting
Eq.\ (\ref{b20}) into Eq.\ (\ref{b18}).  We get the Klein-Gordon
equation for the isospin currents $j_{\lambda I} ^\mu $
\begin{eqnarray}
&&-\partial^\nu \partial_\nu j_{\lambda I} ^\mu   = m_{\lambda I}^2 
j_{\lambda I}^\mu , 
\end{eqnarray}
and similarly  for the isospin gauge fields $A_{\lambda I}$
\begin{eqnarray}
&&-\partial^\nu \partial_\nu A_{\lambda I} ^\mu   = m_{\lambda I}^2 
A_{\lambda I}^\mu , 
\end{eqnarray}
which corresponds to the occurrence of a boson of collective isospin
excitation of the quark-QCD-QED medium, with a mass $m_{\lambda I}$
given by \cite{Won10,Won20}
\begin{eqnarray}
(m_{\lambda I})^2&&= \frac{(g^{\lambda}_{ I})^2 }{\pi}
 =  \frac{ (g^\lambda)^2}{\pi} \left [ \frac{Q_u^\lambda + (-1)^I Q_d^\lambda }{\sqrt{2} }\right ]^2,
\nonumber\\
&&
=\biggl [ \frac{Q_u^\lambda+(-1)^IQ_d^\lambda }{\sqrt{2}} \biggr ]^2\,
\frac{4\alpha_{{}_{\rm \{QCD,QED\}}}}{ \pi R_T^2} ,
\nonumber\\
&&
=\biggl [\sum_f D^\lambda_{If} Q^\lambda_f \biggr ]^2\,
\frac{4\alpha_{{}_{\rm \{QCD,QED\}}}}{ \pi R_T^2} .
\end{eqnarray} 

We note that in the massless limit, the mass of QCD and QED mesons
with $(I,I_3)$=(1,0) is given by
\begin{eqnarray}
(m_{\lambda I})^2 \!\!  = \!\! \frac{(g^\lambda)^2}{\pi}\!\! \left [ \frac{Q^{\{\qcdu,\qedu\}}_u \!+\!(-1)^I Q^{\{ \qcdu,\qedu \}}_d }{\sqrt{2}} \right ]^2\!\!.
\end{eqnarray}
For QCD, we have $Q^{\qcdu}_u $=$Q^{\qcdu}_d$=1, and thus, in the
massless limit for $\pi^0$ with $(I,I_3)$=(1,0), the mass of $\pi^0$
is zero.  Such a vanishing of the $\pi^0$ mass is consistent with the
common concept that $\pi^0$ is a Goldstone boson in QCD.  In this
description, the mass of $\pi^0$ comes only from the spontaneous
chiral symmetry breaking, which contribute a term to the square
$\pi^0$ mass \cite{Won20}
 \begin{eqnarray}
m_\pi^2 && = \Delta m_\pi^2 = \overline  m_f \langle \bar \psi \psi \rangle_{_{\rm QCD}}
\nonumber\\
&&
=
\sum_f  m_f (D^\qcdu_ {If})^2 \langle \bar \psi \psi \rangle_{_{\rm QCD}}
\nonumber\\
&&
=\frac{m_u+m_d}{2}\langle \bar \psi \psi \rangle_{_{\rm QCD}},
\label{eq111}
\end{eqnarray}
where $\langle \bar \psi \psi \rangle_{_{\rm  QCD}}$ is the chiral condensate.  We
can use the pion mass to calibrate the chiral condensate $\langle \bar
\psi \psi \rangle_{_{\rm QCD}}$.  Therefore the masses of neutral QCD mesons is
given by
\begin{eqnarray}
m_{\lambda I}^2
&&
=\biggl [\sum_f D^\lambda_{If} Q^\lambda_f \biggr ]^2\,
\frac{4\alpha_\lambda }{ \pi R_T^2} 
+m_\pi^2  \frac{\sum_f  m_f (D^\lambda_ {If})^2}{m_{ud}},
\label{112}
\end{eqnarray}
where $m_{ud}$=$(m_u+m_d)/2$.  The chiral condensate depends on the
interaction type $\lambda$, specifically on the coupling constant.  We
note that the chiral current anomaly in the chiral current depends on
the coupling constant square,  $e^2=g^2 $, as gives in Eq. (19.108) of
\cite{Pes95}
\begin{eqnarray}
\partial _\mu j^{\mu 5 3}=-\frac{e^2}{32\pi} \epsilon^{\alpha \beta \gamma\delta } F_{\alpha \beta} F_{\gamma \delta},
\end{eqnarray}
which shows that the degree non-conservation  of the chiral current is
proportional to $e^2$.  It is therefore reasonable to infer that 
the chiral condensate term  scales as  the coupling
constant square as $g^2$ or $\alpha$, just as the first term.  Hence, we have \cite{Won20}
\begin{eqnarray}
m_{\lambda I}^2&&= 
\left [\sum_{f=1}^{N_f} 
D^\lambda _{If} Q^\lambda_f  \right]^2 
\frac{4\alpha_{{}_{\rm \{QCD,QED\}}}}{ \pi R_T^2}
\nonumber\\
&&~~~~
+ m_\pi^2\frac{\alpha_{{}_{\rm \{QCD,QED\}}}}{\alpha_\qcdd} 
 \frac{\sum_f^{N_f}  m_f (D^\lambda_ {If})^2}{m_{ud}}.
\label{qed1111}
\end{eqnarray}
Although the above mass formula is  given for two flavors, we have written
the above result in a general form that it is also applicable to QCD
mesons with three flavors \cite{Won20}.  For the $\pi^0,\eta,$ and
$\eta'$ the degree of admixture is known as stated in the Particle
Data Book.  Explicitly, it is given for $\Phi^{\qcdu}_I$=$\sum_f^{N_f}
D_{If} |f \bar f\rangle$, and $\Phi_1=\pi^0$, $\Phi_2=\eta$, and
$\Phi_3=\eta'$, by
\begin{eqnarray}
\begin{pmatrix}
 \!\small \Phi_1 \!\!\\  \!\small \Phi_2 \!\!\\ \! \small \Phi_3  \!\!
\end{pmatrix}
=D
\begin{pmatrix}
 |u \bar u\rangle  \!\!\\  |d \bar d\rangle  \!\!\\ |s \bar s\rangle  \!\!
\end{pmatrix},
\end{eqnarray}
where the $D$ matrix is 
\begin{eqnarray}
D=
\!\!\begin{pmatrix} \frac{1}{\sqrt{2} } & - \frac{1}{\sqrt{2} } &
0  \\
\frac{  \cos \theta_P \!-\!\sqrt{2}  \sin\theta_P}{\sqrt{6}} & \frac{ \cos \theta_P \!-\!\sqrt{2} \sin \theta_P}{\sqrt{6}} 
& \frac{-2\cos \theta_P \!-\! \sqrt{2}\sin\theta_P}{\sqrt{6}} \\
 \!\frac{\sin \theta_P \!+\! \sqrt{2}\cos \theta_P}{\sqrt{6}} & \frac{\sin \theta_P \!+\! \sqrt{2}\cos \theta_P}{\sqrt{6}}
 & \frac{-2\sin \theta_P\! +\! \sqrt{2}\cos \theta_P}{\sqrt{6}}\\
\end{pmatrix}.
\nonumber
\end{eqnarray}
From the tabulation in PDG \cite{PDG19}, we find $\theta_P=-24.5^o$
and ${m_s}/m_{ud}$= 27.3$_{-1.3}^{+0.7}$.  Using the $\pi^0$ mass as a
calibration of the chiral condensate, we search for the flux tube
radius $R_T$ and the QCD coupling constant $\alpha_{\qcdd}$ that can
describe well the masses of $\eta$ and $\eta'$.  We find
$\alpha_\qcd$=0.68$\pm$0.08, and $R_T$=0.40$\pm$0.04 fm.  By
extrapolating to the QED mesons with $\alpha_\qedd$=1/137, we find an
open string isoscalar $I(J^\pi)$=$0(0^-)$ QED meson state at
17.9$\pm$1.6 MeV and an isovector $(I(J^\pi)$=$1(0^-), I_3=0)$ QED
meson state at 36.4$\pm$4.8 MeV.  The predicted masses of the
isoscalar and isovector QED mesons are close to the mass of the hypothetical X17
and E38 particles, making them good candidates for the hypothetical
X17 \cite{Kra16} and E38 \cite{Abr19} particles observed recently.

\vspace*{0.1cm}
\begin{table}[H]
\centering
\caption{The experimental and theoretical masses of neutral, $I_3$=0,
  QCD and QED mesons, obtained with the semi-empirical mass formula
  (\ref{qed1111}) for QCD and QED mesons. }
  \vspace{0.2cm}\hspace*{-0.3cm}
\begin{tabular}{|c|c|c|c|c|c|}

\cline{3-5} \multicolumn{2}{c|}{}& &Experimental&Mass\\ 
\multicolumn{2}{c|}{}&$[I(J^\pi$)] & mass & formula
 \\ 
  \multicolumn{2}{c|}{}& & & Eq.\ (\ref{qed1111})  \\ 
 \multicolumn{2}{c|}{}& & (MeV) & (MeV)  \\ 
\hline
QCD&$\pi^0$ &[1(0$^-$)] &\!\!134.9768$\pm$0.0005\!\!& 134.9$^\ddagger$  \\ 
\!\!meson\!\!& $\eta$ &[0(0$^-$)] &\!\!547.862$\pm$0.017\!\!&498.4$\pm$39.8~~  \\ 
& $\eta'$  &[0(0$^-$)] & 957.78$\pm$0.06& 948.2$\pm$99.6~\\ \hline
QED& X17&[0(0$^-$)] &\!\!16.94$\pm$0.24$^\#$& 17.9$\pm$1.5  \\ 
\!\!meson\!\!&E38&[1(0$^-$)] &37.38$\pm$0.71$^\oplus$ & 36.4$\pm$3.8\\ 
\hline
\end{tabular}

\vspace*{0.1cm}
\hspace*{-2.55cm}$^\ddagger$ Calibration mass~~~~~~~~~~~~~~~~~~~~~~~\\
\vspace*{-0.1cm}\hspace*{-0.90cm}$^\#$A. Krasznahorkay $et~al.$, arxiv:2104.10075 ~~~~\\
\hspace*{0.30cm}$^\oplus$\,K. Abraamyan $et~al.$, EPJ Web Conf       204,08004(2019)\\
\label{tb1}
\end{table}

For the quark-QED system, we will need future experimental information
to ascertain whether there is an isospin symmetry.  If the quark-QED
system does not have an isospin symmetry and allows flavor to be a
good quantum number, then there will be a $|u\bar u\rangle$ QED meson
and a $|d\bar d \rangle $ QED meson with masses given in the massless
quark limits by Eq.\ (\ref{b15}), where $g^\qedu_u$=
$g^\qedu$$|Q^\qedu_u|$=$g^\qedu$2/3 and $g^\qedu_d$=
$g^\qedu$$|Q^\qedu_u|$=$g^\qedu/3$.  For a flux tube radius of 0.4 fm,
the masses of the $|u\bar u\rangle$ and $|d\bar d \rangle $ QED meson
masses can be estimated to be 31.7 MeV and 15.85 MeV, respectively, in
the massless quark limit.  When the quark rest mass corrections with
the chiral condensate term are taken into account as given in
Eq.\ (\ref{qed1111}), the masses of the $|u\bar u\rangle$ and $|d\bar
d \rangle $ pure flavor QED meson states are modified to be 34.7 MeV
and 21.2 MeV, respectively.  In the low-mass region, there can be 
 four possible QED meson states:  the $I=0$ and $I=1$ states possessing good isospin quantum number $I$, a  $|uu\rangle$ state, and a 
$|dd \rangle$ state.  

\section{ Summary, conclusions and discussions}

Quarks carry color and electric charges and they interact in QCD and
QED.  In quark-QCD-QED systems, there are stable collective
excitations which arise from a quark and an antiquark interacting
predominantly in QCD, with the QED interaction as a perturbation.  They
show up as QCD mesons such as $\pi^0$, $\eta$, $\eta'$.  In addition
to QCD mesons, we inquire here whether there may also be stable
low-energy collective excitations arising from a quark and an
antiquark interacting in QED alone.  If there are such stable
collective excitations, would they show up experimentally as neutral,
bound and confined QED mesons?  What would be their masses?

We have been motivated to study such a problem on account of the
observations of the anomalous soft photons
\cite{Chl84,Bot91,Ban93,Bel97,Bel02,DEL06,DEL08,Per09,DEL10}, the hypothetical X17
particle \cite{Kra16,Kra19,Kra21,Jai07}, and the hypothetical E38 particle
\cite{Abr12,Abr19,Bev11}, with masses in the region of many tens of
MeV.  They are anomalous particles because they lie outside the domain
of known Standard Model particle families.  Many speculations have
been proposed to describe these anomalous particles.  In particular,
it was proposed that quarks and antiquarks interacting in QED alone,
without the QCD interaction, may lead to neutral, confined $q\bar q$
QED mesons with masses in the region of many tens of MeV, when we
apply the well-known Schwinger's mass-(coupling constant) relation,
$m=g_\2d/\sqrt{\pi}$, to quarks interacting in QED in (1+1)D.  A
phenomenological open-string model using the bosonization method
reveals that the anomalous particles can be appropriately described as open string
composite $q\bar q$ states interacting in QED in (1+1)D
\cite{Won10,Won20}.

The proposed open-string $q\bar q$ states reside in the
(1+1)D$_{\{x^3,x^0\}}$ space-time, whereas the physical world is the
(3+1)D$_{\{x^1,x^2,x^3,x^0\}}$ space-time.  We need to find out the
physical basis how the dynamics in (3+1)D$_{\{x^1,x^2,x^3,x^0\}}$ may
be idealized as the dynamics in (1+1)D$_{\{x^3,x^0\}}$.  We also wish
to study the dynamics of quarks and gauge fields in the lowest-energy
states and the collective excitations of the quark-QCD-QED systems in
(3+1)D.

The QED mesons proposed in Refs.\ \cite{Won10,Won20} involve the
hypothesis that a quark and an antiquark can interact in QED alone
without the QCD interaction.  There are experimental circumstances in which a
quark and an antiquark pair can be produced with $m_q + m_{\bar q} <
\sqrt{s} < m_\pi$ and they can interact in QED alone \cite{Won22c}.  Furthermore, there is no theorem nor basic
physical principle that forbids a quark and an antiquark to interact
in QED alone.  What is not forbidden is allowed, in accordance with
Gell-Mann's Totalitarian Principle \cite{Gel56}.  We
introduce such a hypothesis because it has the prospect of
linking the perplexing anomalous soft photon, the hypothetical X17 particle, and
the hypothetical E38 particle together within a consistent Standard Model framework
\cite{Won20}.

The idealization of the flux tube in (3+1)D as a string in (1+1)D is
a central ingredient of  the phenomenological open string model of
\cite{Won10,Won20}.  For the QCD interaction, the flux tube
configuration in (3+1)D is well established and its idealization as a
string in (1+1)D is generally accepted to be a reasonable concept
\cite{Ven68,Nam70,Nam74,Got71,tho74a,tho74b,Cas74,Pol77,Pol81,Pol87,Art74,And83,Won94}.  Such is
however not the case for a $q$ and a $\bar q$ interacting in QED in
(3+1)D alone, without the QCD interaction.   Although the proposed phenomenological (1+1)D
open-string model in \cite{Won10,Won20} suggests the confinement of a
quark and an antiquark in QED in (3+1)D as supported by the
experimental observations of anomalous particles, present-day
lattice gauge calculations indicate on the contrary that a static
quark and a static antiquark are not confined in compact QED in (3+1)D
because they belong to the weak-coupling de-confinement regime
\cite{Wil74,Cas74,Kog75,Man75,Pol77,Pol81,Pol87,Ban77,Gli77,Pes78,Dre79,Gut80,Kon98,Arn03,Lov21,Mag20}.

The conclusion on the deconfinement of a quark and an antiquark in
QED in (3+1)D comes from lattice gauge
calculations for a static fermion charge and a static antifermion
charge as applied to quarks.  The important Schwinger's longitudinal
confinement effect for dynamical light quarks in QED
\cite{Sch62,Sch63} has not been included.  The deconfinement
conclusion from lattice gauge calculations may not be definitive for
light quarks in compact QED in (3+1)D because it 
contradicts the experimental absence of fractional charges
and is in variance with phenomenological QED-confined description of the observed anomalous particles 
\cite{Won10,Won20}.
 A definitive conclusion must
await future lattice gauge calculations with the inclusion of  the Schwinger confinement mechanism.
In this regard, there have been many recent advances in the
development of efficient methods of lattice gauge calculations in
compact QED with dynamical fermions in (3+1)D using the tensor network
\cite{Mag20}, dual presentation \cite{Ben20}, magnetic-field
digitization \cite{Bau21}, and regulating magnetic fluctuations
\cite{Kap20}.  There are also other recent advances in the studies of
$q\bar q$ flux tubes in lattice gauge theories in compact U(1) QED and
SU(3) QCD \cite{Ama13,Cos17,Car13,Bic17,Bic18}.  It will be of great
interest if they will be utilized to study the question of confinement
of dynamical quarks in compact U(1) QED in (3+1)D.

In the presence of two opposing conclusions regarding quark
confinement in QED in (3+1)D, we construct here a ``stretch (2+1)D''
flux tube model to study the important Schwinger's longitudinal
confinement mechanism  for light quarks in QED in (3+1)D.  In our study,
we note that quark confinement in (3+1)D consists of transverse
confinement and longitudinal confinement.  The transverse confinement
of light quarks is a necessary condition for the full
confinement of light quarks in QED.  
The transverse confinement property of quarks in QED depends on the
compact or non-compact nature of the QED interaction, which is related
to the quantization and the commensurable properties of the quark
electric charges \cite{Yan70}.  Compact QED interactions are confining
for opposite charges under appropriate conditions \cite{Pol77,Pol87},
whereas non-compact QED interactions are always unconfined.  Because
the quark electric charges are quantized with rational number units,
the underlying QED gauge interaction between a quark and an antiquark
should be the compact type, for which the gauge field $A^\mu$ is an
angular variable with periodic properties.
 
Polyakov \cite{Pol77,Pol87} showed previously that a static fermion
charge and a static antifermion charge in compact QED are transversely
confined in (2+1)D where the transverse confinement arises from the
periodic property of the action and the associated self-interacting
property of the gauge field $A^\mu$ as a transverse angular variable.
As confirmed by Drell $et~al.$ \cite{Dre79}, Polyakov's transverse
confinement of the fermion-antifermion pair in compact QED in (2+1)D
occurs for all non-vanishing coupling strengths, no matter how weak
\cite{Pol77,Pol87}.  Applying Polyakov's transverse confinement in
compact QED in (2+1)D for fermions to quarks, we reach
the conclusion that quarks are transversely confined in compact QED in
(2+1)D.

We can utilize the Polyakov's transverse confinement for quarks
in (2+1)D to construct a ``stretch (2+1)D'' flux tube model in (3+1)D
by duplicating the initial transversely-confined gauge
fields along the longitudinal direction. We can then study the transverse and longitudinal confinement
properties of quarks and antiquark in the constructed flux tube
structure.  
By the duplication of the initial transversely-confined
gauge fields along the longitudinal direction, we obtain a
longitudinal tube with a cylindrical symmetry and transverse
confinement in (3+1)D.  We show in the stretch (2+1)D model that the
longitudinal $\bb B$ field, which is present initially to confine the
quarks and the antiquarks on the transverse plane at their birth,
continues to confine the quarks and the antiquarks transversely,
because of the Landau level dynamics.  As a consequence, quarks and
antiquarks will be transversely confined in the stretch (2+1)D flux
tube model.

Having thus prepared a flux tube in the stretch (2+1)D model, 
we idealize the flux tube in the (3+1)D space-time as a
one-dimensional string in the (1+1)D space-time.  Schwinger's
longitudinal confining solution for massless charges in QED in (1+1)D
can be applied to our problem of the quark-QED system.
We thus find that quarks are confined  when they interact in QCD and
compact QED.  The transverse
confinement arises from the duplication of the transversely-confined
configuration of Polyakov's transverse confinement along the flux
tube.  Longitudinal confinement arises from the Schwinger's
longitudinally confinement effect of QED in (1+1)D for an idealized
flux tube.  Therefore, in the stretch (2+1)D flux tube model, there can be stable
collective excitations of the quark-QCD-QED systems involving quarks in 
the QCD or the QED interaction in (3+1)D, leading to stable QCD mesons and QED
mesons whose masses depend on their coupling constants.  
They
correspond to collective dynamics of the quark-QCD-QED medium
executing motion in the color-singlet current and the color-octet
current respectively in (3+1)D.  The validity of the stretch (2+1)D flux tube
model in compact QED can be tested by confronting its consequences
with experiments.  A phenomenological analysis of the lowest-energy
states in the flux tube model yields agreement with the observed QCD
and QED spectra \cite{Won20}, lending support to the proposed hypotheses
of quark-antiquark confinement in the QED interaction.
Our present study  with the constructed stretch (2+1)D model may not solve the problem completely, but it
may bring us one step closer to answer, at least partially, the central
question whether a quark and an antiquark are confined in QED in
(3+1)D.  Future theoretical lattice gauge calculations of QED in (3+1)D with the  inclusion of the Schwinger confinement mechanism
and  further experimental confirmation of  the  hypothetical X17 and E38 particles 
will shed definitive light
on the dynamics of quarks in the QED gauge fields in (3+1)D.

\appendix

\vspace*{0.4cm}
\noindent{\bf Appendices}

\section{ Separation of quark transverse and longitudinal
  equations of motion in (3+1)D}

To separate out transverse and the longitudinal degrees of freedom, we
follow methods used previously by Wang, Pavel, Brink, Wong, and many
others \cite{Wan88,Sai90,Sch90,Pav91,Won95}.  We employ
the Weyl representation of the gamma matrices in Eq.\ (\ref{wf}) and
write the quark field in terms of the transverse functions
$G_{\{1,2\}}({\bb r}_\perp )$ and the longitudinal functions $f_\pm(X
)$ as
\begin{eqnarray}
\Psi_\4d=\Psi_\4d(X)
&&= \begin{pmatrix}
  G_1   ({\bb r}_\perp ) f_+ (X )  \\
  G_2  ({\bb r}_\perp)  f_- (X ) \\
 G_1  ({\bb r}_\perp)  f_- (X ) \\
-G_2  ({\bb r}_\perp)  f_+(X ) \\
\end{pmatrix}   .
\end{eqnarray}
where $\bb r_\perp$=$\{x^1,x^2\}$, $X$=$\{x^3,x^0\}$.  Using the Weyl
representation of the gamma matrices as given in Eq.\ (\ref{n14}), we
obtain
 \begin{eqnarray}
\bar \Psi_\4d  \gamma^0 \Pi^0 \Psi_\4d &&\!\!\!\!=( G_1^* G_1 + G_2^* G_2 ) (  f_+^* \Pi^0f_+  + f_-^* \Pi^0f_-),
\nonumber\\
\bar \Psi_\4d  \gamma^3 \Pi^3 \Psi_\4d 
&&\!\!\!\!=  ( G_1^* G_1 + G_2^* G_2 )(  f_+^*\Pi^3 f_+  -   f_-^*\Pi^3    f_-),
\nonumber\\
 ~~~~~~\bar \Psi_\4d  \Psi_\4d &&\!\!\!\!=(G_1^*G_1 - G_2^*  G_2 )(  f_+^* f_-+   f_-^* f_+),
 \nonumber\\
\bar \Psi_\4d  \gamma^1   \Pi^1 \Psi_\4d 
&&\!\!\!\!=  ( G_1^* \Pi^1 G_2 + G_2^*\Pi^1  G_1 )(  f_+^* f_-  +   f_-^*    f_+),
\nonumber\\
\bar \Psi_\4d  \gamma^2 \Pi^2 \Psi_\4d   &&\!\!\!\!=  ( -G_1^* \,i\Pi^2 G_2 + G_2^*i\Pi^2G_1 )(  f_+^* f_-  \!+\!   f_-^*    f_+).
\nonumber
\end{eqnarray}
The quark  Lagrangian density becomes
\begin{eqnarray}
{\cal L}&&\!\!\!= \!\!\bar \Psi_\4d  \!\gamma^0\! \Pi^0\! \psi_\4d\! \!-\! \bar \Psi_\4d  \!\gamma^3 \Pi^3 \!\Psi_\4d 
\!\!\!- \!\bar \Psi_\4d  \gamma^1  \! \Pi^1 \Psi_\4d\! \! -\! \bar \Psi_\4d \! \gamma^2 \Pi^2 \Psi_\4d  
\nonumber\\
&&\hspace*{0.3cm}
- m  \bar \psi_\4d  \Psi_\4d \nonumber\\
&&\hspace*{-0.2cm}= ( G_1^* G_1 + G_2^* G_2 ) (  f_+^* \Pi^0f_+  + f_-^* \Pi^0f_-)
\nonumber\\
&&\hspace*{-0.1cm}
 - ( G_1^* G_1 + G_2^* G_2 )(  f_+^*\Pi^3 f_+  -   f_-^*\Pi^3    f_-)
\nonumber\\
&&\hspace*{-0.1cm} - ( G_1^* \Pi^1 G_2 + G_2^*\Pi^1  G_1 )(  f_+^* f_-  +   f_-^*    f_+)
\nonumber\\
&&\hspace*{-0.1cm}
 -  ( -G_1^* ~i\Pi^2~G_2 + G_2^*~i\Pi^2~ G_1 )(  f_+^* f_-  +   f_-^*    f_+)
\nonumber\\
&&\hspace*{-0.1cm} - m (G_1^*G_1 - G_2^*  G_2 )(  f_+^* f_-+   f_-^* f_+).
\end{eqnarray}
where $\Pi^\mu=p^\mu + g A^\mu(x)$.
The minimization of the action integral by  variations with respect to $f_\pm^*$ and  $G_{\{1,2\}}^*$ leads to
 \begin{eqnarray}
\delta^2 {\cal L}/\delta f_+^* \delta G_1^*
\nonumber\\
&&\hspace*{-1.8cm}
=\!( \Pi^0 \!\!-\!\Pi^3) G_1 f_+  \!\!-\! m G_1   f_- \!\! -\!(p^1\!\!- \!ip^2) G_2  f_- \!=\!0,~~
\label{a}
\\
\delta^2 {\cal L}/\delta f_+^* \delta G_2^*
\nonumber\\
&&\hspace*{-1.95cm}
=( \Pi^0 \!\!\!-\!\Pi^3) G_2 f_+  \!\!+\! m G_2   f_-\! \!- \! (\Pi^1\!+\!i\Pi^2) G_1  f_- \!=\!0,~~
\label{b}
\\
\delta^2 {\cal L}/\delta f_-^* \delta G_1^*
\nonumber\\
&&\hspace*{-1.95cm}
= 
  ( \Pi^0\!\!+\!  \Pi^3)G_1    f_- \!\! - \! m G_1  f_+ \!\! -\! (\Pi^1\!\!- \! i\Pi^2)G_2     f_+ \!=\!0 ,~~
  \label{c}
\\
 \delta^2 {\cal L}/\delta f_-^* \delta G_2^*
 \nonumber\\
&&\hspace*{-1.95cm}
= (\Pi^0\! \!+\!\Pi^3) G_2    f_- \! \!+\! m   G_2  f_+ 
 \!- \!(\Pi^1 \! \!  +  \! i\Pi^2) G_1      f_+  \!= \!0. ~~
 \label{d}
\end{eqnarray}
We sum over $G_1^*(\bb r_\perp)$$\times$(\ref{a})+$G_2^*(\bb
r_\perp)$$\times$(\ref{b}) and perform an integration over the
transverse coordinates.  Similarly, we sum over $G_1^*(\bb
r_\perp)$$\times$(\ref{c})+$G_2^*(\bb r_\perp)$$\times$(\ref{d}) and
perform an integration over the transverse coordinates.  We get
\begin{subequations}
  \begin{eqnarray}
&&\int d\bb r_\perp \left\{ G_1^*( \Pi^0 -\Pi^3) G_1 +  G_2^*( \Pi^0 -\Pi^3) G_2 \right\} f_+ \nonumber \\
&&\hspace{0.4cm} - \int d\bb r_\perp \biggl \{ G_1^*m G_1  - G_2^*m G_2 +G_1^*(\Pi^1- i\Pi^2) G_2   
\nonumber\\
&&\hspace*{1.5cm} 
+ G_2^*(\Pi^1+i\Pi^2) G_1 \biggr \} f_-=0,
\label{A8a}
\\
&&\text{and}
\nonumber\\
&&
\int d\bb r_\perp \left\{ G_1^*( \Pi^0 +\Pi^3) G_1 +  G_2^*( \Pi^0 +\Pi^3) G_2 \right\} f_- \nonumber \\
&& - \int d\bb r_\perp \biggl \{ G_1^*m G_1  - G_2^*m G_2  +G_1^*(\Pi^1- i\Pi^2) G_2 
\nonumber \\
&&\hspace*{1.5cm} 
  + G_2^*(\Pi^1+i\Pi^2) G_1 \biggr \} f_+=0.
\label{A8b}
\end{eqnarray}
\end{subequations}

From the second terms in each of the above two equation, we note that
we can separate out the longitudinal and transverse equations by
introducing the separation constant $m_T$ defined by
\begin{eqnarray}\label{aaa}
&&m_T=\int d \bb r_\perp \biggl  \{  m( |G_1 |^2 - |G_2 |^2)  
\nonumber\\
&&\hspace{0.6cm}+  G_1^*(\Pi^1- i\Pi^2) G_2)  + G_2^* (\Pi^1+ i\Pi^2) G_1  \biggr  \}.
\label{A9}
\end{eqnarray}
Because of the normalization condition 
Eq.\ (\ref{Norm-cond})
for $G_{\{1,2\}}$, the above
equation can be rewritten as
\begin{eqnarray}
&&\int d \bb r_\perp \biggl  \{ G_1^*[  -  m_T G_1+  
 m G_1  + (\Pi^1- i\Pi^2) G_2 ]
\nonumber \\
&&\hspace*{1.0cm} 
 +G_2^*[  -  m_T G_2 
 - m G_2  + (\Pi^1+i\Pi^2) G_1  ]
 \biggr  \} = 0
 \nonumber
 \end{eqnarray}
The above is zero if the two terms in the integrand are zero, and we
obtain the eigenvalue equations for the transverse functions
$G_{\{1,2\}}(\bb r_\perp)$,
\begin{eqnarray}
&&\hspace{-0.0cm} \!-\!  m_T G_1 (\bb r_\perp\! )\!+ \! 
  m G_1(\bb r_\perp\! )    \!   + (\! \Pi^1\! - \!  i\Pi^2) G_2  (\bb r_\perp\! ) \!=0,
 \nonumber\\
 &&\hspace{-0.0cm} 
   \! - \!  m_T G_2 (\bb r_\perp\! )\!-\!m G_2  (\bb r_\perp\!) \!+  (\Pi^1 + i\Pi^2) G_1 (\bb r_\perp \!)  \!=0.
\end{eqnarray}
We apply $ (\Pi^1+i\Pi^2)$ on the first equation and $ (p^1+ip^2)$ on the
second equation and we get
\begin{eqnarray}
&& \hspace*{-0.2cm} (\Pi^1\!\!+\!i\Pi^2)(\Pi^1\!\!-\! i\Pi^2) G_2 (\bb r_\perp\!)\! =\!
  (m_T\!\!-\!m) (m_T\!\!+\!m)  G_2  (\bb r_\perp\!)
 \nonumber\\
 && \hspace*{-0.2cm}
   (\Pi^1\!\!- \! i\Pi^2\!)  (\Pi^1\!\!+\! i\Pi^2) G_1(\bb r_\perp\!) \!= \!( m_T\!\! - \!m) (m_T\!\!+\!m) G_1(\bb r_\perp).
\nonumber
\end{eqnarray}
We have then
\begin{subequations}
\label{a12}
\begin{eqnarray}
&& \left \{ \Pi_T^2 + i [\Pi^2,\Pi^1] + m^2 -  m_T^2\right \} G_2 (\bb r_\perp) =  0
\label{A11a} 
 \\
 &&
 \left \{ \Pi_T^2 + i [\Pi^1,\Pi^2]+ m^2 - m_T^2 \right \} G_1(\bb r_\perp)=  0
\end{eqnarray}
\end{subequations}
where $\Pi_T^2=(\Pi^1)^2+(\Pi^2)^2$ is the square of the transverse momentum.  
 Upon the introduction of the constant of separation $m_T$,
 Eqs.\ (\ref{A8a}) and (\ref{A8b}) become
 \begin{subequations}
  \begin{eqnarray}
&&\hspace*{-0.3cm}\int\! \!d\bb r_\perp (G_1^*(\bb r_\perp\!)G_1 (\bb r_\perp\!)
\!+ \!G_2^* (\bb r_\perp\!)G_2(\bb r_\perp\!) )( \Pi^0 \!-\!\Pi^3) f_+ (X\!)
\nonumber\\
&&\hspace*{0.8cm}
- m_T f_- (X)=0,\label{A13a}
 \\
&&\hspace*{-0.3cm}\int \! \! d\bb r_\perp ( G_1^*(\bb r_\perp\!)G_1 (\bb r_\perp\!)\!+ \!G_2^* (\bb r_\perp\!)G_2(\bb r_\perp\!) )( \Pi^0 \!+\!\Pi^3)f_- (X\!)
\nonumber\\
&&\hspace*{0.8cm}
- m_T f_+(X) =0 . 
\label{A13b}
\end{eqnarray}
\end{subequations}
In the above two equations, the operators $\Pi^\mu$ with $\mu=0,3$ are
actually $\Pi_\4d^\mu $=$p^\mu + g_\4d A_\4d^\mu(\bb r_\perp,X)$ where
the gauge field $A_\4d^\mu$ is a function of the (3+1)D coordinates
$(\bb r_\perp,X) $ and $g_\4d$ is the dimensionless coupling constant
in (3+1)D space-time.  Upon carrying out the integration over the
transverse coordinates $\bb r_\perp$, the above two equations can be
cast into the Dirac equation for a particle with a mass $m_T$ in
(1+1)D space time of $X$ coordinates by identifying the terms
integrated over the transverse coordinates as the corresponding gauge
fields $A_\2d^\mu$ in the (1+1)D space-time with the coupling constant
$g_\2d$
\begin{eqnarray}
&&g_\4d \!\int \!d\bb r_\perp (G_1^*(\bb r_\perp\!)G_1(\bb r_\perp\!) \!+ \!G_2^* (\bb r_\perp\!)G_2(\bb r_\perp\!) )A_\4d^\mu(\bb r_\perp,X\!)
\nonumber\\
&&\hspace*{0.4cm}  \equiv g_\2d 
A_\2d^\mu(X),
\end{eqnarray}
where $A_\2d(X)$ is the solution of the Maxwell equation for the gauge
fields in (1+1)D space-time.  By comparing the Maxwell equations in
(3+1)D and (1+1)D space time in Appendix B, it can be shown in
Appendix B that the coupling constant $g_\2d$ is then determined by
\begin{eqnarray}
g_\2d^2\!\!= \! g_\4d^2 \!\int \!\!d \bb r_\perp (G_1^*(\bb r_\perp\!)G_1(\bb r_\perp\!) + G_2^* (\bb r_\perp\!)G_2(\bb r_\perp\!) )^{2}\!.
\end{eqnarray}
We note that the coupling constant $g_\2d$ acquires the dimension of a
mass.  By such an introduction of $A_\2d^\mu$ and $g_\2d$,
Eqs.\ (A13a) and (A13b) become
\begin{subequations}
\label{a16}
\begin{eqnarray}
&&( \Pi_\2d^0 -  \Pi_\2d ^3) f_+ (X) - m_T f_-(X)=0, 
\\
&&( \Pi_\2d^0+  \Pi_\2d^3) f_- (X) - m_T f_+(X)=0,
\end{eqnarray}
\end{subequations}
where $\Pi_\2d^\mu=p^\mu+g_\2d A_\2d^\mu(X)$, with $\mu=0,3$.
Equations (\ref{a12}), and (\ref{a16}) (which correspond to
Eqs.\ (\ref{21}) and (\ref{26a}) in Section III) are the set of
transverse and longitudinal equations of motion for the quark field
$\Psi$ in (3+1)D space for our problem.

\setcounter{equation}{0}

\section{Relation between quantities in (1+1)D and (3+1)D}

We shall examine the relation between various quantities in (1+1)D and
(3+1)D.  The time-like component of the quark current is the quark
density, and the density in (3+1)D involves the transverse spatial
distribution while the density in (1+1)D does not involve the
transverse spatial distribution.

We consider the quasi-Abelian approximation of the non-Abelian QCD gauge field in so that  the QED and QCD gauge
fields can be represented in terms of only the commuting $\tau^0$ and
$\tau^1$ components, as discussed in Section (\ref{sec4c}).  To make
the problem simple, we consider lowest-energy state systems with
cylindrical symmetry so that we can write our wave function of the
quark fields in the form of Eq.\ (\ref{wf}), and the transverse
currents $j^1$ and $j^2$ are given in terms of the basic functions
$G_{\{1,2\}}$ and $f_\pm$ by
\begin{subequations}
\begin{eqnarray}
j^1_\4d\!=\bar \Psi_\4d  \!\gamma^1    \Psi_\4d \!
&&\!\!\!=\!  ( G_1^*  G_2 \!+ \!G_2^* G_1 \!)(  f_+^* f_- \! + \!  f_-^*    f_+\!),
\\
j^{2}_\4d\!=\bar \Psi_\4d \! \gamma^2  \Psi_\4d \!  &&\!\!\!=\!  ( \!-G_1^*G_2 \!+ \!G_2^*G_1\! )(  f_+^* f_- \! \!+ \!  f_-^*    f_+\!).
\end{eqnarray}
\label{b1}
\end{subequations}
For our problem, we choose to examine the quark-QED system in which
the quarks and antiquarks reside in the zero mode with $m_T=m$, and
with the quark wave functions given by Eq.\ (66). For
these zero mode states, $G_{\{1,2\}}$ are spinors aligned in the
longitudinal direction.  As a consequence, when $G_1(\bb r_\perp)$ is
non-zero, $G_2(\bb r_\perp)$ is zero, and vice-versa.  The product
$G_1^* G_2$ and $G_2^*G_1$ are always zero for quark and antiquark in
the lowest energy zero mode states.  Therefore for these lowest energy
states the currents in the direction of $x^1$ and $x^2$ are zero.  We
assume that the transverse gauge fields $A^{\{1,2\}}$ are determined
by the transverse dynamics of their own to yield the solutions of the
transverse stationary states $G_{\{1,2\}}(\bb r_\perp)$.  Within the
cylindrical flux tube, the gauge fields are only a weak function of
the transverse coordinates so that it is necessary to consider the
Maxwell equation only for $\nu, \mu$=0,3 in (3+1)D with
\begin{eqnarray}
\partial _\mu F_\4d^{\mu \nu}&&= \partial _\mu \partial^\mu A_\4d ^{\nu}(\bb r_\perp,X) -  \partial^\nu \partial_\mu A_\4d^{\mu}(\bb r_\perp,X)
\nonumber\\
&&= - g_\4d j_\4d ^\nu (\bb r_\perp,X). ~~~
\label{ap1}
\end{eqnarray} 
With the quark field $\Psi_\4d$ as given by Eq.\ (\ref{wf}), the quark
currents in (3+1)D are given by
\begin{eqnarray}
&&\hspace*{-0.1cm} j_\4d ^0\!\!  (\bb r_\perp,X\!)\! =\!\bar \Psi_\4d \!\! \gamma^0  \Psi_\4d\!\!=\! ( G_1^* G_1 \!+\! G_2^* G_2 \!) (  f_+^* f_+  \!+\! f_-^* f_-\!),
\nonumber\\
\\
&&\hspace*{-0.1cm}  j_\4d ^3\!\! (\bb r_\perp,X\!)\!=\!\bar \Psi_\4d  \!\!\gamma^3  \Psi_\4d \!\!
= \! ( G_1^* G_1 \!+\! G_2^* G_2 \!)(  f_+^* f_+ \! - \!  f_-^*   f_-\!).
\nonumber\\
\end{eqnarray}
From the 2D action ${\cal A_\2d}$ in Eq.\ (\ref{eq34}), we can derive the Maxwell
equation in (1+1)D  as given by
\begin{eqnarray}
&&\partial _\mu F_\2d^{\mu \nu} = \partial _\mu \partial^\mu A_\2d ^{\nu}(X) -\partial^\nu \partial_\mu A_\2d^{\mu}(X)= - g_\2d j_\2d ^\nu (X),
\nonumber\label{ap2}
\\
\end{eqnarray}
for  $\nu,\mu=0,1$ in (1+1)D.
With the quark field $\psi_\2d$ as given by Eq.\ (\ref{22}), the quark
currents $j_\2d ^\nu (X)$ in (1+1)D are given by
\begin{eqnarray}
&& j_\2d ^0 (\bb r_\perp,X) =\bar \psi_\2d  \gamma^0  \psi_\2d= (  f_+^* f_+  + f_-^* f_-),
\\
&& j_\2d ^3 (\bb r_\perp,X)=\bar \psi_\2d  \gamma^3  \Psi_\2d 
=  (  f_+^* f_+  -   f_-^*   f_-).
\end{eqnarray}
Comparison of the quark currents in (3+1)D and (1+1)D in Eqs.\ (B3), (B4) (B6), and (B7) gives
\begin{eqnarray}
 j_\4d ^\mu (\bb r_\perp,X) = ( G_1^*(\bb r_\perp) G_1 (\bb r_\perp)+ G_2^*(\bb r_\perp) G_2 (\bb r_\perp) ) j_\2d ^\mu (X).
\nonumber
\label{B5}
\hspace*{-0.3cm}\\
\end{eqnarray}
Then, substituting the above equation into Eq.(B2), we obtain
for $\nu,\mu=0,3$  in (3+1)D
\begin{eqnarray}
 &&\hspace*{-0.3cm}\partial _\nu \partial^\nu A_\4d ^{\mu}(\bb r_\perp,X) -  \partial^\mu \partial_\nu A_\4d^{\nu}(\bb r_\perp,X)
\nonumber\\
&&\hspace*{0.3cm}
= - g_\4d 
 ( G_1^*(\bb r_\perp) G_1 (\bb r_\perp)+ G_2^*(\bb r_\perp) G_2 (\bb r_\perp) ) j_\2d ^\mu (X).~~~
\nonumber\\
\end{eqnarray} 
The above equation in (3+1)D would be consistent with the (1+1)D
Maxwell equation (B5) if for  $\mu=0,3$, 
\begin{eqnarray}
 A_\4d^{\mu}\!(\bb r_\perp,X\!)\!=\!\frac{g_\4d}{g_\2d} \!(G_1^*(\bb r_\perp\!)G_1 (\bb r_\perp\!)\!+\! G_2^*(\bb r_\perp\!) G_2 (\bb r_\perp\!) ) A_\2d^\mu \!(X\!).
\nonumber\hspace*{-0.3cm}\\
\end{eqnarray}
On the other hand, the equation of motion in Eqs. (A14) would require
that $g_\4d$ and $g_\2d$ are related by
\begin{eqnarray}
&&g_\4d \int d\bb r_\perp (|G_1(\bb r_\perp)|^2 + |G_2(\bb r_\perp)|^2) A_\4d^\mu(\bb r_\perp,X)
\nonumber\\
&& \hspace*{0.4cm}= g_\2d A_\2d^\mu(X).
\end{eqnarray}
The consistency of both equations (B10) and (B11) requires $g_\4d$ and $g_\2d$
to satisfy the relation
\begin{eqnarray}
g_\2d^2\!\! =\! g_\4d^2\!\! \int \!d\bb r_\perp [G_1^*(\bb r_\perp\!) G_1 (\bb r_\perp\!)\!+\! G_2^*(\bb r_\perp\!) G_2  (\bb r_\perp\!) ]^2,
\end{eqnarray}
which is Eq. (\ref{g4-2}).

\vspace*{0.4cm}\noindent
{\bf Acknowledgments}
\vspace*{0.4cm}

The
authors would like to thank Prof. Y. Jack Ng for helpful
communications.  CYW's research was supported in part by the Division
of Nuclear Physics, U.S. Department of Energy under Contract
DE-AC05-00OR22725 with UT-Battelle, LLC

\end{document}